\documentclass{JHEP3}
\usepackage[english]{babel}
\usepackage{cite}

\usepackage{epsf}
\usepackage{amsmath}
\usepackage{amsfonts}
\usepackage{amssymb}
\usepackage{psfrag,epsfig,graphicx,graphics}

\providecommand{\U}[1]{\protect\rule{.1in}{.1in}}

\def\slashchar#1{\setbox0=\hbox{$#1$}
   \dimen0=\wd0
   \setbox1=\hbox{/} \dimen1=\wd1
   \ifdim\dimen0>\dimen1
      \rlap{\hbox to \dimen0{\hfil/\hfil}}
      #1
   \else
      \rlap{\hbox to \dimen1{\hfil$#1$\hfil}}
      /
   \fi}

\def\tr{{\rm tr}}

\def\no{\noindent}

\def\bei{\begin{itemize}}
\def\ei{\end{itemize}}

\def\beeq{\begin{eqnarray}} 
\def\beqa{\begin{eqnarray}}
\def\bea{\begin{eqnarray}}

\def\eea{\end{eqnarray}}
\def\eqa{\end{eqnarray}}
\def\eeeq{\end{eqnarray}}

\def\eqar{\end{array}}
\def\beqar{\begin{array}}

\def\beas{\begin{eqnarray*}}
\def\beqas{\begin{eqnarray*}}

\def\eqas{\end{eqnarray*}}
\def\eeas{\end{eqnarray*}}

\def\beq{\begin{equation}} 
\def\be{\begin{equation}}

\def\ee{\end{equation}}
\def\eq{\end{equation}}
\def\eeq{\end{equation}}

\def\beqd{\begin{displaymath}}
\def\eeqd{\end{displaymath}}
\def\eqd{\end{displaymath}}

\def\beeq{\begin{eqnarray}} \def\eeeq{\end{eqnarray}}

\newcommand{\fin}{\end{document}}

\title{On the one loop $\gamma^{(*)}\to q\bar{q}$ impact factor and the exclusive diffractive cross sections for the production of two or three jets}

\author{R.~Boussarie\\
LPT, Universit{\'e} Paris-Sud, CNRS, Universit\'e Paris-Saclay, 91405, Orsay, France\\
Email : \email{renaud.boussarie@th.u-psud.fr}}
\author{A.~V.~Grabovsky\\
Budker Institute of Nuclear Physics, 11, Lavrenteva avenue, 630090, Novosibirsk, Russia {\em \&} \\
Novosibirsk State University, 630090, 2, Pirogova street, Novosibirsk, Russia {\em \&}\\
Department of Physics, University of Minnesota, Duluth, Minnesota, 55812\\
Email: \email{A.V.Grabovsky@inp.nsk.su}}
\author{ L. Szymanowski\\
National Centre for Nuclear Research (NCBJ), Warsaw, Poland\\
Email: \email{Lech.Szymanowski@ncbj.gov.pl}}
\author{S. Wallon\\
LPT, Universit{\'e} Paris-Sud, CNRS, Universit\'e Paris-Saclay, 91405, Orsay, France {\em \&} \\
UPMC Univ. Paris 06, facult\'e de physique, 4 place Jussieu, 75252 Paris Cedex 05, France\\
Email: \email{wallon@th.u-psud.fr}}

\abstract{
We present the calculation of the impact factor for the $\gamma^{(*)}\to q\bar{q}$ transition with one loop accuracy  in
arbitrary kinematics. The calculation was done within Balitsky's high energy operator expansion. Together with our previous result
for the  $\gamma^{(*)}\to q\bar{q} g$ Born impact factor it allows one to derive cross-sections for 2- (one loop) and 3-jet
(Born) difractive electroproduction. We write such cross sections for the  2 and 3 jet exclusive diffractive electroproduction
off a proton in terms of hadronic matrix elements of Wilson lines. For the 2-jet cross section we  demonstrate the cancellation of
IR, collinear and rapidity singularities. Our result can be directly exploited to describe the recently analyzed data on
exclusive dijet production at HERA and used for the study of jet photoproduction in ultraperipheral proton or nuclear scattering.}

\begin{document}

\pagestyle{empty}
\newpage

\mbox{}

\pagestyle{plain}

\setcounter{page}{1}
\section{Introduction}

For several decades  
diffraction has been one of the most intriguing phenomena of strong interaction. The HERA research program has shown for the first time that diffractive processes in the semi-hard regime can be measured and studied based on QCD, giving one of the main tools to access the internal dynamics of the nucleon 
in a regime of very high gluon densities\footnote{For reviews, see Refs.~\cite{Wusthoff:1999cr,Wolf:2009jm}}.
One of the most important legacies of HERA is the fact that almost 10~\%  of the $\gamma^* p \to X$ deep inelastic scattering (DIS) events contain
 a rapidity gap between the proton remnants 
and the hadrons 
coming from the fragmentation region of the initial virtual photon. This
subset of events, called diffractive deep inelastic scattering (DDIS), thus looks like
$\gamma^* p \to X  \, Y$~\cite{Aktas:2006hx,Aktas:2006hy,Chekanov:2004hy,Chekanov:2005vv,Aaron:2010aa,Aaron:2012ad,Chekanov:2008fh,Aaron:2012hua}, where $Y$ is the outgoing proton or one of its low-mass excited states, and $X$ is the diffractive final state. Apart from the inclusive DDIS data, one can further focus on more specific interesting observables, like  diffractive jet(s) production, or  exclusive meson production. 

Due to the existence of a rapidity gap between $X$ and $Y$, it is natural to describe diffraction through a Pomeron exchange in the $t-$channel between these $X$ and $Y$ states. This is a common concept underlying the various approaches to describe diffraction within perturbative QCD. 

In the collinear framework, justified by the existence of a hard scale (the photon virtuality $Q^2$ of DIS),  a QCD factorization theorem was derived~\cite{Collins:1997sr}. Similarly to DIS on a proton, here one introduces diffractive structure functions, which are convolutions of
coefficient functions with diffractive parton distributions. In this resolved Pomeron model, those distributions describe the partonic content of the Pomeron, similarly to the usual parton distribution functions for proton in DIS. 

At high energies, it is natural to model the diffractive events by a {\em direct} Pomeron contribution involving the coupling of a Pomeron with the diffractive state $X$ of invariant mass $M.$ For low values of $M^2$, $X$ can be modeled by a  $q \bar{q}$ pair, while for larger values of $M^2,$
the cross-section with an additional produced gluon, i.e. $X=q \bar{q} g,$
is enhanced. Based on such a model, with a simplified 
 two-gluon exchange picture for the Pomeron,  a good description of HERA data for diffraction could be achieved
\cite{Bartels:1998ea}. Interestingly, the $q \bar{q}$ component with a longitudinally polarized photon plays a crucial role in the region of small
diffractive mass $M$, although it is a
twist-4 contribution. It is a typical signature of such models.

In this direct Pomeron contribution, the $q \bar{q} g$ diffractive state has been studied in two particular limits. First, at large $Q^2$, a collinear approximation can be used, based on the fact that the transverse momentum of the gluon is much smaller than the transverse momentum of the emitter~\cite{Wusthoff:1995hd,Gotsman:1996ix,Wusthoff:1997fz}.
Second, for very large $M^2$, contributions with a strong ordering of longitudinal momenta are enhanced~\cite{Bartels:1999tn,Bartels:2002ri}.
These two limiting results were combined in a single model, and  applied to HERA data for DDIS in Ref.~\cite{Marquet:2007nf}.

In the present article, we pursue our program in order to get a complete next-to-leading order (NLO) description of the direct coupling of the Pomeron
to the diffractive $X$ state. 
To be more specific, the ``Pomeron'' should be here understood as  
a
QCD shockwave, in the spirit of Balitsky's high energy operator expansion~\cite{Balitsky:1995ub, Balitsky:1998kc, Balitsky:1998ya, Balitsky:2001re}.

In our previous study~\cite{Boussarie:2014lxa},
we 
computed the $\gamma^{(*)} \to q \bar{q} g$ impact factor and rederived the $\gamma^{(*)} \to q \bar{q}$ impact factor, both at tree level\footnote{Here the photon can be either on-shell of off-shell, hence the notation $\gamma^{(\ast)}$}. In the present article\footnote{Partial results of the present study were already presented in Refs.~\cite{Boussarie:2015qet,Boussarie:2015acw}.},  we complete the results of Ref.~\cite{Boussarie:2014lxa} by a study of the virtual contributions,
and compute the one-loop $\gamma^{(*)} \to q \bar{q}$ impact factor.
We emphasize that in these results, the impact factors are computed without any soft or collinear approximation for the emitted gluon, in contrast with the results reported in the literature. This thus presents an important step towards a consistent description of  inclusive DDIS, or exclusive two--jet diffractive production, in the fragmentation region of the scattered photon, i.e. in
the forward rapidity region, with NLO precision. Since the results we derive are obtained in the QCD shock-wave approach, and depend on the total available center-of-mass energy, the present framework
is rather general and can have many applications. Indeed, below the saturation regime, one might describe 
the $t-$channel exchanged state in the linear BFKL regime~\cite{Fadin:1975cb, Kuraev:1976ge, Kuraev:1977fs, Balitsky:1978ic}, here with NLO precision~\cite{Fadin:1998py,Ciafaloni:1998gs}.
At higher energies, beyond the saturation limit, 
the Wilson-line operators, whose matrix element describes the $t-$channel exchanged state,  
 evolve with respect to rapidity according to the Balitsky hierarchy. In the case of a dipole operator, it reduces to the Balitsky-Kovchegov (BK) 
equation~\cite{Balitsky:1995ub, Balitsky:1998kc, Balitsky:1998ya, Balitsky:2001re, Kovchegov:1999yj, Kovchegov:1999ua} in the large $N_c$ limit. 

In the present paper we calculate the matrix element for the $\gamma^{(*)}\to q\bar{q}$ transition in the shockwave background of
the target. It depends on the target via the matrix elements of two Wilson line operators $\tr(U_1 U_2^\dag)$ and
$\tr(U_1 U_3^\dag) \, \tr(U_3 U_2^\dag) - N_c \tr(U_1 U_2^\dag)$ between the $in$ and $out$ target states. The Wilson lines are functions of the rapidity
 which separates the gluons belonging to the impact factor and the gluons from the Wilson lines. For hadron targets these matrix
elements are to be described by some models. For example for the former one there are several saturation models, inspired by the Golec-Biernat and W\"usthoff  model~\cite{GolecBiernat:1998js,GolecBiernat:1999qd}, while for the latter, to the best of our
knowledge, we are not aware about any such model. These Wilson line operators can also be calculated as solutions of the NLO BK and the LO double dipole evolution
equations with the initial conditions at the rapidity of the target. In the linear limit (BFKL) for forward scattering these
solutions are known analytically with a running coupling \cite{Chirilli:2013kca,Grabovsky:2013gta}. In addition, in the low density regime one can always linearize the second Wilson line operator
and write the cross section in terms of matrix elements of color dipoles only. 

Here we will focus on the detail of the coupling of these Wilson-line operators to the diffractive state.
The various possible regimes and the related appropriate projections which are required for phenomenological applications will be the subject of a future study.  This choice is motivated by the fact that one of the technical difficulties in this framework is to prove explicitly that the various infrared (IR) and ultraviolet (UV) singularities cancel properly.

Next, motivated by the phenomenological importance of our results,
we study in detail the cross-section for exclusive dijet  production in diffraction, as was recently reported by ZEUS ~\cite{Abramowicz:2015vnu}, to show how these cancellations occur in a detailed way.
For that,
we derived the differential cross section for the $\gamma^* p \to q\bar{q} p'$ transition. Taking the
corresponding matrix element from Ref.~\cite{Boussarie:2014lxa} we also calculated the $\gamma^* p \to q\bar{q}g p'$ cross section. Combining them
we wrote the $\gamma^* p \to 2jets\, p'$ exclusive cross section canceling the soft and collinear singularities in the
small cone approximation. Besides, outside the jet cones one can use the  $\gamma^* p \to
q\bar{q}g p'$ cross section to study the electroproduction of 3 jets as well.

The paper is organized as follows. The next section contains the definitions
and necessary results. Section 3 briefly introduces the basic notations and reproduces the LO $\gamma^{(*)}\rightarrow
q\bar{q}$ impact factor. Section 4 gives the general expression for the
$\gamma^{(*)}\rightarrow q\bar{q}$ impact factor at one-loop accuracy. 
Section 5 gives the $\gamma^{(*)}\rightarrow q\bar{q} g$ impact factor at Born order in arbitrary dimensions.
Section 6 gives the $\gamma^{(*)} P \to q\bar{q} g P'$ cross-section at leading and next-to-leading order.
Section 7 gives the $\gamma^{(*)} P \to q\bar{q} g P'$ cross-section at leading order.
Section 8 gives the final result for 
exclusive $\gamma^{*} P \to dijet P'$ transition, showing explicitly the cancellation of divergencies, based on the two previous sections.
Section 9 concludes the paper. 
The appendices comprise the necessary technical details.

\section{Definitions and building blocks}

We introduce the light cone vectors $n_{1}$ and $n_{2}$ as follows :
\begin{equation}
n_{1} \equiv \left(  1,0_{\bot},1\right)  ,\quad n_{2} \equiv \frac{1}{2}\left(  1,0_{\bot
},-1\right)  ,\quad n_{1}^{+}=n_{2}^{-}=(n_{1} \cdot n_{2})=1 .
\end{equation}
For any vector $p$ we note%
\begin{equation}
p^{+} = p_{-} \equiv (p \cdot n_{2})=\frac{1}{2}\left(  p^{0}+p^{3}\right)  ,\qquad p_{+}%
=p^{-} \equiv (p \cdot n_{1})=p^{0}-p^{3},
\end{equation}%
\begin{equation}
p=p^{+}n_{1}+p^{-}n_{2}+p_{\bot},
\end{equation}%
so that
\begin{equation}
(p\,\cdot k)=p^{\mu}k_{\mu}=p^{+}k^{-}+p^{-}k^{+}+(p_\bot \cdot k_\bot)=p_{+}%
k_{-}+p_{-}k_{+}-(\vec{p} \cdot \vec{k}).
\end{equation}
For the moment, we will consider the open production of partons, the conversion into jets will be discussed later in this paper. We denote the initial photon vector as $p_{\gamma},$ and the outgoing quark
and antiquark vectors as $p_{q}$, and $p_{\bar{q}}$. In the real correction, an additional external gluon is emitted. Its momentum will be denoted as $p_g.$ We will focus on diffraction off a proton $P$ which remains intact after the interaction. We denote the initial and final proton momenta as $p_{0}$ and
$p_{0}^{\prime}.$ Our calculation can be used for other processes later on with minor modifications. We consider semihard kinematics with the hard scale
\begin{equation}
s=(p_{\gamma}+p_{0})^{2}\gg|p_{\gamma}^{2}|,\text{ }M_{P}^{2},\text{
}|p_{00^{\prime}}^{2}|.\qquad
\end{equation}
Here and throughout this paper we use the notation $p_{ij}=p_{i}-p_{j}$ for two given vectors $p_i$ and $p_j$.
$M_{P}$ is the proton mass. The semihard scale comes
from either the photon virtuality $|p_{\gamma}^{2}|,$ the momentum transfer
$|p_{00^{\prime}}^{2}|,$ or the invariant mass of the produced jets. As a result one can write
\begin{equation}
s\simeq2p_{\gamma}^{+}p_{0}^{-},
\end{equation}
and choose the reference frame where
\begin{equation}
p_{\gamma}^{+},p_{0}^{-}\sim\sqrt{s}.
\end{equation}
In the case of our process, one can write
\begin{equation}
p_{\gamma}^{+}\sim p_{q}^{+}\sim p_{\bar{q}}^{+}\gg p_{0}^{+},p_{0}^{\prime
+},\quad p_{0}^{-}\gg p_{\gamma}^{-},p_{\bar{q}}^{-},p_{q}^{-}.
\end{equation}
The longitudinal momentum fractions of the $q\bar{q}$ pair
are defined by
\begin{equation}
\frac{p_{q}^{+}}{p_{\gamma}^{+}} \equiv x_q,\quad\frac{p_{\bar{q}}^{+}}%
{p_{\gamma}^{+}}\equiv x_{\bar{q}}. \label{xs}%
\end{equation}
For simplicity we consider a forward photon with virtuality $Q$ and no transverse momentum :
\begin{equation}
\vec{p}_{\gamma}=0,\quad p_{\gamma}^{\mu}=p_{\gamma}^{+}n_{1}^{\mu}%
+\frac{p_{\gamma}^{2}}{2p_{\gamma}^{+}}n_{2}^{\mu},\quad-p_{\gamma}^{2}%
\equiv Q^{2}>0. \label{photonk}%
\end{equation}
We will denote its transverse polarization $\varepsilon_T$. Its longitudinal polarization vector reads%
\[
\varepsilon_{L}^{\alpha}=\frac{1}{\sqrt{-p_{\gamma}^{2}}}\left(  p_{\gamma
}^{+}n_{1}^{\alpha}-\frac{p_{\gamma}^{2}}{2p_{\gamma}^{+}}n_{2}^{\alpha
}\right)  ,\quad\varepsilon_{L}^{+}=\frac{p_{\gamma}^{+}}{Q},\quad
\varepsilon_{L}^{-}=\frac{Q}{2p_{\gamma}^{+}}.
\]
We work in the light-cone gauge $\mathcal{A} \cdot n_{2}=0$. In this gauge, the bare gluon
propagator is given by
\begin{equation}
\hat{G}_{0}^{\mu\nu}\left(  p\right)  =\frac{-id^{\mu\nu}\left(  p\right)  }%
{p^{2}+i0},
\end{equation}
where
\begin{equation}
d^{\mu\nu}\left(  p\right)  =d_{0}^{\mu\nu}\left(  p\right)  -\frac{n_{2}%
^{\mu}n_{2}^{\nu}p^{2}}{(p^{+})^{2}},\quad d_{0}^{\mu\nu}\left(  p\right)
=g_{\bot}^{\mu\nu}-\frac{p_{\bot}^{\mu}n_{2}^{\nu}+p_{\bot}^{\nu}n_{2}^{\mu}%
}{p^{+}} - \frac{n_{2}^{\mu}n_{2}^{\nu}{\vec{p}^{\,\,2}}}{(p^{+})^{2}}.
\end{equation}
The bare fermion propagator reads%
\begin{equation}
G_{0}\left(  p\right)  =\frac{i(\hat{p}+m)}{p^{2}-m^{2}+i0}.
\end{equation}
We will need the propagators and the external lines in the shockwave
background \cite{Balitsky:1995ub}, \cite{Boussarie:2014lxa}.%
\begin{align}
&  \overline{u}(p,y)|_{0>y^{+}}=\langle q_{p}|T(\overline{\psi}\left(
y\right)  e^{i\int\mathcal{L}_{i}\left(  z\right)  dz})|0\rangle_{sw},\quad
v(p,y)|_{0>y^{+}}=\langle\bar{q}_{p}|T(\psi\left(  y\right)  e^{i\int
\mathcal{L}_{i}\left(  z\right)  dz})|0\rangle_{sw},\\
&  G_{\mu\nu}^{ab}(x,y)=\langle0|T(A_{\mu}^{a}\left(  x\right)  A_{\nu}%
^{b}\left(  y\right)  e^{i\int\mathcal{L}\left(  z\right)  dz})|0\rangle
_{sw},\hat{G}(x,y)=\langle0|T(\psi\left(  x\right)  \bar{\psi}\left(  y\right)
e^{i\int\mathcal{L}_{i}\left(  z\right)  d^{D}z})|0\rangle_{sw}.
\end{align}
The external line for a particle with momentum $p$ which steams from the coordinate $y$ with lightcone time $y^+<0$ can be written as
\begin{align}
\overline{u}(p,y)|_{0>y^{+}}  &  =\frac{\theta(p^{+})}{\sqrt{2p^{+}}}%
e^{ip^{+}y^{-}}\int\frac{dp_{2\bot}^{D-2}}{\left(  2\pi\right)  ^{D-2}%
}e^{i(p_{2\bot} \cdot y_\bot)-i\frac{y^{+}}{2p^{+}}(p_{2\bot}^{\,\,2}-m^{2}%
+i0)}\nonumber\\
&  \times\overline{u}_{p}\gamma^{+}U(p_{\bot}-p_{2\bot})\frac{\left[
\gamma^{-}p^{+}+\hat{p}_{2\bot}+m\right]  }{2p^{+}}, \label{ubar}%
\end{align}%
in the case of a quark,
\begin{align}
v(p,y)|_{0>y^{+}}  &  =\frac{\theta(p^{+})}{\sqrt{2p^{+}}}e^{ip^{+}y^{-}}%
\int\frac{dp_{2\bot}^{D-2}}{\left(  2\pi\right)  ^{D-2}}e^{i(p_{2\bot} \cdot y_\bot)-i\frac{y^{+}}{2p^{+}}(p_{2\bot}^{\,\,2}-m^{2}+i0)}\nonumber\\
&  \times\frac{\left[  \gamma^{-}p^{+}+\hat{p}_{2\bot}-m\right]  }{2p^{+}%
}U^{\dag}(p_{2\bot}-p_{\bot})\gamma^{+}v_{p} \, , \label{v}%
\end{align}%
in the case of an antiquark, and
\begin{eqnarray}
\left[\epsilon_{\nu}^{*}\left(p,\, y\right)\right]_{0>y^{+}}^{ab} & = & \frac{\theta\left(p^{+}\right)}{\sqrt{2p^{+}}}e^{ip^{+}y^{-}}\int\frac{d^{D-2}p_{2\perp}}{\left(2\pi\right)^{D-2}}e^{i\left(p_{2\perp}\cdot y_{\perp}\right)-i\frac{y^{+}}{2p^{+}}\left(p_{2\perp}^{2}+i0\right)}\\ \nonumber
 & \times & \epsilon_{p\perp\sigma}^{*}\left[g_{\perp\nu}^{\sigma}-\frac{p_{2\perp}^{\sigma}}{p^{+}}n_{2\nu}\right]U^{ab}\left(p_{\perp}-p_{2\perp}\right) \, ,
\end{eqnarray}
in the case of a gluon. \\
The propagators read 
\begin{align}
&  \hat{G}(y,x)|_{y^{+}>0>x^{+}} \nonumber \\
& =\int\frac{dp_{1}^{+}d^{D-2}p_{1\bot}}{(2\pi)^{D-1}%
}\int\frac{dp_{2}^{+}d^{D-2}p_{2\bot}}{(2\pi)^{D-1}}e^{-iy^{-}p_{2}%
^{+}-i (p_{2\bot} \cdot y_{\bot})}e^{ix^{-}p_{1}^{+}+i(p_{1\bot}\cdot x_{\bot})}2\pi
\delta(p_{12}^{+})\theta(p_{2}^{+})\nonumber\\
&  \times\theta(-x^{+})\theta(y^{+})e^{iy^{+}\frac{p_{2\bot}^{2}-m^{2}%
+i0}{2p_{2}^{+}}}e^{-ix^{+}\frac{p_{1\bot}^{2}-m^{2}+i0}{2p_{1}^{+}}}%
\frac{\gamma^{-}p_{2}^{+}+\hat{p}_{2\bot}+m}{2p_{2}^{+}}\gamma^{+}U(p_{21\bot
})\frac{\gamma^{-}p_{1}^{+}+\hat{p}_{1\bot}+m}{2p_{1}^{+}}, \label{propq}
\end{align}%
for a quark,
\begin{align} \nonumber
&  \hat{G}(x,y)|_{y^{+}>0>x^{+}} \\ \nonumber 
& = -\int\frac{dp_{1}^{+}d^{D-2}p_{1\bot}}{(2\pi
)^{D-1}}\int\frac{dp_{2}^{+}d^{D-2}p_{2\bot}}{(2\pi)^{D-1}}e^{-iy^{-}p_{2}%
^{+}-i(p_{2\bot}\cdot y_{\bot})}e^{ix^{-}p_{1}^{+}+i(p_{1\bot} \cdot x_{\bot})}2\pi
\delta(p_{12}^{+})\theta(p_{2}^{+})\nonumber\\
&  \times\theta(-x^{+})\theta(y^{+})e^{iy^{+}\frac{p_{2\bot}^{2}-m^{2}%
+i0}{2p_{2}^{+}}}e^{-ix^{+}\frac{p_{1\bot}^{2}-m^{2}+i0}{2p_{1}^{+}}}%
\frac{\gamma^{-}p_{1}^{+}+\hat{p}_{1\bot}-m}{2p_{1}^{+}}\gamma^{+}U^{\dag
}(p_{12\bot})\frac{\gamma^{-}p_{2}^{+}+\hat{p}_{2\bot}-m}{2p_{2}^{+}} \label{propqbar},
\end{align}%
for an antiquark, and
\begin{align}
G_{\mu\nu}(x,y)|_{x^{+}>0>y^{+}}  &  =-\int\frac{dp_{1}^{+}d^{D-2}p_{1\bot}%
}{(2\pi)^{D-1}}\int\frac{dp_{2}^{+}d^{D-2}p_{2\bot}}{(2\pi)^{D-1}}%
e^{-ip_{2}^{+}x^{-}+ip_{1}^{+}y^{-}}e^{-i(p_{2\bot}\cdot x_{\bot})+i(p_{1\bot}\cdot y_{\bot})%
}\nonumber\\
& \hspace{-.5cm} \times\pi\frac{\delta(p_{12}^{+})\theta(p_{2}^{+})}{p_{1}^{+}}%
e^{i\frac{p_{2\bot}^{2}+i0}{2p_{2}^{+}}x^{+}-i\frac{p_{1\bot}^{2}+i0}%
{2p_{1}^{+}}y^{+}}d_{0\mu\alpha}(p_{2}^{+},p_{2\bot})U(p_{21\bot})g_{\bot
}^{\alpha\delta}d_{0\delta\nu}(p_{1}^{+},p_{1\bot}) \, , \label{propg}
\end{align}
for a gluon. \\
In these formulas $D \equiv 2+d \equiv 4+2\epsilon$ is the space-time dimension and
\begin{equation}
U(p_{\bot})=\int d^{D-2}r_{\bot}e^{i(p_\bot \cdot r_\bot)}U_{r_{\bot}}\text{ and
}U^{\dag}(p_{\bot})=\int d^{D-2}r_{\bot}e^{-i(p_\bot \cdot r_\bot)}U_{r_{\bot}}^{\dag}%
\end{equation}
are the Fourier transforms of the Wilson lines%
\begin{equation}
U_{r}=U\left(  \vec{r},\eta\right)  =Pe^{ig\mu^{\epsilon
}\int_{-\infty}^{+\infty}b_{\eta}^{-}(r^{+} \! , \,\vec{r})dr^{+}}
\label{WL}
\end{equation}
For convenience we will write $U_i \equiv U_{r_i}$ for a point $r_i$. The external shock-wave field $b_{\eta}%
^{-}$ is built from the gluons that are slow in the asymmetric boosted frame where the non-perturbative dynamics only occur in the proton. Its form in coordinate space can be written in a simple way, as
\begin{equation}
b_{\eta}^{-}=\int\frac{d^{4}p}{\left(  2\pi\right)  ^{4}}e^{-i(p \cdot z)}b^{-}\left(
p\right)  \theta(e^{\eta}-|\frac{p^{+}}{p_{\gamma}^{+}}|),\quad b^{\mu}\left(
z\right)  =b^{-}(z^{+} \! , \, \vec{z})n_{2}^{\mu}=\delta(z^{+})B\left(  \vec
{z}\right)  n_{2}^{\mu}. \label{cutoff}%
\end{equation}
Here $\eta$ is the rapidity divide, which separates the slow gluons in the
Wilson lines and the fast ones in the impact factors. In most cases $\eta$ will be the typical rapidity of a target remnant. \\
To construct the cross section after calculating the impact factor one has to
integrate w.r.t. the field $b$ generated by the proton. Technically it means
that one has to treat the field $b$ as an operator and use the matrix element
of the total Wilson operator between the proton states
\begin{equation}
U_{i}\dots\rightarrow\langle P_{p_{0}^{\prime}}^{\prime}|T(U_{i}%
\dots)|P_{p_{0}}\rangle. \label{ExplicitSW}%
\end{equation}
For simplicity of the notations we will still use the operator $U$ instead of its matrix element during the calculation of the
impact factor, and return to the matrix element later on.

We introduced the regularization scale $\mu$ in (\ref{WL}) because in
dimensional regularization the coupling constant is a dimensional quantity%
\begin{equation}
g_{0}=g\mu^{-\epsilon},\quad\alpha_{s0}=\alpha_{s}\mu^{-2\epsilon}.
\label{alphas}%
\end{equation}
We also introduce a regularization cutoff $\alpha$ for the spurious light cone singularity $p^+_g \rightarrow 0$. Evolving the operators $U(r,\rho)$ from $\rho = \alpha$ to $\rho = e^\eta$ with the help of the BK equation will allow us to cancel such singularities, as shown in section 3.2.2. 

In the following we will need the BK equation in $d$ dimensions. It reads
\cite{Caron-Huot:2013fea}%
\begin{align}
\frac{\partial tr(U_{1}U_{2}^{\dag})}{\partial\eta}  &  =\frac{\alpha_{s}%
\mu^{2-d}}{2\pi^{d}}\Gamma^{2}(\frac{d}{2}) \int \! d^{d}r_{3\bot} \left[ tr(U_{1}%
U_{3}^{\dag})tr(U_{3}U_{2}^{\dag})-N_{c}tr(U_{1}U_{2}^{\dag}) \right] \nonumber\\
&  \times\left[  \frac{2(r_{13\bot} \cdot r_{23\bot})}{\left(  -r_{13\bot}%
^{2}+i0\right)  ^{\frac{d}{2}}\left(  -r_{23\bot}^{2}+i0\right)  ^{\frac{d}%
{2}}}+\frac{1}{\left(  -r_{13\bot}^{2}+i0\right)  ^{d-1}}+\frac{1}{\left(
-r_{23\bot}^{2}+i0\right)  ^{d-1}}\right]  \, ,
\end{align}%
in coordinate space, and
\begin{align}
&  \frac{\partial tr(U(p_{1\bot})U^{\dag}(-p_{2\bot}))}{\partial\eta}%
=\delta(k_{1\bot}+k_{2\bot}+k_{3\bot}-p_{1\bot}-p_{2\bot})2\alpha_{s}\mu
^{2-d}\nonumber\\
& \times \int\frac{d^{d}k_{1\bot}d^{d}k_{2\bot}d^{d}k_{3\bot}}{(2\pi)^{2d}%
} \left[  -\frac{2 (k_{1\bot}-p_{1\bot}) \cdot (k_{2\bot}-p_{2\bot})}{(k_{1}-p_{1}%
)_{\bot}^{2}(k_{2}-p_{2})_{\bot}^{2}} \right. \\ \nonumber 
&+ \left. \frac{\pi^{\frac{d}{2}}\Gamma(1-\frac
{d}{2})\Gamma(\frac{d}{2})^{2}}{\Gamma(d-1)}\left(  \frac{\delta(k_{2\bot
}-p_{2\bot})}{(-(k_{1}-p_{1})_{\bot}^{2})^{1-\frac{d}{2}}}+\frac
{\delta(k_{1\bot}-p_{1\bot})}{(-(k_{2}-p_{2})_{\bot}^{2})^{1-\frac{d}{2}}%
}\right)  \right] \nonumber\\
&  \times \left[ tr(U_{1}U_{3}^{\dag})tr(U_{3}U_{2}^{\dag})-N_{c}tr(U_{1}U_{2}^{\dag
})\right] (k_{1\bot,}k_{2\bot},k_{3\bot}) \label{BKMOM}
\end{align}
in momentum space.
Here we introduced the Fourier transform of the operator%
\begin{align}
&  \left[ tr(U_{1}U_{3}^{\dag})tr(U_{3}U_{2}^{\dag})-N_{c}tr(U_{1}U_{2}^{\dag
}) \right] (k_{1\bot,}k_{2\bot
},k_{3\bot}) \nonumber\\
&  =\int d^{d}r_{1\bot}d^{d}r_{2\bot}d^{d}r_{3\bot}%
e^{i [(r_{1\bot } \cdot k_{1\bot}) +(r_{2\bot} \cdot k_{2\bot})+ (r_{3\bot}.\cdot k_{3\bot})]} \left[ tr(U_{1}U_{3}^{\dag}%
)tr(U_{3}U_{2}^{\dag})-N_{c}tr(U_{1}U_{2}^{\dag})\right] . \label{trtrtr}%
\end{align}

\section{Impact factor for $\gamma\rightarrow q\bar{q}$
transition}%


\begin{figure}
[tbh]
\begin{center}
\includegraphics[
height=6.9869cm,
width=9.072cm
]%
{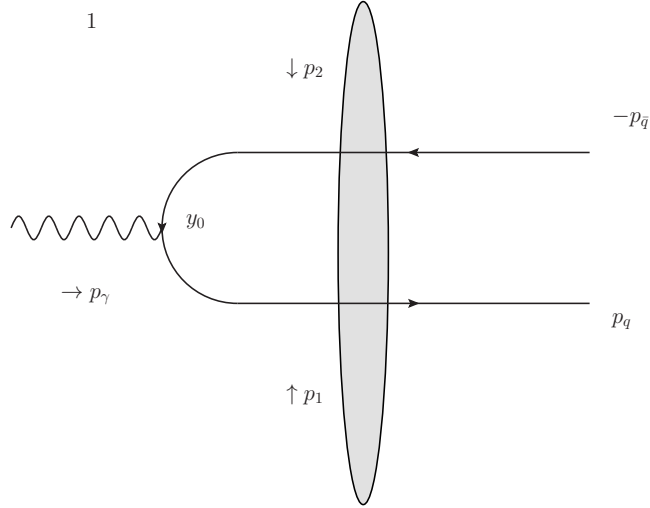}%
\caption{LO impact factor. The momenta $p_{1}$ and $p_{2}$ go from the
shockwave to the quark and antiquark.}%
\label{loi}%
\end{center}
\end{figure}

The matrix element for the EM current in the shockwave background reads%
\begin{equation}
\tilde{M}^{\alpha}=-ie_{q}\int d^{D}y_{0}\frac{e^{-i (p_{\gamma} \cdot y_{0})}}%
{\sqrt{2p_{\gamma}^{+}}}\frac{\delta_{l}^{n}}{\sqrt{N_{c}}}\langle
0|T(b_{p_{\bar{q}}}^{l}(a_{p_{q}})_{n}\overline{\psi}\left(  y_{0}\right)
\gamma^{\alpha}\psi\left(  y_{0}\right)  e^{i\int\mathcal{L}_{i}\left(
z\right)  dz})|0\rangle_{sw}.
\end{equation}
Here $a$ and $b$ are the quark and antiquark annihilation operators, $e_{q}$
is the quark electric charge, and $\frac{\delta_{l}^{n}}{\sqrt{N_{c}}}$ is the projector on the color singlet. To shorten the notation we will work with the reduced
matrix element $\tilde{T}$
\begin{equation}
\tilde{M}^{\alpha} \equiv \frac{-ie_{q}}{\sqrt{2p_{\gamma}^{+}}}\frac{-i\delta
(p_{q}^{+}+p_{\bar{q}}^{+}-p_{\gamma}^{+})}{\sqrt{N_{c}}\left(  2\pi\right)
^{D-3}\sqrt{2p_{\bar{q}}^{+}}\sqrt{2p_{q}^{+}}}\tilde{T}. \label{Mtilde}%
\end{equation}

\subsection{LO impact factor}

Its expression at LO is obtained with the help of (\ref{ubar}) and\ (\ref{v}). The result
can be written as%
\begin{equation}
\tilde{T}_{0}^{\alpha}=\int d^{d}p_{1\bot}d^{d}p_{2\bot}\delta(p_{q1\bot
}+p_{\bar{q}2\bot}-p_{\gamma\bot}) \, tr [ U(p_{1\bot})U^{\dag
}(-p_{2\bot}) ] \Phi_{0}^{\alpha} .
\end{equation}
After subtraction of the noninteracting part one gets%
\begin{equation}
T_{0}^{\alpha}=\int d^{d}p_{1\bot}d^{d}p_{2\bot}\delta(p_{q1\bot}+p_{\bar
{q}2\bot}-p_{\gamma\bot})\Phi_{0}^{\alpha}[tr(U_{1}U_{2}^{\dag})-N_{c}%
](p_{1\bot},p_{2\bot}). \label{LOi}%
\end{equation}
Here%
\begin{equation}
[tr(U_{1}U_{2}^{\dag})-N_{c}](p_{1\bot},p_{2\bot})=\int \! d^d r_{1\bot}
d^d r_{2\bot}e^{i(p_{1\bot} \cdot r_{1\bot})+i(p_{2\bot} \cdot r_{2\bot})}[tr(U_{1}%
U_{2}^{\dag})-N_{c}] \label{subtractedtrdef}%
\end{equation}
is the dipole operator. The function
\begin{equation}
\Phi_{0}^{\alpha}\equiv\Phi_{0}^{\alpha}(p_{1\bot},p_{2\bot})
\end{equation}
is the LO impact factor and we will often suppress its dependence on variables
for brevity. Its components have the form (\ref{LOifplus}-\ref{LOifi}), in which $x\equiv x_q$ and $\bar{x} = 1-x = x_{\bar{q}}$ :
\begin{align}
\Phi_{0}^{+}  &  =-\frac{p_{\gamma}^{+}}{p_{\gamma}^{-}}\Phi_{0}^{-}%
=\frac{2x\bar{x}p_{\gamma}^{+}}{\vec{p}_{q1}^{\, 2} +x\bar{x}Q^{2}} ( \overline
{u}_{p_{q}}\gamma^{+}v_{p_{\bar{q}}} ) ,\label{LOifplus}\\
\Phi_{0}^{i}  &  =\frac{\overline{u}_{p_{q}%
}(\hat{p}_{q1\bot}\gamma^{i}-2xp_{q1\bot}^{i})\gamma
^{+}v_{p_{\bar{q}}}}{\vec{p}_{q1}^{\, 2} +x\bar{x}Q^{2}}. \label{LOifi}%
\end{align}
The first equality in (\ref{LOifplus}) holds thanks to the electromagnetic gauge invariance,
which allows us to calculate only the $^{+}$ component of the impact factor.

\subsection{NLO impact factor}

At NLO the substracted matrix element generalizing expression (\ref{LOi}) can be split into two terms, depending on the type of Wilson line operators involved. Expressing all the Wilson operators in the fundamental representation, we can show that one term involves a single dipole operator $\mathcal{U}_{12}=tr(U_1 U_2^\dagger)-N_c$, while the other one involves the double-dipole operator $tr(U_{1}U_{3}^{\dagger})tr(U_{3}U_{2}^{\dagger})-N_{c}tr(U_{1}U_{2}^{\dagger}) = \mathcal{U}_{13} \, \mathcal{U}_{32}+N_{c}\left(\mathcal{U}_{13}+\mathcal{U}_{32}-\mathcal{U}_{12}\right)
$ :
\begin{align}
T_{1}^{\alpha}  &  =\alpha_{s}\frac{\Gamma(1-\epsilon)}{\left(  4\pi\right)
^{1+\epsilon}}\int d^{d}p_{1\bot}d^{d}p_{2\bot}\left\{  \delta(p_{q1\bot
}+p_{\bar{q}2\bot}-p_{\gamma\bot})\Phi_{1}^{\alpha}\frac{N_{c}^{2}-1}{N_{c}%
}[tr(U_{1}U_{2}^{\dag})-N_{c}](p_{1\bot},p_{2\bot})\right. \nonumber\\
& \hspace{-.5cm}  +\left.  \int\frac{d^{d}p_{3\bot}}{(2\pi)^{d}}\delta(p_{q1}+p_{\bar{q}%
2}-p_{\gamma\bot}-p_{3\bot})\Phi_{2}^{\alpha}[tr(U_{1}U_{3}^{\dag
})tr(U_{3}U_{2}^{\dag})-N_{c}tr(U_{1}U_{2}^{\dag})](p_{1\bot,}p_{2\bot
},p_{3\bot})\right\}  . \label{NLOi}%
\end{align}

The Wilson operators here are defined in (\ref{trtrtr}),
(\ref{subtractedtrdef}) and the dependence of the coupling constant on the
regularization scale (\ref{alphas}) is included in the definition of
$\Phi_{1}^{\alpha}$ and $\Phi_{2}^{\alpha}.$ 


\begin{figure}
[tbh]
\begin{center}
\includegraphics[
height=9.0591cm,
width=16.3709cm
]%
{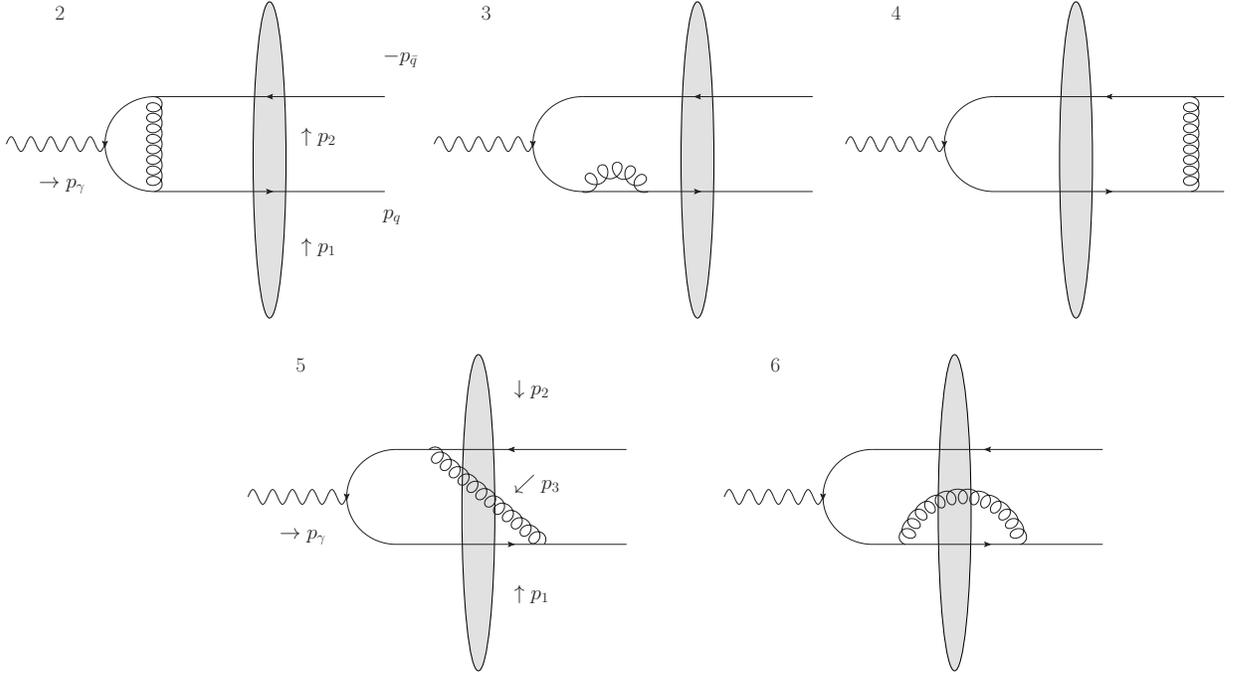}%
\caption{One-loop diagrams for the $\gamma\rightarrow q\bar{q}$ transition.
The momenta $p_{1},$ $p_{2},$ and $p_{3}$ go from the shockwave to the quark,
antiquark and gluon. }%
\label{diags}%
\end{center}
\end{figure}

There are 8 one-loop diagrams contributing to the matrix element $T_1.$ Five of
them are presented in figure \ref{diags}. The remaining ones can be obtained from
diagrams 3, 5 and 6 via the substitution $p_{q}\leftrightarrow p_{\bar{q}},$ $\bar{u}_{q}\leftrightarrow v_{\bar{q}},$ $p_{1}\leftrightarrow p_{2}$ and the reversal of the order of the gamma matrices, which we will note $(q\leftrightarrow\bar{q}).$ Diagrams 2, 3 and 4
contribute only to the dipole term $\Phi_{1}$ while diagrams 5 and 6 contribute to both
$\Phi_{1}$ and the double-dipole term $\tilde{\Phi}_{2}.$ Indeed, after projecting on the color singlet state in the $t$-channel and substracting the non-interacting part, it is straightforward to see that diagrams 2, 3 and 4 involve $tr(U_1 U_2^\dagger) -N_c$, while diagrams 5 and 6 involve the Fourier transform of an operator which can be decomposed as follows : 
\begin{equation}
tr(t^{a}U_{1}t^{b}U_{2}^{\dag})U_{3}^{ab}-\frac{N_{c}^{2}-1}{2}=\frac{1}%
{2}[tr(U_{1}U_{3}^{\dag})tr(U_{3}U_{2}^{\dag})-N_{c}tr(U_{1}U_{2}^{\dag
})]+\frac{N_{c}^{2}-1}{2N_{c}}[tr(U_{1}U_{2}^{\dag})-N_{c}].
\end{equation}

\subsubsection{Method of calculation of the NLO corrections}
Due to the presence of the rapidity singularity $p^+ \rightarrow 0$ in lightcone gauge, we cannot integrate directly in $D$ dimensions with the usual Feynman integration methods. Dimensional regularization can be used for the transverse components with dimension $d = D-2$, while the longitudinal divergences will be regularized by the cutoff $\alpha$. \\
We will now present the essential steps required to compute the diagrams in figure 2. For this purpose, we will consider the simple case of diagram 3, the contribution of the other diagrams being obtained in a similar way.
The initial expression for this diagram is the projection on the color singlet state of the following expression :

\begin{eqnarray}
\tilde{M}^{\alpha}|_{3} &=& \frac{ig^2e_q}{\sqrt{N_c}} \int d^{D}y_{2}d^{D}y_{1}d^{D}y_{0}\bar{u}\left(p_{q},\, y_{2}\right)\gamma^{\mu}\hat{G}_0\left(y_{21}\right)\gamma^{\nu}\hat{G}_0\left(y_{10}\right)\gamma^{\alpha}\frac{\epsilon_{\alpha}}{\sqrt{2p_{\gamma}^{+}}}\\ \nonumber 
&\times & \theta\left(p_{\gamma}^{+}\right)\theta\left(-y_{2}^{+}\right)\theta\left(-y_{1}^{+}\right)\theta\left(-y_{0}^{+}\right)v\left(p_{\bar{q}},\, y_{0}\right)G_{\mu\nu}\left(y_{21}\right)e^{-i(p_{\gamma} \cdot y_{0})} .
\end{eqnarray}

Using the building blocks defined by Eqs. (\ref{ubar}) to (\ref{propg}) and projecting on the singlet, we get :

\begin{eqnarray}
\tilde{M}^{\alpha}|_{3} & = & \frac{ig^2e_q}{\sqrt{N_c}} \int \! d^{D}y_{2}d^{D}y_{1}d^{D}y_{0}\int\frac{d^{D}q_{2}}{\left(2\pi\right)^{D}}\frac{d^{D}q_{1}}{\left(2\pi\right)^{D}}\frac{d^{D}l}{\left(2\pi\right)^{D}}\frac{N_{c}^{2}-1}{2N_{c}} \\ \nonumber 
&\times & tr\left(U_{p_{1\perp}}U_{-p_{2\perp}}^{\dagger}\right) \theta\left(-y_{2}^{+}\right)\theta\left(-y_{1}^{+}\right)\theta\left(-y_{0}^{+}\right) \\ \nonumber
 & \times & \frac{\theta\left(p_{q}^{+}\right)}{\left(2\pi\right)^{D-2}\sqrt{2p_{q}^{+}}}e^{ip_{q}^{+}y_{2}^{-}}\bar{u}_{p_{q}}\gamma^{+}\frac{p_{q}^{+}\gamma^{-}+\hat{p}_{q1\perp}}{2p_{q}^{+}}e^{i\left(p_{q1\perp}\cdot y_{2\perp}\right)-i\frac{y_{2}^{+}}{2p_{q}^{+}}\left(p_{q1\perp}^{2}+i0\right)}\\ \nonumber
 & \times & e^{-i\left(q_{2}+l\right)\cdot y_{21}-i\left(q_{1}\cdot y_{10}\right)}\gamma^{\mu}\hat{G}_0\left(q_{2}\right)\gamma^{\nu}\hat{G}_0\left(q_{1}\right)\gamma^{\alpha}\frac{\epsilon_{\alpha}}{\sqrt{2p_{\gamma}^{+}}}e^{-i\left(p_{\gamma}\cdot y_{0}\right)}\theta\left(p_{\gamma}^{+}\right)G_{\mu\nu}\left(l\right) \\ \nonumber
 & \times & \frac{\theta\left(p_{\bar{q}}^{+}\right)}{\left(2\pi\right)^{D-2}\sqrt{2p_{\bar{q}}^{+}}}e^{ip_{\bar{q}}^{+}y_{0}^{-}+i\left(p_{2\bar{q}\perp}\cdot y_{0\perp}\right)-i\frac{y_{0}^{+}}{2p_{\bar{q}}^{+}}\left(p_{2\bar{q}\perp}^{2}+i0\right)}\frac{p_{\bar{q}}^{+}\gamma^{-}+\hat{p}_{\bar{q}2\perp}}{2p_{\bar{q}}^{+}}\gamma^{+}v_{p_{\bar{q}}} .
\end{eqnarray}

We now apply the following procedure :
\begin{itemize}
\item Integrating w.r.t. the $^-$ and transverse components of the coordinates $y_2$, $y_1$ and $y_0$ gives relations for the conservation of $^+$ and transverse momentum components, such as \[ \delta(p_q^+ -q_2^+ -l^+) \delta(p_{q\bot}-q_{2\bot}-l_\bot) . \] 
\item The $^+$ and transverse momentum integrations are now taken, as trivial $\delta$ integrations.
\item We integrate the $^-$ momenta by pole integration.
\end{itemize}
This way, one obtains

\begin{eqnarray} 
&& \tilde{M}^{\alpha}|_{3} = - \frac{ig^2e_q}{\sqrt{N_c}} \frac{tr\left(U_{p_{1\perp}}U_{-p_{2\perp}}^{\dagger}\right)}{\left(2\pi\right)^{2D-4}}\left(\frac{N_{c}^{2}-1}{2N_{c}}\right)\frac{\theta\left(p_{q}^{+}\right)\theta\left(p_{\bar{q}}^{+}\right)\theta\left(p_{\gamma}^{+}\right)\epsilon_{\alpha}}{\sqrt{2p_{\gamma}^{+}2p_{q}^{+}2p_{\bar{q}}^{+}}} \\ \nonumber
& \times & \int dy_{2}^{+}dy_{1}^{+}dy_{0}^{+}\theta\left(-y_{2}^{+}\right)\theta\left(-y_{1}^{+}\right)\theta\left(-y_{0}^{+}\right)\\ \nonumber
 & \times & \int dl^{+}d^{D-2}l_{\perp}\delta\left(p_{q}^{+}+p_{\bar{q}}^{+}-p_{\gamma}^{+}\right)\delta\left(p_{q1\perp}+p_{2\bar{q}\perp}-p_{\gamma\perp}\right)\bar{u}_{p_{q}}\gamma^{+}\left[\frac{p_{q}^{+}\gamma^{-}+\hat{p}_{q1\perp}}{2p_{q}^{+}2\left(p_{q}^{+}-l^{+}\right)}\right]\gamma^{\mu}\\ \nonumber
 & \times & \left(\left[\left(p_{q}^{+}-l^{+}\right)\gamma^{-}-\frac{\left(p_{q1\perp}-l_{\perp}\right)^{2}}{2\left(p_{q}^{+}-l^{+}\right)}\gamma^{+}+\left(\hat{p}_{q1\perp}-\hat{l}_{\perp}\right)\right] \right. \\ \nonumber 
 &\times & \left.  \left[\theta\left(p_{q}^{+}-l^{+}\right)\theta\left(y_{21}^{+}\right)-\theta\left(l^{+}-p_{q}^{+}\right)\theta\left(y_{12}^{+}\right)\right]+i\delta\left(y_{21}^{+}\right) \gamma^{+}\right) \\ \nonumber
 & \times &\gamma^{\nu}\left(\left[p_{q}^{+}\gamma^{-}-\frac{p_{q1\perp}^{2}}{2p_{q}^{+}}\gamma^{+}+\hat{p}_{q1\perp}\right]\theta\left(y_{10}^{+}\right)+i\delta\left(y_{10}^{+}\right)\gamma^{+}\right)\gamma^{\alpha}\left[\frac{p_{\bar{q}}^{+}\gamma^{-}+p_{\bar{q}2\perp}}{2p_{q}^{+}2p_{\bar{q}}^{+}}\right]\gamma^{+}v_{p_{\bar{q}}}\\ \nonumber
 & \times & \frac{1}{2l^{+}}\left(\left[g_{\perp\mu\nu}-\frac{l_{\perp\mu}n_{2\nu}+l_{\perp\nu}n_{2\mu}}{l^{+}}+\frac{n_{2\mu}n_{2\nu}}{\left(l^{+}\right)^{2}}l_{\perp}^{2}\right]\left[\theta\left(l^{+}\right)\theta\left(y_{21}^{+}\right)-\theta\left(-l^{+}\right)\theta\left(y_{12}^{+}\right)\right] \right. \\ \nonumber
 & -& \left. 2i\frac{n_{2\mu}n_{2\nu}}{l^{+}}\delta\left(y_{21}^{+}\right)\right) \exp\left[ -i\left(p_{\gamma}^{-}+\frac{p_{2\bar{q}\perp}^{2}+i0}{2p_{\bar{q}}^{+}}+\frac{p_{q1\perp}^{2}+i0}{2p_{q}^{+}}\right)y_{0}^{+} \right] \\ \nonumber
 & \times & \exp\left[i\left(\frac{l_{\perp}^{2}+i0}{2l^{+}}+\frac{\left(p_{q1\perp}-l_{\perp}\right)^{2}+i0}{2\left(p_{q}^{+}-l^{+}\right)}-\frac{p_{q1\perp}^{2}+i0}{2p_{q}^{+}}\right)y_{21}^{+}\right] .
\end{eqnarray}
The link with old-fashioned perturbation theory, as used in the dipole picture~\cite{Mueller:1993rr}, can easily be made from this expression. For example the second last line is the expression for the gluon propagator in this formalism. The first term with $\theta(y_{21}^+)$ would correspond to the case when the gluon moves forward in time, the second term $\theta(y_{12}^+)$ would be backwards in time and the last term $\delta(y_{21}^+)$ would be the instantaneous gluon contribution. Note that ill-defined quantities such as $\theta(y^+)\delta(y^+)$ will cancel. In this example, one can easily see that the contributions from the terms with $\delta(y_{21}^+)$ or $\theta(y_{21}^+)$ give zero, due to either the gamma structure or $^+$-momentum ordering. One can also show that the term with $\delta(y_{10}^+)$ cancels, since after shifting $l_\bot$ to make the $l_\bot$ exponential an exact gaussian, the expression for this term becomes odd for $l_\bot \rightarrow -l_\bot$. \\
The integration w.r.t. the $^+$ components of the coordinates finally gives

\begin{eqnarray}
&& \tilde{M}^{\alpha}|_{3} =  i\frac{ie_q g^2}{\sqrt{N_c}}\frac{tr\left(U_{p_{1\perp}}U_{-p_{2\perp}}^{\dagger}\right)}{\left(2\pi\right)^{2D-4}}\left(\frac{N_{c}^{2}-1}{2N_{c}}\right) \\ \nonumber
&\times &  \frac{\theta\left(p_{q}^{+}\right)\theta\left(p_{\bar{q}}^{+}\right)\theta\left(p_{\gamma}^{+}\right)\epsilon_{\alpha}}{\sqrt{2p_{\gamma}^{+}2p_{q}^{+}2p_{\bar{q}}^{+}}}\frac{\delta\left(p_{q}^{+}+p_{\bar{q}}^{+}-p_{\gamma}^{+}\right)\delta\left(p_{q1\perp}+p_{2\bar{q}\perp}\right)}{\left(p_{\gamma}^{-}+\frac{p_{\perp}^{2}+i0}{2p_{q}^{+}}+\frac{p_{\perp}^{2}+i0}{2p_{\bar{q}}^{+}}\right)^{2}}\\ \nonumber
 & \times & \bar{u}_{p_{q}}\gamma^{+}\left[\frac{p_{q}^{+}\gamma^{-}+\hat{p}_{\perp}}{2p_{q}^{+}}\right]\int_{\alpha p_\gamma^+}^{p_{q}^{+}}\frac{dl^{+}}{\left(2p_{q}^{+}\right)^{2}}\int\frac{d^{D-2}l_{\perp}}{\left[\left(l_{\perp}-\frac{l^{+}}{p_{q}^{+}}p_{\perp}\right)^{2}+\frac{2l^{+}\left(p_{q}^{+}-l^{+}\right)}{p_{q}^{+}}\left(p_{\gamma}^{-}+\frac{p_{\gamma}^{+}}{2p_{q}^{+}p_{\bar{q}}^{+}}p_{\perp}^{2}\right)\right]}\\ \nonumber
 & \times & \gamma^{\mu}\left[\left(p_{q}^{+}-l^{+}\right)\gamma^{-}-\frac{\left(p_{\perp}-l_{\perp}\right)^{2}}{2\left(p_{q}^{+}-l^{+}\right)}\gamma^{+}+\left(\hat{p}_{\perp}-\hat{l}_{\perp}\right)\right]\gamma^{\nu}\\ \nonumber
 & \times & \left[g_{\perp\mu\nu}-\frac{l_{\perp\mu}n_{2\nu}+l_{\perp\nu}n_{2\mu}}{l^{+}}+\frac{n_{2\mu}n_{2\nu}}{\left(l^{+}\right)^{2}}l_{\perp}^{2}\right] \left[p_{q}^{+}\gamma^{-}-\frac{p_{\perp}^{2}}{2p_{q}^{+}}\gamma^{+}+\hat{p}_{\perp}\right]\gamma^{\alpha}\left[\frac{p_{\bar{q}}^{+}\gamma^{-}-p_{\perp}}{2p_{\bar{q}}^{+}}\right]\gamma^{+}v_{p_{\bar{q}}} ,
\end{eqnarray}
where we introduced the regularization cutoff $\alpha$ for the divergences which will emerge from the longitudinal integration. \\
The transverse momentum integration can be performed straightforwardly with the Feynman parameter or Schwinger representation method. We then decompose the result as a divergent term in $\epsilon = \frac{D-4}{2}$ and a constant term. Thus :

\begin{eqnarray} \nonumber
\tilde{M}^{\alpha}|_{3} & = & -\frac{e_qg^2}{\sqrt{N_c}}\frac{\Gamma\left(1-\epsilon\right)}{\left(16\pi^{3}\right)^{1+\epsilon}}  tr\left(U_{p_{1\perp}}U_{-p_{2\perp}}^{\dagger}\right)\left(\frac{N_{c}^{2}-1}{2N_{c}}\right)\frac{\theta\left(p_{q}^{+}\right)\theta\left(p_{\bar{q}}^{+}\right)\theta\left(p_{\gamma}^{+}\right)\epsilon_{\alpha}}{\sqrt{2p_{\gamma}^{+}2p_{q}^{+}2p_{\bar{q}}^{+}}}\\ \nonumber
 & \times & \frac{\delta\left(p_{q}^{+}+p_{\bar{q}}^{+}-p_{\gamma}^{+}\right)\delta\left(p_{q1\perp}+p_{2\bar{q}\perp}\right)}{\left(4p_{\gamma}^{+}\right)\left(\vec{p}_{q1}^{2}+x\bar{x}Q^{2}\right)}\\ \nonumber
 & \times & \int_{\alpha}^{x}\frac{dz}{x^{2}}\left[\frac{1}{\epsilon}+\ln\left(\frac{\vec{p}_{q1}^{2}+x\bar{x}Q^{2}}{x\bar{x}\mu^{2}}\right)+\ln\left(\frac{z\left(x-z\right)}{x}\right)\right]\left(dz+4\frac{x\left(x-z\right)}{z}\right)\\
 & \times & \bar{u}_{p_{q}}\left[\left(\gamma^{+}\hat{p}_{q1\perp}\right)+p_{q}^{+}\left(\gamma^{+}\gamma^{-}\right)\right]\gamma^{\alpha}\left(p_{\bar{q}}^{+}\gamma^{-}-p_{q1\perp}\right)\gamma^{+}v_{p_{\bar{q}}} \, .
\end{eqnarray}

To regularize the integration over parameter $z$, one will always write 
\begin{eqnarray}
\int_\alpha^{z_0} dz \, \phi(z) & = & \int_\alpha^{z_0} dz \, \phi_0(z) +  \int_\alpha^{z_0} dz \, [\phi(z)-\phi_0(z)] \\ \nonumber
&\equiv & \int_\alpha^{z_0} dz \, \phi_0(z) +  \int_\alpha^{z_0} dz \, \left[ \phi(z) \right]_+ \label{zregularization} \, ,
\end{eqnarray}
where $\phi(z) = \phi_0(z)+O(1)$ for $z\rightarrow 0$, so that $\phi_0$ contains the divergence of $\phi$.\\
In the case of diagram 3, both these terms can be computed analytically to the end. However it is not the case for some other diagrams, although the first term in the right hand side of (\ref{zregularization}) is always straightforward to obtain. \\
The method for the computation of other NLO virtual diagrams is similar although more elaborate. We will not give the details for them in the body text and relegate the details of the results to the appendices.
We will only give the divergent part, by which we mean the first part of (\ref{zregularization}) and the $\frac{1}{\epsilon}$ term in the second part of (\ref{zregularization}).

\subsubsection{Double dipole contribution $\Phi_2$}

For reader's convenience, we will only write explicitly the divergent part of these diagrams here. The expressions for the constant parts can be found in appendix A.

The contributions of diagram 5 to $\tilde{\Phi}_{2}$, including the $(q\leftrightarrow\bar{q})$ terms, reads :%
\begin{align}
\Phi_{2}^{+}|_{5}=  &  2p_{\gamma}^{+} (\bar{u}_{p_{q}}\gamma
^{+}v_{p_{\bar{q}}})\frac{x\bar{x}(\vec{p_{3}}{}^{2}-{}\vec{p}_{\bar{q}2}%
^{\,\,2}-\vec{p_{q1}}{}^{2}-2x\bar{x}Q^{2})}{(\vec{p}_{\bar{q}%
2}^{\,\,2}+x\bar{x}Q^{2})\left( \vec{p_{q1}}{}^{2}+ x\bar{x} Q^{2}\right) - x\bar{x}Q^{2}%
\vec{p_{3}}{}^{2}}\nonumber\\
\times &  \ln\left(  \frac{x\bar{x}}{\alpha^{2}}\right)  \ln\left(  \frac
{(\vec{p}_{\bar{q}2}^{\,\,2}+x\bar{x}Q^{2})\left(  \vec{p_{q1}}{}^{2}%
+x\bar{x}Q^{2}\right)  }{x\bar{x}Q^{2}\vec{p_{3}}{}^{2}}\right)  +(\bar{u}_{p_{q}%
}C_{2\Vert}^{5}v_{p_{\bar{q}}})
\end{align}%
for a longitudinal photon, or
\begin{align}
\Phi_{2}^{i}|_{5}  &  =\ln\left(  \frac{x\bar{x}}{\alpha^{2}}\right)
\left\{  \overline{u}_{p_{q}}(p_{q1\bot}^{i}(1-2x)+\frac{1}{2}[\hat
{p}_{q1\bot},\,\gamma_{\bot}^{i}])\gamma^{+}v_{p_{\bar{q}}}\right. \nonumber\\
&  \times\left[  \frac{1}{\vec{p_{q1}}^{2}}\ln\left(  \frac{\vec{p_{q1}}^{2}+x\bar{x}Q^2}{x\bar{x}Q^{2}}\right)  -
\frac{\vec{p}_{\bar{q}2}^{\,\,2}+x\bar{x}Q^{2}%
}{\left( \vec{p_{q1}}^{2}+x\bar{x} Q^{2}\right)  (\vec{p}_{\bar{q}2}^{\,\,2}+x\bar{x}Q^{2})-x\bar{x}Q^{2}\vec{p_{3}}^{2}}\right. \nonumber\\
&  \times\left.  \left.  \ln\left(  \frac{\left(\vec{p_{q1}}^{2}+x\bar{x}  Q^{2}\right)  \left( \vec{p}_{\bar{q}2}^{\,\,2}+x\bar{x} Q^{2}\right)
}{x\bar{x}Q^{2}\vec{p_{3}}{}^{2}}\right)  \right]  +(q\leftrightarrow\bar
{q})\right\}  +\bar{u}_{p_{q}}C_{2\bot}^{5i}v_{p_{\bar{q}}}.
\end{align}%
for a transverse photon. The contribution of diagram 6 reads :
\begin{equation}
\Phi_{2}^{+}|_{6}=-\left(  \frac{x\bar{x}p_{\gamma}^{+}(\bar{u}_{p_{q}%
}\gamma^{+}v_{p_{\bar{q}}})}{Q^{2}+x\bar{x}\vec{p}_{q1}^{\,\,2}}\left[
4\ln\left(  \frac{\bar{x}}{\alpha}\right)  \left(  \frac{1}{\epsilon}+\ln\left(
\frac{\vec{p_{3}}{}^{2}}{\mu^{2}}\right)  \right)  -\frac{3}{\epsilon}\right]
+\left(  q\leftrightarrow\bar{q}\right)  \right)  +\bar{u}_{p_{q}}C_{2\Vert
}^{6}v_{p_{\bar{q}}}.
\end{equation}
for a longitudinal photon, or 

\begin{align}
\Phi_{2}^{i}|_{6}  &  =-\left(  \frac{\overline{u}_{p_{q}}(p_{q1\bot}^{i}(1-2x)+\frac{1}{2}[\hat
{p}_{q1\bot},\,\gamma_{\bot}^{i}])\gamma^{+}v_{p_{\bar{q}}}}{Q^{2}+x\bar{x}\vec{p}_{q1}^{2}}\left[
2\ln\left(  \frac{\bar{x}}{\alpha}\right)  \left(  \frac{1}{\epsilon}+\ln\left(
\frac{\vec{p_{3}}{}^{2}}{\mu^{2}}\right)  \right)  -\frac{3}{2\epsilon
}\right]  \right. \nonumber\\
&  +\left.  \frac{{}}{{}}\left(  q\leftrightarrow\bar{q}\right)  \right)
+\bar{u}_{p_{q}}C_{2\bot}^{6i}v_{p_{\bar{q}}}
\end{align}
for a transverse photon. \\
In these expressions the functions $C$ do not contain singularities. Their exact
form will be given in Appendix A. The remaining (divergent) part contains a rapidity divergence of the form $\ln{\alpha}$. Such terms have to be absorbed into the renormalized Wilson operators with the help of the BK equation. Indeed the LO contribution as defined in (\ref{LOi}) involves the Wilson line operators at rapidity $\ln{\alpha}$. We thus have to use the BK evolution from $\alpha$ to $e^\eta$, by writing
\begin{equation}
U(x,\alpha) = U(x,e^\eta)+\int_\alpha^{e^\eta} \, d\rho \left( \frac{\partial U(x,\rho)}{\partial\rho}  \right) d\rho . \label{BKconv}
\end{equation}
Let us note that the BK equation is of order $\alpha_s$ so in the NLO contributions $U(x,\alpha)$ can be directly replaced by $U(x,e^\eta)$ without concern. \\
Plugging this (\ref{BKconv}) into (\ref{LOi}) and using the explicit BK equation (\ref{BKMOM}) allows one to evolve the LO dipole contribution into an NLO double-dipole contribution. This contribution reads :

\begin{align}
&  \langle T_{0}^{\alpha}\rangle^{\eta}=\int d^{d}p_{1\bot}d^{d}p_{2\bot
}\delta(p_{q1\bot}+p_{\bar{q}2\bot}-p_{\gamma\bot})\Phi_{0}^{\alpha}(p_{1\bot
},p_{2\bot})\nonumber\\
&  \times\ln(\frac{e^{\eta}}{\alpha})\delta(k_{1\bot}+k_{2\bot}+k_{3\bot
}-p_{1\bot}-p_{2\bot})2\alpha_{s}\mu^{2-d}\int\frac{d^{d}k_{1\bot}%
d^{d}k_{2\bot}d^{d}k_{3\bot}}{(2\pi)^{2d}}\nonumber\\
&  \times\left[  -\frac{2(k_{1\bot}-p_{1\bot})\cdot(k_{2\bot}-p_{2\bot})}{(k_{1}-p_{1}%
)_{\bot}^{2}(k_{2}-p_{2})_{\bot}^{2}} \right. \\ \nonumber &+ \left. \frac{\pi^{\frac{d}{2}}\Gamma(1-\frac
{d}{2})\Gamma(\frac{d}{2})^{2}}{\Gamma(d-1)}\left(  \frac{\delta(k_{2\bot
}-p_{2\bot})}{(-(k_{1}-p_{1})_{\bot}^{2})^{1-\frac{d}{2}}}+\frac
{\delta(k_{1\bot}-p_{1\bot})}{(-(k_{2}-p_{2})_{\bot}^{2})^{1-\frac{d}{2}}%
}\right)  \right] \\ \nonumber
&  \times[tr(U_{1}U_{3}^{\dag})tr(U_{3}U_{2}^{\dag})-N_{c}tr(U_{1}U_{2}^{\dag
})](k_{1\bot,}k_{2\bot},k_{3\bot}).
\end{align}
After integrating w.r.t. $p_{2\bot}$ and renaming the variables, we get%
\begin{align}
&  \langle T_{0}^{\alpha}\rangle^{\eta}=\int\frac{d^{d}p_{1\bot}d^{d}p_{2\bot
}d^{d}p_{3\bot}}{(2\pi)^{d}}\delta(p_{q1\bot}+p_{\bar{q}2\bot}-p_{3\bot
}-p_{\gamma\bot})\nonumber\\
&  \times\ln(\frac{e^{\eta}}{\alpha})2\alpha_{s}\mu^{2-d}\int\frac
{d^{d}p_{\bot}}{(2\pi)^{d}}\Phi_{0}^{\alpha}(p_{1\bot}+p_{\bot},p_{2\bot
}+p_{3\bot}-p_{\bot})\nonumber\\
&  \times\left[  \frac{2(p_\bot.(p_\bot-p_{3\bot}))}{p_{\bot}^{2}(p-p_{3})_{\bot}^{2}%
}+\frac{\pi^{\frac{d}{2}}\Gamma(1-\frac{d}{2})\Gamma(\frac{d}{2})^{2}}%
{\Gamma(d-1)}\left(  \frac{\delta(p_{\bot}-p_{3\bot})}{(-p_{\bot}%
^{2})^{1-\frac{d}{2}}}+\frac{\delta(p_{\bot})}{(-(p-p_{3})_{\bot}%
^{2})^{1-\frac{d}{2}}}\right)  \right] \nonumber\\
&  \times[tr(U_{1}U_{3}^{\dag})tr(U_{3}U_{2}^{\dag})-N_{c}tr(U_{1}U_{2}^{\dag
})](p_{1\bot,}p_{2\bot},p_{3\bot}).
\end{align}
Integrating w.r.t. $p_\bot$, one can get the contribution from this convolution :
\begin{eqnarray}
\Phi_{BK}^+\left( p_{1\bot}, p_{2\bot}, p_{3\bot}\right) & = & -4x\bar{x}p_\gamma^+ (\bar{u}_{p_q}\gamma^+ v_{p_{\bar{q}}} ) \ln \left( \frac{e^\eta}{\alpha} \right) \\ \nonumber
& \times & \left[  \left( \ln \left( \frac{  \vec{p}_3^{\, \, 2}}{\mu^2} \right) + \frac{1}{\epsilon} \right)  \left( \frac{-1}{\vec{p}_{q1}^{\, \, 2}+x\bar{x}Q^2} + \frac{-1}{\vec{p}_{\bar{q}2}^{\, \, 2} +x\bar{x}Q^2} \right) \right. \\ \nonumber
& + & \left. \frac{\vec{p}_3^{\, \, 2} - \vec{p}_{q1}^{\, \, 2} - \vec{p}_{\bar{q}2}^{\, \, 2} -2x\bar{x}Q^2}{(\vec{p}_{q1}^{\, \, 2}+x\bar{x}Q^2) (\vec{p}_{\bar{q2}}^{\, \, 2} +x\bar{x}Q^2) - x\bar{x}\vec{p}_3^{\, \, 2}Q^2}  \ln \left(  \frac{(\vec{p}_{q1}^{\, \, 2}+x\bar{x}Q^2) (\vec{p}_{\bar{q}2}^{\, \, 2}+x\bar{x}Q^2)}{x\bar{x}\vec{p}_3^{\, \, 2}Q^2}  \right)    \right] \, ,
\end{eqnarray}
in the longitudinal case, and
\begin{eqnarray}
\Phi_{BK}^i & = & -2\ln \left( \frac{e^\eta}{\alpha} \right) \left\{ \bar{u}_{p_q} \left[ (\hat{p}_{q1\bot} \gamma_\bot^i) -2x p_{q1\bot}^i \right] \gamma^+ v_{p_{	\bar{q}}} \right.  \\ \nonumber
& \times & \left[ \frac{-1}{\vec{p}_{q1}^{\, \, 2}+x\bar{x}Q^2} \left( \ln \left( \frac{\vec{p}_3^{\, \, 2}}{\mu^2} \right) +\frac{1}{\epsilon}  \right)  + \frac{1}{\vec{p}_{q1}^{\, \, 2}} \ln \left( \frac{\vec{p}_{q1}^{\, \, 2}+x\bar{x}Q^2}{x\bar{x}Q^2} \right) \right. \\ \nonumber
& - & \left. \left. \frac{\vec{p}_{\bar{q}2}^{\, \, 2}+x\bar{x}Q^2}{(\vec{p}_{q1}^{\, \, 2}+x\bar{x}Q^2) (\vec{p}_{\bar{q}2}^{\, \, 2}+x\bar{x}Q^2) - x\bar{x}\vec{p}_3^{\, \, 2}Q^2} \right. \right. \\ \nonumber
&\times & \left. \left. \ln \left( \frac{(\vec{p}_{q1}^{\, \, 2}+x\bar{x}Q^2) (\vec{p}_{\bar{q}2}^{\, \, 2} +x\bar{x}Q^2)}{x\bar{x}\vec{p}_3^{\, \, 2}Q^2} \right) \right] + (q\leftrightarrow \bar{q}) \right\} \, ,
\end{eqnarray}
in the transverse case. \\
Combining this substraction term with the results from diagrams 5 and 6 just before, we can cancel the rapidity divergence
in $\ln\alpha$ and obtain the actual double-dipole part $\Phi_2^\prime = \Phi_2 + \Phi_{BK}$ of the impact factor :
\begin{align}
\Phi_{2}^{\prime+}  &  =2p_{\gamma}^{+}(\bar{u}_{p_{q}}\gamma^{+}v_{p_{\bar{q}%
}})\left\{  \frac{x\bar{x}(\vec{p_{3}}{}^{2}-{}\vec{p}_{\bar{q}2}^{\,\,2}%
-\vec{p_{q1}}{}^{2}-2x\bar{x}Q^{2})}{(\vec{p}_{\bar{q}2}^{\,\,2}+x\bar{x}Q^{2})\left(\vec{p_{q1}}^{2}+x\bar{x}Q^{2}\right)  -x\bar{x}Q^{2}\vec
{p_{3}}{}^{2}}\right. \nonumber\\
&  \times\ln\left(  \frac{x\bar{x}}{e^{2\eta}}\right)  \ln\left(  \frac{(\vec
{p}_{\bar{q}2}^{\,\,2}+x\bar{x}Q^{2})\left(  \vec{p_{q1}}^{2}+x\bar{x}Q^{2}\right)  }{x\bar{x}Q^{2}\vec{p_{3}}{}^{2}}\right) \nonumber\\
&  -\left.  \left(  \frac{x\bar{x}}{Q^{2}+x\bar{x}\vec{p}_{q1}^{\,\,2}}\left[
2\ln\left(  \frac{\bar{x}}{e^{\eta}}\right)  \left(  \frac{1}{\epsilon}+\ln\left(
\frac{\vec{p_{3}}{}^{2}}{\mu^{2}}\right)  \right)  -\frac{3}{2\epsilon
}\right]  +\left(  q\leftrightarrow\bar{q}\right)  \right)  \right\}
\nonumber\\
&  +\bar{u}_{p_{q}}(C_{2\Vert}^{5}+C_{2\Vert}^{6})v_{p_{\bar{q}}}
\end{align}
in the longitudinal case, or 
\begin{align}
\Phi_{2}^{\prime i}  &  =\left\{ \overline{u}_{p_{q}}(p_{q1\bot}^{i}(1-2x)+\frac{1}{2}[\hat
{p}_{q1\bot},\,\gamma_{\bot}^{i}])\gamma^{+}v_{p_{\bar{q}}}\right.  \left(  \frac{-1}{Q^{2}+x\bar{x}\vec{p}_{q1}^{\, \, 2}%
}\right. \nonumber\\
&  \times\left[  2\ln\left(  \frac{\bar{x}}{e^{\eta}}\right)  \left(  \frac
{1}{\epsilon}+\ln\left(  \frac{\vec{p_{3}}{}^{2}}{\mu^{2}}\right)  \right)
-\frac{3}{2\epsilon}\right]  +\ln\left(  \frac{x\bar{x}}{e^{2\eta}}\right)
\nonumber\\
&  \times\left[  \frac{1}{\vec{p_{q1}}^{2}}\ln\left(  \frac{\vec{p_{q1}}^{2}+x\bar{x}Q^2}{x\bar{x}Q^{2}}\right)  -\frac{\vec{p}_{\bar{q}2}^{\,\,2}+x\bar{x}Q^{2}}{\left( \vec{p}_{q1}^{\, \, 2}+x\bar{x} Q^{2}\right)  (\vec{p}_{\bar{q}2}^{\,\,2}+x\bar{x}Q^{2}-x\bar{x}Q^{2}\vec{p_{3}}^{2})}\right. \nonumber\\
&  \times\left.  \left.  \left.  \ln\left(  \frac{\left( \vec
{p}_{q1}^{\, \, 2}+x\bar{x} Q^{2}\right)  \left( \vec{p}_{\bar{q}2}^{\,\,2} +x\bar{x}Q^{2}%
\right)  }{x\bar{x}Q^{2}\vec{p_{3}}{}^{2}}\right)  \right]  \right)
+(q\leftrightarrow\bar{q})\right\}  +\bar{u}_{p_{q}}(C_{2\bot}^{5i}+C_{2\bot
}^{6i})v_{p_{\bar{q}}}.
\end{align}
in the transverse case. \\
These impact factors still contain $\frac{1}{\epsilon}$ terms, although by construction
they should not have any IR, UV or collinear singularity. These poles are artificial UV poles and already appear in the momentum representation of the BK equation (\ref{BKMOM}). They originate from the fact that when we transform the Wilson line operator (\ref{trtrtr}) into its momentum space representation straightforwardly, we do not take into account its property of vanishing when $r_{3}=r_{2}$ or $r_{3}=r_{1}.$ This property reveals in the convolution of the impact factor and the operator (\ref{trtrtr})
killing all the artificial singularities. Indeed, the divergent terms depend
only on $\vec{p}_{1}$ and are independent of $\vec{p}_{3}$ and $\vec{p}_{2}$
(up to a (1$\leftrightarrow$2) permutation). Writing those terms as $F(p_{1\bot})$ and covoluting them as in (\ref{NLOi}) gives
\begin{align}
&  \int d^{d}p_{1\bot}d^{d}p_{2\bot}d^{d}p_{3\bot}\delta(p_{q1}+p_{\bar{q}%
2}-p_{\gamma\bot}-p_{3\bot})F\left(  p_{1\bot}\right) \nonumber\\
&  \times[tr(U_{1}U_{3}^{\dag})tr(U_{3}U_{2}^{\dag})-N_{c}tr(U_{1}U_{2}^{\dag
})](p_{1\bot,}p_{2\bot},p_{3\bot})\nonumber\\
=  &  \int d^{d}p_{1\bot}d^{d}p_{3\bot}d^{d}r_{1\bot}d^{d}r_{2\bot}d^{d}r_{3\bot} \nonumber\\
&  \times F\left(  p_{1\bot}\right)  e^{i(r_{1\bot}\cdot p_{1\bot})+ir_{2\bot
}\cdot(p_{q1\bot}+p_{\bar{q}\gamma\bot})+i(p_{3\bot}\cdot r_{32\bot})}[tr(U_{1}U_{3}^{\dag
})tr(U_{3}U_{2}^{\dag})-N_{c}tr(U_{1}U_{2}^{\dag})]\nonumber\\
\sim &  \int d^{d}p_{1\bot}d^{d}r_{1\bot}d^{d}r_{2\bot} F\left(  p_{1\bot}\right) e^{i(r_{1\bot} \cdot p_{1\bot})+ir_{2\bot}\cdot (p_{q1\bot}+p_{\bar{q}\gamma\bot})
}\nonumber\\
&  \times\int d^{d}r_{3\bot}\delta(r_{32\bot})[tr(U_{1}U_{3}^{\dag})tr(U_{3}%
U_{2}^{\dag})-N_{c}tr(U_{1}U_{2}^{\dag})]=0.
\end{align}
Thus the artificially divergent part
\begin{equation}
F(p_{1\bot})=\frac{x\bar{x}}{Q^{2}+x\bar{x}\vec{p}_{q1}^{\,\,2}}\left[  2\ln\left(
\frac{\bar{x}}{e^{\eta}}\right)    \frac{1}{\epsilon}   -\frac{3}{2\epsilon}\right] 
\end{equation}
will cancel once convoluted, so it can be omitted. For a more involved discussion about such terms, see Ref.~\cite{Fadin:2011jg}. \\
The same computation can allow one to omit the $\ln(\mu^2)$ contribution. However we will keep it so that no dimensional log appears, keeping in mind that there is no actual $\mu$ dependence. Therefore hereafter we will use
\begin{align}
\Phi_{2}^{\prime+}  &  =2p_{\gamma}^{+}(\bar{u}_{p_{q}}\gamma^{+}v_{p_{\bar{q}%
}})\left\{  \frac{x\bar{x}(\vec{p_{3}}{}^{2}-{}\vec{p}_{\bar{q}2}^{\,\,2}%
-\vec{p_{q1}}{}^{2}-2x\bar{x}Q^{2})}{(\vec{p}_{\bar{q}2}^{\,\,2}+x\bar{x}Q^{2})\left(\vec{p_{q1}}^{2}+x\bar{x}Q^{2}\right)  -x\bar{x}Q^{2}\vec
{p_{3}}{}^{2}}\right. \nonumber\\
&  \times\ln\left(  \frac{x\bar{x}}{e^{2\eta}}\right)  \ln\left(  \frac{(\vec
{p}_{\bar{q}2}^{\,\,2}+x\bar{x}Q^{2})\left(  \vec{p_{q1}}^{2}+x\bar{x}Q^{2}\right)  }{x\bar{x}Q^{2}\vec{p_{3}}{}^{2}}\right) \nonumber\\
&  \left. +  \left(  \frac{-2x\bar{x}}{Q^{2}+x\bar{x}\vec{p}_{q1}^{\,\,2}}
\ln\left(  \frac{\bar{x}}{e^{\eta}}\right)  \ln\left(
\frac{\vec{p_{3}}{}^{2}}{\mu^{2}}\right)  +\left(  q\leftrightarrow\bar{q}\right)  \right)  \right\}
\nonumber\\
&  +\bar{u}_{p_{q}}(C_{2\Vert}^{5}+C_{2\Vert}^{6})v_{p_{\bar{q}}}
\end{align}%
and
\begin{align}
\Phi_{2}^{\prime i}  &  =\left\{ \overline{u}_{p_{q}}(p_{q1\bot}^{i}(1-2x)+\frac{1}{2}[\hat
{p}_{q1\bot},\,\gamma_{\bot}^{i}])\gamma^{+}v_{p_{\bar{q}}}\right.  \left(  \frac{-2}{Q^{2}+x\bar{x}\vec{p}_{q1}^{\, \, 2}%
}\ln\left(  \frac{\bar{x}}{e^{\eta}}\right)  \ln\left(  \frac{\vec{p_{3}}{}^{2}}{\mu^{2}}  \right) \right. \nonumber\\
&    +\ln\left(  \frac{x\bar{x}}{e^{2\eta}}\right) \left[  \frac{1}{\vec{p_{q1}}^{2}}\ln\left(  \frac{\vec{p_{q1}}^{2}+x\bar{x}Q^2}{x\bar{x}Q^{2}}\right)  -\frac{\vec{p}_{\bar{q}2}^{\,\,2}+x\bar{x}Q^{2}}{\left( \vec{p}_{q1}^{\, \, 2}+x\bar{x} Q^{2}\right)  (\vec{p}_{\bar{q}2}^{\,\,2}+x\bar{x}Q^{2})-x\bar{x}Q^{2}\vec{p_{3}}^{2}}\right. \nonumber\\
&  \times\left.  \left.  \left.  \ln\left(  \frac{\left( \vec
{p}_{q1}^{\, \, 2}+x\bar{x} Q^{2}\right)  \left( \vec{p}_{\bar{q}2}^{\,\,2} +x\bar{x}Q^{2}%
\right)  }{x\bar{x}Q^{2}\vec{p_{3}}{}^{2}}\right)  \right]  \right)
+(q\leftrightarrow\bar{q})\right\}  +\bar{u}_{p_{q}}(C_{2\bot}^{5i}+C_{2\bot
}^{6i})v_{p_{\bar{q}}}.
\end{align}

\subsubsection{Dipole contribution $\Phi_{1}$}

The combined contributions of diagrams 2, 3 and the diagram obtained from 3
via $(q\leftrightarrow\bar{q})$ reads
\begin{align}
&  \Phi_{1}^{+}|_{23}=-\frac{p_{\gamma}^{+}}{p_{\gamma}^{-}}\Phi_{1}^{-}%
|_{23}=\frac{x\bar{x}p_{\gamma}^{+}(\overline{u}_{p_{q}}\gamma^{+}v_{p_{\bar
{q}}})}{\vec{p}_{q1}^{\,\,2}+x\bar{x}Q^{2}}\nonumber\\
&  \times\left[ \left(  2\ln\left(  \frac{x\bar{x}}{\alpha^{2}}\right)
-3\right)  \left(  \ln\left(  \frac{\left(  \vec{p}_{q1}^{\,\,2}%
+x\bar{x}Q^{2}\right)^{2}}{x\bar{x}\mu^{2}Q^{2}}\right)  +\frac{1}{\epsilon
}\right)  +\ln^{2}\left(  \frac{\bar{x}}{x}\right)  -\frac{\pi^{2}}{3}+6\right]  
\end{align}%
for a longitudinal photon and
\begin{align}
\Phi_{1}^{i}|_{23} &  =\frac{\overline{u}_{p_{q}}((1-2x)p_{q1\bot}^{i}%
+\frac{1}{2}[\hat{p}_{q1\bot}, \,\gamma_{\bot}^{i}])\gamma^{+}v_{p_{\bar{q}}}%
}{2(\vec{p}_{q1}^{\,\,2}+x\bar{x}Q^{2})}\left\{  \left(  2\ln\left(
\frac{x\bar{x}}{\alpha^{2}}\right)  -3\right)  \right.  \nonumber\\
&  \times\left(  \ln\left(  \frac{\vec{p}_{q1}^{\,\,2}+x\bar{x}Q^{2}}{\mu^{2}%
}\right)  +\frac{x\bar{x}Q^{2}}{\vec{p}_{q1}^{\,\,2}}\ln\left(  \frac
{x\bar{x}Q^{2}}{\vec{p}_{q1}^{\,\,2}+x\bar{x}Q^{2}}\right)  +\frac{1}{\epsilon
}\right)  \nonumber\\
&  \left.  +\ln^{2}\left(  \frac{\bar{x}}{x}\right)  -\frac{\pi^{2}}{3}+6\right\}
\end{align}
for a transverse photon. \\
This set of diagrams is electromagnetically gauge invariant. \\
The contribution of diagram 4 reads%
\begin{align}
&  \Phi_{1}^{+}|_{4}=\frac{x\bar{x}p_{\gamma}^{+}(\overline{u}_{p_{q}}%
\gamma^{+}v_{p_{\bar{q}}})}{\vec{p}_{q1}^{\,\,2}+x\bar{x}Q^{2}}\left\{  \ln
^{2}\left(  \frac{x\bar{x}}{\alpha^{2}}\right)  -\ln^{2}\left(  \frac{\bar{x}}%
{x}\right)  \right.  \nonumber\\
&  +\left.  2\ln\left(  \frac{x\bar{x}}{\alpha^{2}}\right)  \left(  \ln\left(
\frac{\left(  \vec{p}_{q1}^{\,\,2}+x\bar{x}Q^{2}\right)^{2}}{Q^{2}(x\vec
{p}_{\bar{q}}-\bar{x}\vec{p}_{q})^{2}}\right)  +i\pi\right)  \right\}
+\overline{u}_{p_{q}}C_{\Vert}^{4}v_{p_{\bar{q}}}
\end{align}
for a longitudinal photon and

\begin{align}
\Phi_{1}^{i}|_{4} &  =\frac{\overline{u}_{p_{q}}[(\hat{p}_{q1\bot}\gamma_{\bot}^{i}) -2xp_{q1\bot}^{i}]\gamma^{+}v_{p_{\bar{q}}}%
}{\vec{p}_{q1}^{\,\,2}+x\bar{x}Q^{2}}\left\{  \frac{1}{2}\ln^{2}\left(
\frac{x\bar{x}}{\alpha^{2}}\right)  -\frac{1}{2}\ln^{2}\left(  \frac{\bar{x}}%
{x}\right)  \right.  \nonumber\\
&  +\ln\left(  \frac{x\bar{x}}{\alpha^{2}}\right)  \left(  \frac{Q^{2}%
x\bar{x}}{\vec{p}_{q1}^{\,\,2}}\ln\left(  \frac{x\bar{x}Q^{2}}{\vec{p}_{q1}^{\,\,2}+x\bar{x}Q^{2}
}\right)  \right.  \nonumber\\
&  +\left.  \left.  \ln\left(  \frac{x\bar{x}\left(\vec{p}%
_{q1}^{\,\,2}+x\bar{x}Q^{2}\right)  }{(x\vec{p}_{\bar{q}}-\bar{x}\vec{p}_{q})^{2}}\right)
+i\pi\right)  \right\}  +\overline{u}_{p_{q}}C_{\bot}^{4i}v_{p_{\bar{q}}}
\end{align}
for a transverse photon. \\
The contribution of diagram 5 reads%
\begin{align}
\Phi_{1}^{+}|_{5}= &  \frac{x\bar{x}p_{\gamma}^{+} (\bar{u}_{p_{q}}\gamma
^{+}v_{p_{\bar{q}}})}{\vec{p}_{q1}^{\, \, 2}+x\bar{x}Q^{2}}\left(  2\ln\left(
\frac{x\bar{x}}{\alpha^{2}}\right)  \ln\left(  \frac{x\bar{x}Q^{4}}{\left(
\vec{p}_{q1}^{\, \, 2}+x\bar{x}Q^{2}\right)^{2}}\right)  \right.  \nonumber\\
&  -\left.  \ln^{2}\left(  \frac{x\bar{x}}{\alpha^{2}}\right)  +\ln^{2}\left(
\frac{\bar{x}}{x}\right)  \right)  +\bar{u}_{p_{q}}C_{1\Vert}^{5}v_{p_{\bar{q}}} \, ,
\end{align}%
for a longitudinal photon, and 
\begin{align}
&  \Phi_{1}^{i}|_{5}=-\frac{\overline{u}_{p_{q}}[(\hat{p}_{q1\bot}\gamma_{\bot}^{i})-2xp_{q1\bot}^{i}]\gamma^{+}v_{p_{\bar{q}}}%
}{\vec{p}_{q1}^{\,\,2}+x\bar{x}Q^{2}}\left(  \frac{1}{2}\ln^{2}\left(
\frac{x\bar{x}}{\alpha^{2}}\right)  -\frac{1}{2}\ln^{2}\left(  \frac{\bar{x}}%
{x}\right)  \right.  \nonumber\\
&  +\left.  \ln\left(  \frac{x\bar{x}}{\alpha^{2}}\right)  \left[  \frac{2x\bar{x}Q^{2}%
}{\vec{p}_{q1}^{\, \, 2}}\ln\left(  \frac{x\bar{x}Q^{2}}{\vec{p}_{q1}^{\, \, 2}+x\bar{x}Q^{2}}\right)  +\ln(x\bar{x})\right]  \right)  +\bar{u}_{p_{q}%
}C_{1\bot}^{5i}v_{p_{\bar{q}}} \, ,
\end{align}
for a transverse photon. \\
The contribution of diagram 6 reads%
\begin{align}
\Phi_{1}^{+}|_{6}= &  -\frac{x\bar{x}p_{\gamma}^{+}(\bar{u}_{p_{q}}\gamma
^{+}v_{p_{\bar{q}}})}{Q^{2}+x\bar{x}\vec{p}_{q1}^{\,\,2}}\left(  \ln\left(
\frac{x\bar{x}}{\alpha^{2}}\right)  \left[  \frac{4}{\epsilon}-2\ln\left(
\frac{x\bar{x}\mu^{4}}{(\vec{p}_{q1}^{\,\,2}+x\bar{x}Q^{2})^{2}}\right)
\right]  \right.  \nonumber\\
&  -\left.  \ln^{2}\left(  \frac{x\bar{x}}{\alpha^{2}}\right)  +\ln^{2}\left(
\frac{\bar{x}}{x}\right)  -\frac{6}{\epsilon}\right)  +\bar{u}_{p_{q}}C_{1\Vert
}^{6}v_{p_{\bar{q}}} \, ,
\end{align}%
for a longitudinal photon and
\begin{align}
\Phi_{1}^{i}|_{6}= &  -\frac{\overline{u}_{p_{q}}[(\hat{p}_{q1\bot}\gamma_{\bot}^{i})-2xp_{q1\bot}^{i}]\gamma^{+}v_{p_{\bar{q}}}%
}{x(1-x)Q^{2}+\vec{p}_{q1}^{\,\,2}}\left(  \frac{1}{2}\ln^{2}\left(
\frac{\bar{x}}{x}\right)  -\frac{1}{2}\ln^{2}\left(  \frac{x\bar{x}}{\alpha^{2}%
}\right)  \right.  \nonumber\\
&  +\left.  \ln\left(  \frac{x\bar{x}}{\alpha^{2}}\right)  \left[  \ln\left(
\frac{\left(  \vec{p}_{q1}^{\, \, 2}+x\bar{x}Q^{2}\right)^{2}}{x\bar{x}\mu^{4}}\right)
+\frac{2}{\epsilon}\right]  -\frac{3}{\epsilon}\right)  +\bar
{u}_{p_{q}}C_{1\bot}^{6i}v_{p_{\bar{q}}} \, ,
\end{align}
for a transverse photon. \\
As in the previous section, the $C$ functions do not contain singularities. They will be presented in Appendix A. \\
Summing the contributions from all diagrams finally gives :

\begin{equation}
\Phi_{1}^{\alpha}=\frac{S_{V}}{2}\Phi_{0}^{\alpha}+\Phi_{1R}^{\alpha} \, ,\label{Phi1p}%
\end{equation}
where the singular term reads%
\begin{align}
\frac{S_{V}}{2} &  =\left[  \ln\left(  \frac{x\bar{x}}{\alpha^{2}}\right)
-\frac{3}{2}\right]  \left[  \ln\left(  \frac{x\bar{x}\mu^{2}}{(x\vec{p}%
_{\bar{q}}-\bar{x}\vec{p}_{q})^{2}}\right)  -\frac{1}{\epsilon}\right]  +i\pi
\ln\left(  \frac{x\bar{x}}{\alpha^{2}}\right)  +\frac{1}{2}\ln^{2}\left(  \frac{x\bar{x}}{\alpha^{2}}\right)  -\frac{\pi
^{2}}{6}+3 \, , \label{SV}%
\end{align}
and the regular terms read
\begin{equation}
\Phi_{1R}^{+}=\frac{3}{2}\Phi_{0}^{+}\ln\left(  \frac{x^2\bar{x}^{2}\mu
^{4}Q^{2}}{(x\vec{p}_{\bar{q}}-\bar{x}\vec{p}_{q})^{2}\left(  \vec{p}_{q1}^{\, \, 2}+x\bar{x}Q^{2}\right)^{2}}\right)  +\bar{u}_{p_{q}}(C_{\Vert}%
^{4}+C_{1\Vert}^{5}+C_{1\Vert}^{6})v_{p_{\bar{q}}} \, ,
\end{equation}%
and
\begin{align}
\Phi_{1R}^{i} &  =\frac{3}{2}\Phi_{0}^{i}\left[  \ln\left(  \frac
{x\bar{x}\mu^{4}}{(x\vec{p}_{\bar{q}}-\bar{x}\vec{p}_{q})^{2}(\vec{p_{q1}}^{2}+x\bar{x}Q^{2})}\right)
 -\left.  \frac{x\bar{x}Q^{2}}{\vec{p}_{q1}^{\, \, 2}}\ln\left(  \frac
{x\bar{x}Q^{2}}{\vec{p}_{q1}^{\, \, 2}+x\bar{x}Q^{2}}\right)  \right]   \right.  \nonumber\\
&   +\bar{u}%
_{p_{q}}(C_{\bot}^{4i}+C_{1\bot}^{5i}+C_{1\bot}^{6i})v_{p_{\bar{q}}%
} \, .\label{Phi1l}%
\end{align}
Note that the $i\pi \ln\left(  \frac{x\bar{x}}{\alpha^{2}}\right)$ term will never contribute, since in the cross sections $\frac{S_V}{2}$ will actually always appear as $\frac{1}{2}(S_V + S_V^\ast)$.

\section{$\gamma\rightarrow q\bar{q}g$ impact factor}

We will now derive the $\gamma \rightarrow q\bar{q}g$ impact factor. In the body of this paper, it will be used to construct a well defined cross section for dijet production, free of the soft and the collinear singularities. The IR finiteness is discussed in details. The full expression for the $\gamma \rightarrow q\bar{q}g$ cross section is included in appendix B. \\
The computation of this impact factor in dimension 4 was already presented in \cite{Boussarie:2014lxa}.
For the purpose of the present study we need its divergent part in dimension $D$, therefore
we will rewrite our results for an arbitrary value of D. The corresponding matrix element for the EM current in the
shockwave background reads%
\begin{equation}
\tilde{M}^{\prime\alpha}=-ie_{q}\int d^{D}y_{0}\frac{e^{-i ( p_{\gamma} \cdot y_{0})}%
}{\sqrt{2p_{\gamma}^{+}}}\sqrt{\frac{2}{N_{c}^{2}-1}}(t^{r})_{l}^{n}%
\langle0|T(b_{p_{\bar{q}}}^{l}(a_{p_{q}})_{n}c_{p_{g}}^{r}\overline{\psi
}\left(  y_{0}\right)  \gamma^{\alpha}\psi\left(  y_{0}\right)  e^{i\int
\mathcal{L}_{i}\left(  z\right)  dz})|0\rangle_{sw},
\end{equation}
where $c$ is the gluon annihilation operator and $\sqrt{\frac{2}{N_{c}^{2}-1}%
}(t^{r})_{l}^{n}$ is the projector on the color singlet. We label the emitted
gluon momentum as
\begin{equation}
p_{g}^{\mu}=zp_{\gamma}^{+}n_{1}^{\mu}+\frac{-p_{g\bot}^{2}}{2zp_{\gamma}^{+}%
}n_{2}^{\mu}+p_{g\bot}^{\mu}.
\end{equation}
Again, we will work with the reduced matrix element $\tilde{T}^{\prime}$
\begin{equation}
\tilde{M}^{\prime\alpha}=\frac{-ie_{q}}{\sqrt{2p_{\gamma}^{+}}}\sqrt{\frac
{2}{N_{c}^{2}-1}}\frac{-i\delta(p_{q}^{+}+p_{\bar{q}}^{+}+p_{g}^{+}-p_{\gamma
}^{+})}{\left(  2\pi\right)  ^{D-3}\sqrt{2p_{\bar{q}}^{+}}\sqrt{2p_{q}^{+}%
}\sqrt{2p_{g}^{+}}}\tilde{T}^{\prime\alpha},
\end{equation}
which after subtraction of the noninteracting part can be parametrized as%
\begin{eqnarray}
T^{\prime\alpha} & = & g\mu^{-\epsilon}\int d^{d}p_{1\bot}d^{d}p_{2\bot
}\left\{  \delta(p_{q1\bot}+p_{\bar{q}2\bot}+p_{g\gamma\bot})\Phi_{3}^{\alpha
}\frac{N_{c}^{2}-1}{N_{c}}[tr(U_{1}U_{2}^{\dag})-N_{c}](p_{1\bot},p_{2\bot
})\right. \nonumber\\
&+& \left.  \int\frac{d^{d}p_{3\bot}}{(2\pi)^{d}}\delta(p_{q1\bot}+p_{\bar
{q}2\bot}+p_{g\gamma\bot}-p_{3\bot})\Phi_{4}^{\alpha} \right. \nonumber \\ 
&\times & \left. [tr(U_{1}U_{3}^{\dag
})tr(U_{3}U_{2}^{\dag})-N_{c}tr(U_{1}U_{2}^{\dag})](p_{1\bot,}p_{2\bot
},p_{3\bot})\right\}  . 
\end{eqnarray}
The expressions for the impact factors in $D$-dimensional space with a longitudinal photon read
\begin{equation}
\Phi_{4}^{+}=\frac{p_{\gamma}^{+}\overline{u}_{p_{q}}[2x_{q}g_{\bot}%
^{\mu\nu}+z(\gamma_{\bot}^{\nu}\gamma_{\bot}^{\mu})]\gamma
^{+}v_{p_{\bar{q}}}\varepsilon_{g\bot\mu}^{\ast}(zp_{q1\nu\bot}-x_{q}%
p_{g3\nu\bot}{})}{x_{q}z(x_{q}+z)\left(  Q^{2}+\frac{\vec{p}_{\bar{q}2}^{\, \, 2}}{x_{\bar{q}}(1-\bar{x})}\right)  \left(  Q^{2}+\frac{\vec{p}_{q1}^{\, \, 2}}{x_{q}}+\frac{\vec{p}_{\bar{q}2}^{\, \, 2}}{x_{\bar{q}}}+\frac{\vec{p}_{g3}^{\, \, 2}}{z}%
\right)  }-(q\leftrightarrow\bar{q}) \, ,
\end{equation}%
and
\begin{equation}
\Phi_{3}^{+}=\Phi_{4}^{+}|_{p_{3}=0}+\left( - \frac{zp_{\gamma}^{+}\overline
{u}_{p_{q}}\hat{\varepsilon}_{g}^{\ast}(\hat{p}_q+\hat{p}_g)\gamma^{+}v_{p_{\bar{q}}}}{x_{q}(\vec{p}_{g}%
-\frac{z}{x_{q}}\vec{p}_{q})^2\left(  Q^{2}+\frac{\vec{p}_{\bar{q}2}^{\, \, 2}}{x_{\bar{q}}(x_{q}+z)}\right)  }-(q\leftrightarrow\bar{q})\right)  .
\end{equation}
For a transverse photon, they read
\begin{eqnarray*}
\Phi_{4\perp}^{i} & = & \frac{\epsilon_{g\perp\mu}^{*}\bar{u}_{p_{q}}\gamma^{+}}{2x_{q}\bar{x}_{q}\left(x_{q}+z\right)\left(Q^{2}+\frac{\vec{p}_{\bar{q}2}^{2}}{x_{\bar{q}} \left(1 - x_{\bar{q}}\right) }\right)\left(Q^{2}+\frac{\vec{p}_{q1}^{2}}{x_{q}}+\frac{\vec{p}_{\bar{q}2}^{2}}{x_{\bar{q}}}+\frac{\vec{p}_{g3}^{2}}{z}\right)}\\
 &  & \left[x_{q}x_{\bar{q}}Q^{2}(\gamma_{\perp}^{\mu}\gamma_{\perp}^{i})+(\hat{p}_{q1\perp}\gamma_{\perp}^{\mu}\gamma_{\perp}^{i}\hat{p}_{\bar{q}2\perp})+4\frac{x_{q}}{z}p_{\bar{q}2\perp}^{i}(x_{\bar{q}}p_{g3\perp}^{\mu}-zp_{\bar{q}2\perp}^{\mu})\right.\\
 &  & \left.-2\frac{x_{q}}{z}p_{g3\perp}^{\mu}(\gamma_{\perp}^{i}\hat{p}_{\bar{q}2\perp})+2x_{q}p_{\bar{q}2\perp}^{i}(\hat{p}_{\bar{q}2\perp}\gamma_{\perp}^{\mu})-2x_{\bar{q}}p_{\bar{q}2\perp}^{i}(\hat{p}_{q1\perp}\gamma_{\perp}^{\mu})\right]v_{p_{\bar{q}}}-\left(q\leftrightarrow\bar{q}\right)
\end{eqnarray*}

and

\begin{equation}
\Phi_{3}^{i}=\Phi_{4}^{i}|_{p_{3}=0}+\left( - \frac{z\overline{u}_{p_{q}}%
\hat{\varepsilon}_{g}^{\ast}(\hat{p}_q + \hat{p}_g ) \gamma^{+}(\gamma_{\bot}^{i}\hat{p}_{\bar
{q}2\bot}-2x_{\bar{q}}p_{\bar{q}2\bot}^{i})v_{p_{\bar{q}}}}{2x_{q}x_{\bar{q}%
}\left( 1 -  x_{\bar{q}} \right)  (\vec{p}_{g}-\frac{z}{x_{q}}\vec{p}_q)^{2}\left(
Q^{2}+\frac{\vec{p}_{\bar{q}2}^{\, \, 2}}{x_{\bar{q}}\left( 1-  x_{\bar{q}}\right)
}\right)  }-(q\leftrightarrow\bar{q})\right)  .
\end{equation}

\section{Construction of the $\gamma P\rightarrow q\bar{q}P^{\prime}$ cross
section}
Let us define the reduced matrix element $A_3$ such that the $\gamma P\rightarrow q\bar{q}P^{\prime}$ cross section reads%
\begin{equation}
d\sigma=\frac{1}{4s}(2\pi)^{D}\delta^{(D)}(p_{\gamma}+p_{0}-p_{q}-p_{\bar{q}%
}-p_{0}^{\prime})|A_{3}|^{2}d\rho_{3}.
\end{equation}
We will need the parametrization of the proton matrix elements in the shockwave background%
\begin{equation}
\langle P^{\prime}(p_{0}^{\prime})|(tr(U_{\frac{z_{\bot}}{2}}U_{-\frac
{z_{\bot}}{2}}^{\dag})-N_{c})|P(p_{0})\rangle \equiv 2\pi\delta(p_{00^{\prime}}%
^{-})F_{p_{0\bot}p_{0\bot}^{\prime}}(z_{\bot}) \equiv 2\pi\delta(p_{00^{\prime}}%
^{-})F(z_{\bot}),
\end{equation}%
\begin{align}
&  \langle P^{\prime}(p_{0}^{\prime})|(tr(U_{\frac{z}{2}}U_{x}^{\dag}%
)tr(U_{x}U_{-\frac{z}{2}}^{\dag})-N_{c}tr(U_{\frac{z}{2}}U_{-\frac{z}{2}%
}^{\dag}))|P(p_{0})\rangle\nonumber\\
& \equiv 2\pi\delta(p_{00^{\prime}}^{-})\tilde{F}_{p_{0\bot}p_{0\bot}^{\prime}%
}(z_{\bot},x_{\bot}) \equiv 2\pi\delta(p_{00^{\prime}}^{-})\tilde{F}(z_{\bot}%
,x_{\bot}).
\end{align}
We dropped the dependence on the proton transverse momenta ${p_{0\bot
}p_{0\bot}^{\prime}}$\ for convenience, and we assumed the following proton state normalization :
\begin{equation}
\langle P^{\prime}(p_{0}^{\prime})|P(p_{0})\rangle=(2\pi)^{D-1}\delta
(p_{00^{\prime}}^{-})\delta_{\bot}^{D-2}(p_{00^{\prime}\mathbf{\bot}}%
)\delta_{s_{P}s_{P^{\prime}}}
\end{equation}
The corresponding Fourier transforms
read%
\begin{align}
\int d^{d}z_{\bot}e^{i(z_{\bot} \cdot p_{\bot})}F(z_{\bot})  &  \equiv \mathbf{F}(p_{\bot
}),\\
\int d^{d}z_{\bot}d^{d}x_{\bot}e^{i(p_{\bot} \cdot x_{\bot})+i(z_{\bot} \cdot q_{\bot})}%
\tilde{F}(z_{\bot},x_{\bot})  & \equiv \mathbf{\tilde{F}}(q_{\bot},p_{\bot}).
\end{align}
These hadronic matrix elements naturally appear when we insert the Wilson line
operators between the proton states and we extract the overall momentum
conservation delta functions.
The matrix element for the dipole operator reads
\begin{align}
&  \langle P^{\prime}(p_{0}^{\prime})|(tr(U_{1}U_{2}^{\dag})-N_{c})[p_{1\bot},p_{2\bot
}]|P(p_{0})\rangle\nonumber\\
&  =(2\pi)^{d}\delta(p_{1\bot}+p_{2\bot}+p_{0^{\prime}0\bot})\int d^{d}%
z_{\bot}e^{i\frac{(z_{\bot}\cdot p_{12\bot})}{2}}\langle P^{\prime}(p_{0}^{\prime
})|(tr(U_{\frac{z_{\bot}}{2}}U_{-\frac{z_{\bot}}{2}}^{\dag})-N_{c}%
)|P(p_{0})\rangle.
\end{align}
For the double dipole operator the analogous formula has the form :
\begin{align}
&\langle P^{\prime}(p_{0}^{\prime})|(tr(U_{1}U_{3}^{\dag
})tr(U_{3}U_{2}^{\dag})-N_{c}tr(U_{1}U_{2}^{\dag}))[p_{1\bot,}p_{2\bot
},p_{3\bot}]|P(p_{0})\rangle\\
&  =(2\pi)^{d}\delta(p_{1\bot}+p_{2\bot}+p_{3\bot}+p_{0^{\prime}0\bot
})\nonumber\\
&  \times\int d^{d}z_{\bot}d^{d}x_{\bot}e^{i\frac{(z_{\bot} \cdot p_{12\bot
})}{2}+i(p_{3\bot} \cdot x_{\bot})}\langle P^{\prime}(p_{0}^{\prime})|(tr(U_{\frac{z}{2}%
}U_{x}^{\dag})tr(U_{x}U_{-\frac{z}{2}}^{\dag})-N_{c}tr(U_{\frac{z}{2}%
}U_{-\frac{z}{2}}^{\dag}))|P(p_{0})\rangle. \nonumber
\end{align}
In our kinematics, momentum conservation reads
\begin{align}
\hspace{-.2cm} \delta^{(D)}(p_{\gamma}+p_{0}-p_{q}-p_{\bar{q}}-p_{0}^{\prime})  &
=\delta(p_{00^{\prime}}^{-})\delta(p_{q}^{+}+p_{\bar{q}}^{+}-p_{\gamma}%
^{+})\delta^{(d)}(p_{q\bot}+p_{\bar{q}\bot}-p_{\gamma\bot}+p_{0^{\prime}0\bot
}),
\end{align}
with the phase space measure
\begin{align}
d\rho_{3}  &  =\frac{dp_{q}^{+}d^{d}p_{q\bot}}{2p_{q}^{+}(2\pi)^{d+1}}%
\frac{dp_{\bar{q}}^{+}d^{d}p_{\bar{q}\bot}}{2p_{\bar{q}}^{+}(2\pi)^{d+1}}%
\frac{dp_{0}^{\prime-}d^{d}p_{0\bot}^{\prime}}{2p_{0}^{\prime-}(2\pi)^{d+1}} \, .
\end{align}
The reduced matrix element $A_{3}$ includes the LO and NLO dipole contributions and the NLO double dipole contribution, as defined in section 3. It reads :
\begin{align}
A_{3}  &  =\frac{-2p_{0}^{-}e_{q}\varepsilon_{\alpha}}{\sqrt{N_{c}}\left(
2\pi\right)  ^{D-4}}\int d^{d}p_{1\bot}d^{d}p_{2\bot} \\ \nonumber 
& \times \left[  \delta(p_{q1\bot
}+p_{\bar{q}2\bot}-p_{\gamma\bot})\left\{  \Phi_{0}^{\alpha}+\alpha_{s}%
\frac{\Gamma(1-\epsilon)}{\left(  4\pi\right)  ^{1+\epsilon}}\frac{N_{c}%
^{2}-1}{N_{c}}\Phi_{1}^{\alpha}\right\}  \mathbf{F}\left(\frac{p_{12}{}_{\bot}}%
{2}\right)\right. \nonumber\\
&  +\left.  \alpha_{s}\frac{\Gamma(1-\epsilon)}{\left(  4\pi\right)
^{1+\epsilon}}\int\frac{d^{d}p_{3\bot}}{(2\pi)^{d}}\delta(p_{q1\bot}%
+p_{\bar{q}2\bot}-p_{\gamma\bot}-p_{3\bot})\Phi_{2}^{\alpha}\mathbf{\tilde{F}%
}\left(\frac{p_{12}{}_{\bot}}{2},p_{3\bot}\right)\right] \nonumber  .
\end{align}
Since the photon in the initial state can appear with different polarizations, we need the density matrix constructed from the cross sections
\begin{equation}
d\sigma_{JI}=%
\begin{pmatrix}
d\sigma_{LL} & d\sigma_{LT}\\
d\sigma_{TL} & d\sigma_{TT}%
\end{pmatrix}
,\qquad d\sigma_{TL}=d\sigma_{LT}^{\ast}.
\end{equation}
Each element of this matrix has a LO contribution $d\sigma_{0}$, an NLO contribution $d\sigma_{1}$ involving two dipole operators and an NLO contribution $d\sigma_{2}$ involving a dipole operator and a double-dipole operator.
\begin{equation}
d\sigma_{JI}=d\sigma_{0JI}+d\sigma_{1JI}+d\sigma_{2JI}. \label{sigmaV}
\end{equation}
The leading order cross section can be written as
\begin{eqnarray}  \label{dsigma0} 
d\sigma_{0JI} & = & \frac{\alpha_{\mathrm{em}}Q_{q}^{2}}{\left(2\pi\right)^{4\left(d-1\right)}N_{c}}\frac{\left(p_{0}^{-}\right)^{2}}{2x\bar{x}s^{2}}dxd\bar{x}d^{d}p_{q\perp}d^{d}p_{\bar{q}\perp}\delta\left(1-x-\bar{x}\right)\left(\varepsilon_{I\beta}\varepsilon_{J\gamma}^\ast\right)\\ \nonumber
 & \times & \int d^{d}p_{1\perp}d^{d}p_{2\perp}d^{d}p_{1^{\prime}\perp}d^{d}p_{2^{\prime}\perp}\delta\left(p_{q1\perp}+p_{\bar{q}2\perp}\right)\delta\left(p_{11^{\prime}\perp}+p_{22^{\prime}\perp}\right)\\ \nonumber
 & \times & \Phi_{0}^{\beta}\left(p_{1\perp},\, p_{2\perp}\right)\Phi_{0}^{\gamma*}\left(p_{1^{\prime}\perp},\, p_{2^{\prime}\perp}\right)\mathbf{F}\left(\frac{p_{12\perp}}{2}\right)\mathbf{F^{*}}\left(\frac{p_{1^{\prime}2^{\prime}\perp}}{2}\right) .
\end{eqnarray}

The dipole $\times$ dipole NLO cross section is given by

\begin{eqnarray} \nonumber
d\sigma_{1JI} & = & \alpha_{s}\frac{\Gamma\left(1-\epsilon\right)}{\left(4\pi\right)^{1+\epsilon}}\left(\frac{N_{c}^{2}-1}{N_{c}}\right)\frac{\alpha_{\mathrm{em}}Q_{q}^{2}}{\left(2\pi\right)^{4\left(d-1\right)}N_c}\frac{\left(p_{0}^{-}\right)^{2}}{2x\bar{x}s^{2}}dxd\bar{x}d^{d}p_{q\perp}d^{d}p_{\bar{q}\perp}\delta\left(1-x-\bar{x}\right)\left(\varepsilon_{I\beta}\varepsilon_{J\gamma}^\ast\right)\\ \nonumber
 & \times & \int d^{d}p_{1\perp}d^{d}p_{2\perp}d^{d}p_{1^{\prime}\perp}d^{d}p_{2^{\prime}\perp}\delta\left(p_{q1\perp}+p_{\bar{q}2\perp}\right)\delta\left(p_{11^{\prime}\perp}+p_{22^{\prime}\perp}\right)\mathbf{F}\left(\frac{p_{12\perp}}{2}\right)\mathbf{F^{*}}\left(\frac{p_{1^{\prime}2^{\prime}\perp}}{2}\right)\\  \label{dsigma1}
 & \times & \left[\Phi_{1}^{\beta}\left(p_{1\perp},\, p_{2\perp}\right)\Phi_{0}^{\gamma*}\left(p_{1^{\prime}\perp},\, p_{2^{\prime}\perp}\right)+\Phi_{0}^{\beta}\left(p_{1\perp},\, p_{2\perp}\right)\Phi_{1}^{\gamma*}\left(p_{1^{\prime}\perp},\, p_{2^{\prime}\perp}\right)\right]
\end{eqnarray}

We can separate this cross section into its divergent part and its convergent part. To get the convergent part, one only has to replace $\Phi_1$ by $\Phi_{1R}$ in the previous equation and to set $\epsilon$ to 0. The remaining divergent part reads

\begin{equation}
(d\sigma_{1JI})_{div} = \alpha_s \frac{\Gamma(1-\epsilon)}{(4\pi)^{1+\epsilon}}\left( \frac{N_c^2-1}{2 N_c} \right)(S_V+S_V^\ast) \, d\sigma_{0JI} .
\end{equation}

Replacing $\Phi_2$ by the contribution $\Phi_2^\prime$ which includes the BK evolution (see the discussion in section 3.3), one gets a non-divergent dipole $\times$ double dipole NLO contribution, which reads 

\begin{eqnarray}  \label{dsigma2}
d\sigma_{2JI} & = & \alpha_{s}\frac{\Gamma\left(1-\epsilon\right)}{\left(4\pi\right)^{1+\epsilon}}\frac{\alpha_{\mathrm{em}}Q_{q}^{2}}{\left(2\pi\right)^{4\left(d-1\right)}N_{c}}\frac{\left(p_{0}^{-}\right)^{2}}{2x\bar{x}s^{2}}dxd\bar{x}d^{d}p_{q\perp}d^{d}p_{\bar{q}\perp}\delta\left(1-x-\bar{x}\right)\left(\varepsilon_{I\beta}\varepsilon_{J\gamma}^\ast\right)\\ \nonumber
 &  & \int d^{d}p_{1\perp}d^{d}p_{2\perp}d^{d}p_{1^{\prime}\perp}d^{d}p_{2^{\prime}\perp}\frac{d^{d}p_{3\perp}d^{d}p_{3^{\prime}\perp}}{\left(2\pi\right)^{d}}\delta\left(p_{q1\perp}+p_{\bar{q}2\perp}-p_{3\perp}\right)\delta\left(p_{11^{\prime}\perp}+p_{22^{\prime}\perp}+p_{33^{\prime}\perp}\right)\\ \nonumber
 &  & \left[\Phi_{2}^{\prime\beta}\left(p_{1\perp},\, p_{2\perp},\, p_{3\perp}\right)\Phi_{0}^{\gamma\ast}\left(p_{1^{\prime}\perp},\, p_{2^{\prime}\perp}\right)\mathbf{F}^{\ast}\left(\frac{p_{1^{\prime}2^{\prime}\perp}}{2}\right)\tilde{\mathbf{F}}\left(\frac{p_{12\perp}}{2},\, p_{3\perp}\right)\delta\left(p_{3^{\prime}\perp}\right)\right.\\ \nonumber
 &  & \left.+\Phi_{2}^{\prime \gamma\ast}\left(p_{1^{\prime}\perp},\, p_{2^{\prime}\perp},\, p_{3^{\prime}\perp}\right)\Phi_{0}^{\beta}\left(p_{1\perp},\, p_{2\perp}\right)\mathbf{F}\left(\frac{p_{12\perp}}{2}\right)\tilde{\mathbf{F}}^{\ast}\left(\frac{p_{1^{\prime}2^{\prime}\perp}}{2},\, p_{3^{\prime}\perp}\right)\delta\left(p_{3\perp}\right)\right]
\end{eqnarray}

\subsection{Results for the Born cross section}

Using (\ref{LOifplus}) and (\ref{LOifi}) and summing over helicities, one gets%
\begin{equation}
\sum_{helicities}\! \!  \Phi_{0}^{+}(p_{1\bot},p_{2\bot})\Phi_{0}^{+}(p_{1\bot
}^{\prime},p_{2\bot}^{\prime}) = \frac{32(p_{\gamma}^{+}%
)^{4}x^{3}\bar{x}^3}{(\vec{p}_{q1}^{\, \, 2}+x\bar{x}Q^{2})(\vec{p}_{q1^\prime}^{\, \, 2}+x\bar{x}Q^{2})},
\end{equation}%
\begin{equation}
\sum_{helicities} \! \!  \Phi_{0}^{+}(p_{1\bot},p_{2\bot})\Phi_{0}^{i}(p_{1\bot
}^{\prime},p_{2^{\prime}\bot})^{\ast} = \frac{16(p_{\gamma}%
^{+})^{3}x^{2}\bar{x}^2	p_{q1^{\prime}\bot}^{i}(1-2x)}{(\vec{p}_{q1}^{\, \, 2}+x\bar{x}Q^{2})(\vec{p}_{q1^\prime}^{\, \, 2}+x\bar{x}Q^{2})},
\end{equation}%
and
\begin{align}
\sum_{helicities} \! \!  \Phi_{0}^{i}(p_{1\bot},p_{2\bot})\Phi_{0}^{k}(p_{1\bot
}^{\prime},p_{2^{\prime}\bot})^{\ast} =  \frac{ 8(p_{\gamma}%
^{+})^{2}x\bar{x} [(1-2x)^{2}g_{\bot}^{ri}g_{\bot}^{lk}-g_{\bot}^{rk}g_{\bot
}^{li}+g_{\bot}^{rl}g_{\bot}^{ik}]p_{q1\bot r}p_{q1^{\prime}\bot l} }
{(\vec{p}_{q1}^{\, \, 2}+x\bar{x}Q^{2})(\vec{p}_{q1^\prime}^{\, \, 2}+x\bar{x}Q^{2})}.
\end{align}
As a result, the LO density matrix elements read
\begin{eqnarray}
d\sigma_{0LL}&=&\frac{4\alpha_{\mathrm{em}}Q_{q}^{2}}{(2\pi)^{4}N_{c}}
dxd\bar{x}d^{d}p_{q\bot}d^{d}p_{\bar{q}\bot}\delta(1-x-\bar{x}) x^2\bar{x}^2Q^{2} \\ \nonumber
& \times & \left\vert \int\frac{d^{d}p_{1\bot}}{\vec{p}_{q1}^{\, \, 2}+x\bar{x}Q^{2}} \mathbf{F}(p_{1q\bot}+\frac{p_{q\bar{q}}{}_{\bot}}{2}) \right\vert^{2} \, ,
\end{eqnarray}%

\begin{eqnarray}
d\sigma_{0TL}   &=& \frac{2\alpha_{\mathrm{em}}Q_{q}^{2}}{(2\pi)^{4}N_{c}}
dxd\bar{x}d^{d}p_{q\bot}d^{d}p_{\bar{q}\bot} \delta(1-x-\bar{x}) x\bar{x}(1-2x)Q \\ \nonumber
 &\times & \left[ \int \frac{d^{d}p_{1\bot}}{\vec{p}_{q1}^{\, \, 2}+x\bar{x}Q^{2}} \mathbf{F}(p_{1q\bot}+\frac{p_{q\bar{q}\bot}}{2}) \right] \left[  \int \frac{d^{d}p_{1\bot}^{\prime}(\varepsilon_\bot \cdot p_{q1^{\prime}\bot})}{\vec{p}_{q1^\prime}^{\, \, 2} +x\bar{x}Q^{2}} \mathbf{F}(p_{1^{\prime}q\bot}+\frac{p_{q\bar{q}\bot}}{2}) \right]  ^{\ast},
\end{eqnarray}%

and

\begin{eqnarray}
d\sigma_{0TT}  &=& \frac{\alpha_{\mathrm{em}} Q_{q}^{2}}{(2\pi)^{4}N_{c}}dxd\bar{x}d^{d}%
p_{q\bot}d^{d}p_{\bar{q}\bot}\delta(1-x-\bar{x})[(1-2x)^{2}g_{\bot}^{ki}g_{\bot}^{lj}-g_{\bot
}^{kj}g_{\bot}^{li}+g_{\bot}^{kl}g_{\bot}^{ij}]\\ \nonumber
&\times & \left[ \int \frac{d^{d}p_{1\bot} ( \varepsilon_{\bot i}p_{q1\bot k} )}{\vec{p}_{q1}^{\, \, 2}+x\bar{x}Q^2}\mathbf{F}(p_{1q\bot}+\frac{p_{q\bar{q}\bot}}{2}) \right]  \left[  \int \frac{ d^{d}p_{1\bot}^{\prime} (\varepsilon_{\bot j}p_{q1^{\prime}\bot l})}{\vec{p}_{q1^\prime}^{\, \, 2}+x\bar{x}Q^{2}} \mathbf{F}(p_{1^{\prime}q\bot}+\frac{p_{q\bar{q}\bot}}{2}) \right]^{\ast}.
\end{eqnarray}

\subsection{Dipole - dipole NLO cross section $d\sigma_{1}$}

\subsubsection{LL photon transition}

Combining (\ref{dsigma1}), (\ref{Phi1p}) and (\ref{LOifplus}), and summing over the polarization components $ \varepsilon^+\Phi_0^- + \varepsilon^-\Phi_0^+ $ with the help of the gauge invariance relation $\Phi_0^- = \frac{Q^2}{2\left( p_\gamma^+\right)^2}\Phi_0^+$, we get
\begin{eqnarray} \label{dsigma1LL}
d\sigma_{1LL}  & = & \alpha_{s}\frac{\Gamma(1-\epsilon)}{\left(  4\pi\right)
^{1+\epsilon}} \left( \frac{N_{c}^{2}-1}{2N_{c}} \right) (S_{V}+S_{V}^{\ast
})d\sigma_{0LL} \\ \nonumber 
& + & \frac{\alpha_{s}Q^{2}}{4\pi} \left( \frac{N_{c}^{2}-1}{N_{c}} \right) \frac{\alpha_{\mathrm{em}}Q_{q}^{2}}{\left(  2\pi\right)^{4}N_c}  dxd\bar{x}d^{d}p_{q\bot}d^{d}p_{\bar{q}\bot
} \delta(1-x-\bar{x})\\ \nonumber
& \times & \int d^{d}p_{1\bot} d^dp_{2\bot} d^{d}p_{1\bot}^{\prime} d^dp_{2\bot}^\prime 
\delta(p_{q1\bot}+p_{\bar{q}2\bot}) \frac{\delta(p_{11^\prime\bot}+p_{22^\prime\bot})}{ \vec{p}_{q1^{\prime}}^{\, \, 2} +x\bar{x}  Q^{2}} \mathbf{F}\left( \frac{p_{12\bot}}{2}\right)\mathbf{F}^\ast\left( \frac{p_{1^\prime 2^\prime \bot}}{2}\right)
\\ \nonumber
& \times &  \left[  \frac{6x^{2}\bar{x}^2}{\vec{p}_{q1}^{\, \, 2}+x\bar{x}Q^{2}}\ln\left(  \frac
{x^{2}\bar{x}^2\mu^{4}Q^{2}}{(x\vec{p}_{\bar{q}}-\bar{x}\vec{p}_{q})^{2}(\vec
{p}_{q1}^{\,\,2}+x\bar{x}Q^{2}){}^{2}}\right) + \frac{(p_{0}^{-})^{2}}{s^{2}p_{\gamma}^{+}}tr((C_{\Vert}%
^{4}+C_{1\Vert}^{5}+C_{1\Vert}^{6})\hat{p}_{\bar{q}}\gamma^{+}\hat{p}%
_{q})\right] +h.c.
\end{eqnarray}
We will parametrize the finite contribution of the $C$ functions as :
\begin{equation}
\frac{(p_{0}^{-})^{2}}{s^{2}p_{\gamma}^{+}}tr(C_{||}^{n}\hat{p}_{\bar{q}%
}\gamma^{+}\hat{p}_{q})=\int_{0}^{x}dz\left[(\phi_n)_{LL}\right]_+ +(q\leftrightarrow\bar{q}) \, , \label{phi1LL}
\end{equation}
where $n=4,\, 5 \, \, \mathrm{or} \, \, 6$. The expressions for $(\phi_n)_{LL}$ are given in appendix A. For $n=5 \, \mathrm{or} \, 6$, these expressions must be evaluated at $\vec{p}_3 = \vec{0}$.

\subsubsection{LT photon transition}

Using the same method as for the LL component, we get
\begin{eqnarray}
&& d\sigma_{1TL} = \alpha_{s}\frac{\Gamma(1-\epsilon)}{\left(  4\pi\right)^{1+\epsilon}} \left( \frac{N_{c}^{2}-1}{2N_{c}} \right) (S_{V}+S_{V}^{\ast}) \, d\sigma_{0TL} \label{dsigma1LT} \\ \nonumber
&+& \frac{\alpha_{s}Q}{\left(  4\pi\right)
}\left( \frac{N_{c}^{2}-1}{N_{c}} \right)\frac{\alpha_{\mathrm{em}} Q_{q}^{2}}{(2\pi)^{4}N_c}%
dxd\bar{x}d^{d}p_{q\bot}d^{d}p_{\bar{q}\bot} \delta(1-x-\bar{x}) \varepsilon_{Ti}^{\ast} \\ \nonumber
&\times & \int d^{d}p_{1\bot} d^dp_{2\bot} d^{d}p_{1\bot}^{\prime} d^dp_{2^\prime \bot} \delta(p_{q1\bot}+p_{\bar{q}2\bot})\delta(p_{11^\prime\bot}+p_{22^\prime\bot}) \mathbf{F}\left(\frac{p_{12\bot}}{2}\right)\mathbf{F}^\ast\left(\frac{p_{1^\prime 2^\prime \bot}}{2}\right) \\ \nonumber
&\times & \left[  \frac{(p_{0}^{-})^{2}}{s^{2}}\frac{tr((C_{\bot}^{\prime
4i}+C_{1\bot}^{\prime5i}+C_{1\bot}^{\prime6i})\hat{p}_{\bar{q}}\gamma^{+}%
\hat{p}_{q})^{\dag}}{\vec{p}_{q1}^{2}+x\bar{x}Q^{2}}+\frac
{3x\bar{x}(1-2x)p_{q1^{\prime}\bot}^{i}}{(\vec{p}_{q1}^{\, \,2}+x\bar{x}Q^{2})(\vec{p}_{q1^{\prime}}^{\, \, 2}+x\bar{x}Q^{2})}\right. \nonumber\\
&\times &\left(  \ln\left(  \frac{x^{3}\bar{x}^3\mu^{8}Q^{2}(x\vec{p}_{\bar
{q}}-\bar{x}\vec{p}_{q})^{-4}}{(\vec{p}_{q1}^{\,\,2}+x\bar{x}Q^{2})^{2}(\vec
{p}_{q1^{\prime}}^{\,\,2}+x\bar{x}Q^{2})}\right)  \right.  -\left.
\frac{x\bar{x}Q^{2}}{\vec{p}_{q1^{\prime}}^{\,\,2}}\ln\left(  \frac
{x\bar{x}Q^{2}}{\vec{p}_{q1^{\prime}}^{\,\,2}+x\bar{x}Q^{2}}\right)  \right)
\nonumber\\
&+& \left.  \frac{(p_{0}^{-})^{2}}{s^{2}}\frac{tr((C_{\Vert}^{4}+C_{1\Vert
}^{5}+C_{1\Vert}^{6})\hat{p}_{\bar{q}}(\gamma^{i}\hat{p}_{q1^{\prime}\bot}-2x p_{q1^{\prime}\bot}^{i})\gamma^{+}\hat{p}_{q})}%
{2p_{\gamma}^{+}x\bar{x}\left( \vec{p}_{q1^{\prime}}^{\, \,2} +x\bar{x} Q^{2}\right)
}\right]  .
\end{eqnarray}
Once more we will parametrize the contributions from the $C$ functions, as
\begin{equation}
\frac{(p_{0}^{-})^{2}tr(C_{||}^{n}\hat{p}_{\bar{q}}((1-2x)p_{q1^{\prime}\bot
}^{i}-\frac{1}{2}[\hat{p}_{q1^{\prime}\bot}\gamma_{\bot}^{i}])\gamma^{+}%
\hat{p}_{q})}{s^{2}p_{\gamma}^{+}}=\int_{0}^{x}dz[(\phi_n)_{LT}^i]_+
+(q\leftrightarrow\bar{q}) \, ,
\end{equation}
and
\begin{equation}
\frac{(p_{0}^{-})^{2}}{s^{2}}tr(C_{\bot}^{ni}\hat{p}_{\bar{q}}\gamma^{+}%
\hat{p}_{q})=\int_{0}^{x}dz[(\phi_n)_{TL}^i]_++(q\leftrightarrow\bar{q}) \, ,
\end{equation}
with $n=4, \, 5, \, 6$. The values for $(\phi_n)$ are given in Appendix A, although for $n = 5 \, \mathrm{or} \, 6$ they must be evaluated for $\vec{p}_3=\vec{0}$.
\begin{align}
&  \frac{(p_{0}^{-})^{2}}{s^{2}}tr(C_{1\bot}^{6i}\hat{p}_{\bar{q}}\gamma
^{+}\hat{p}_{q})=\frac{-x\bar{x}(1-2x)p_{q1\bot}^{\, \, i}}{\vec{p}_{q1}^{\,\,2}+
x\bar{x}Q^{2}-\bar{x}\vec{p}_{1}^{\, \, 2}}\ln\left(
\frac{\bar{x}\vec{p}_{1}^{\, \, 2}}{\vec{p}_{q1}^{\,\,2}+x\bar{x}Q^{2}}\right)
\nonumber\\
&  +\frac{x\bar{x}(1-2x)p_{q1\bot}^{i}}{\vec{p}_{q1}^{\,\,2}+x\bar{x}Q^{2}%
}\left[  4\text{Li}_{2}\left(  \frac{\bar{x}\vec{p}_{1}^{\, \, 2}}{\vec{p}_{q1}^{\,\,2}+x\bar{x}Q^{2}%
}+1\right)  +3\ln\left(  \frac{\vec{p_{1}}{}^{2}%
}{\mu^{2}}\right)  -8\right] \nonumber\\
&  +\frac{-x\bar{x}p_{1\bot}^{i}}{3\vec{p}_{1}^{\, \, 2}}\left[  \pi^{2}%
-6\text{Li}_{2}\left(  \frac{\bar{x}\vec{p}_{1}^{\, \, 2}}{\vec{p}_{q1}^{\,\,2}+x\bar{x}Q^{2}}
+1\right)  \right]  + x\bar{x}\left(  xp_{q1\bot}^i-\bar{x}p_{q\bot}^i\right) \nonumber\\
&  \times\left(  \frac{-\bar{x}\vec{p}_{1}^{\, \, 2}}{(\vec{p}_{q1}^{\,\,2}%
+x\bar{x}Q^{2}-\bar{x}\vec{p}_{1}^{\, \, 2}){}^{2}}\ln\left(  \frac{\bar{x}\vec{p}_{1}^{2}}{\vec{p}_{q1}^{\,\,2}+x\bar{x}Q^{2}}\right)  \right. \nonumber\\
+ &  \left.  \frac{1}{ ( \vec
{p}_{q1}^{\,\,2} + x \bar{x} Q^{2}-\bar{x}\vec{p}_{1}^{\, \, 2} ) }\left[  2\ln\left(  \frac{\bar{x}\vec{p}_{1}^{\, \, 2}}%
{\vec{p}_{q1}^{\,\,2}+x\bar{x}Q^{2}}\right)  -1\right]  \right)
+(q\leftrightarrow\bar{q}).
\end{align}

\subsubsection{TT photon transition}

The cross section for the TT transition reads
\begin{eqnarray} \label{dsigma1TT}
d\sigma_{1TT}  &=& \alpha_{s}\frac{\Gamma(1-\epsilon)}{\left(  4\pi\right)
^{1+\epsilon}} \left( \frac{N_{c}^{2}-1}{2N_{c}} \right) (S_{V}+S_{V}^{\ast
}) \, d\sigma_{0TT} \\ \nonumber
&+& \frac{\alpha_{s}}{4\pi} \left( \frac{N_{c}^{2}-1}{N_{c}} \right) \frac{\alpha_{\mathrm{em}}Q_{q}^{2}
}{\left(  2\pi\right)  ^{4}N_c}dx d\bar{x} d^{d}p_{q\bot}d^{d}p_{\bar{q}\bot} \delta(1-x-\bar{x})(\varepsilon_{Ti}%
\varepsilon_{Tk}^{\ast})\nonumber\\
&\times & \int d^{d}p_{1\bot}d^dp_{2\bot}d^{d}p_{1\bot}^{\prime}d^dp^\prime_{2\bot} \delta(p_{q1\bot}+p_{\bar{q}2\bot}) \delta(p_{11^\prime\bot}+p_{22^\prime\bot}) \mathbf{F}\left(\frac{p_{12\bot}}{2}\right)\mathbf{F}^\ast\left(\frac{p_{1^\prime 2^\prime \bot}}{2}\right) \nonumber \\ 
&\times & \left\{  \frac{3}{2}\frac{ p_{q1\bot r}p_{q1^{\prime}\bot l}}{(\vec{p}_{q1}^{\,\,2}+x\bar{x}Q^{2})(\vec{p}_{q1^{\prime}}^{\, \, 2}+x\bar{x}Q^{2})}[(1-2x)^{2}g_{\bot}%
^{ri}g_{\bot}^{lk}-g_{\bot}^{rk}g_{\bot}^{li}+g_{\bot}^{rl}g_{\bot}%
^{ik}] \right. \nonumber\\
&\times & \left[  \ln\left(  \frac{x\bar{x}\mu^{4}}{(x\vec{p}_{\bar{q}}%
-\bar{x}\vec{p}_{q})^{2}(\vec{p}_{q1}^{\,\,2}+x\bar{x}Q^{2})}\right)
-\frac{x\bar{x}Q^{2}}{\vec{p}_{q1}^{\,\,2}}\ln\left(  \frac{x\bar{x}Q^{2}}%
{\vec{p}_{q1}^{\,\,2}+x\bar{x}Q^{2}}\right)  \right] \nonumber\\
&+& \left.  \frac{(p_{0}^{-})^{2}}{2s^{2}x\bar{x}}\frac{tr[(C_{\bot}%
^{4i}+C_{1\bot}^{5i}+C_{1\bot}^{6i})\hat{p}_{\bar{q}}(p_{q1^{\prime}\bot}%
^{j}(1-2x)-\frac{1}{2}[\hat{p}_{q1^{\prime}\bot}\, , \, \gamma^{i}]j\gamma^{+}\hat
{p}_{q}]}{\vec{p}_{q1^{\prime}}^{\,\,2}+x\bar{x}Q^{2}}+h.c.|_{\substack{p_{1}%
\leftrightarrow p_{1}^{\prime}\\i\leftrightarrow k}}\right\} \nonumber.
\end{eqnarray}
The $C$ functions are given by :
\begin{equation}
\frac{(p_{0}^{-})^{2}}{2s^{2}}tr(C_{\bot}^{ni}\hat{p}_{\bar{q}}(p_{q1^{\prime
}\bot}^{j}(1-2x)-\frac{1}{2}[\hat{p}_{q1^{\prime}\bot}\gamma_{\bot}%
^{j}])\gamma^{+}\hat{p}_{q})=\int_{0}^{x} dz \left[ (\phi_n^{ij})_{TT} \right]_+^{ij}+(q\leftrightarrow
\bar{q}) \, ,
\end{equation}%
with $n=4, \, 5 \, ,\, 6$. The values for $\phi_n$ are given in Appendix A, although $\phi_5$ and $\phi_6$ must be evaluated for $\vec{p}_3 = \vec{0}$.
\begin{align*}
&  \frac{(p_{0}^{-})^{2}}{s^{2}}tr(C_{1\bot}^{6i}\hat{p}_{\bar{q}%
}(p_{q1^{\prime}\bot}^{k}(1-2x)-\frac{1}{2}[\hat{p}_{q1^{\prime}\bot}%
\gamma_{\bot}^{k}])\gamma^{+}\hat{p}_{q})\\
&  = x\bar{x}\left[  g_{\bot}^{ik}(\vec{p}_{1} \cdot \vec{p}_{q1^{\prime}})+p_{1\bot
}^{k}{}p_{q1^{\prime}\bot}^{i}{}+(2x-1)p_{1\bot}^{i}{}p_{q1^{\prime}\bot}%
^{k}{}\right] \\
&  \times\left[  \frac{-\bar{x}^2\vec{p}_{1}^{\, \, 2}}{ ( \vec{p}_{q1}^{\,\,2} + x\bar{x}Q^{2}%
-\bar{x}\vec{p}_{1}^{\, \, 2} )^{2}}\ln\left(
\frac{\bar{x}\vec{p}_{1}^{\, \, 2}}{\vec{p}_{q1}^{\,\,2}+x\bar{x}Q^{2}}\right)
\right. \\
&  +\frac{\bar{x}}{\vec{p}_{q1}^{\,\,2} + x \bar{x} Q^{2} - \bar{x}\vec{p}_{1}^{\, \, 2} }\left[  2\ln\left(  \frac{\bar{x}\vec{p}_{1}^{\, \, 2}}{\vec{p}_{q1}^{\,\,2}+x\bar{x}Q^{2}}\right)  -1\right] \\
&  +\left.  \frac{2}{\vec{p}{}_{1}^{\,\,2}}\left(  \frac{\pi^{2}}{6}%
-\text{Li}_{2}\left(  \frac{\vec
{p}_{q1}^{\,\,2}+x\bar{x}Q^{2}-\bar{x}\vec{p}_{1}^{\,\,2}}{\vec
{p}_{q1}^{\,\,2}+x\bar{x}Q^{2}}\right)  \right)  \right]
\end{align*}%
\begin{align}
-  &  x\bar{x} \left[  p_{q1\bot}^{i}{}p_{q1^{\prime}\bot}^{k}{}(1-2x)^{2}%
-g_{\bot}^{ik}(\vec{p}_{q1} \cdot \vec{p}_{q1^{\prime}})-p_{q1\bot}^{k} p_{q1^{\prime}\bot}^{i}\right] \nonumber\\
\times &  \left[  \frac{-1}{\vec{p}_{q1}^{\,\,2} + x \bar{x} Q^{2}- \bar{x} \vec{p}_{1}^{\, \, 2}}\left(  1-3\ln\left(  \frac{\bar{x}\vec{p}_{1}^{\, \, 2}
}{\vec{p}_{q1}^{\,\,2}+x\bar{x}Q^{2}}\right)  \right)  \right. \nonumber\\
-  &  \frac{1}{\vec{p}_{q1}^{\,\,2}+x\bar{x}Q^{2}}\left(  3\ln\left(
\frac{\vec{p_{1}}{}^{2}}{\mu^{2}}\right)  +4\text{Li}_{2}\left(
\frac{\vec{p}_{q1}^{\,\,2}+x\bar{x}Q^{2}-\bar{x}\vec{p}_{1}^{\, \, 2}}{\vec{p}_{q1}^{\,\,2}+x\bar{x}Q^{2}}\right)
-8\right) \nonumber\\
+  &  \left.  \frac{-\bar{x}\vec{p}_{1}^{\, \, 2}}{(\vec{p}_{q1}^{\,\,2} + x \bar{x} Q^{2}- \bar{x} \vec{p}_{1}^{\, \, 2})^2}\ln\left(  \frac
{\bar{x}\vec{p}_{1}^{\, \, 2}}{\vec{p}_{q1}^{\,\,2}+x\bar{x}Q^{2}}\right)  \right]
+(q\leftrightarrow\bar{q}).
\end{align}

\subsection{Dipole - double dipole cross section $d\sigma_{2}$}

\subsubsection{LL photon transition}%

\begin{eqnarray} \label{dsigma2LL}
d\sigma_{2LL}  & = & \frac{\alpha_{s}Q^2}{4\pi}\frac{\alpha_{\mathrm{em}} Q_{q}^{2}}{\left(  2\pi\right)^{4}N_{c}%
}dxd\bar{x}d^{d}p_{q\bot}d^{d}p_{\bar{q}\bot}\delta(1-x-\bar{x}) \nonumber\\
& \times & \int d^dp_{1\bot}d^dp_{2\bot}d^dp^\prime_{1\bot}d^dp_{2\bot}^\prime \int\frac{d^{d}p_{3\bot}}{\left(  2\pi\right)  ^{d}} \delta(p_{q1\bot}+p_{\bar{q}2\bot}-p_{3\bot})\delta(p_{11^\prime\bot}+p_{22^\prime\bot}+p_{3\bot}) \nonumber \\
& \times & \frac{1}{\vec{p}_{q1^{\prime}}^{\, \, 2}+x\bar{x}Q^{2}} \mathbf{\tilde{F}}\left( \frac{p_{12\bot}}{2},p_{3\bot}\right) \int\mathbf{F}^{\ast}\left( \frac{p_{1^\prime 2^\prime\bot}}{2}\right)\nonumber\\
& \times & \left(  4x\bar{x}\left\{  \frac{x\bar{x}(\vec{p_{3}}{}^{2}-{}%
\vec{p}_{\bar{q}2}^{\, \, 2}-\vec{p_{q1}}{}^{2}-2x\bar{x}Q^{2})}%
{(\vec{p}_{\bar{q}2}^{\, \, 2}+x\bar{x}Q^{2})( \vec{p}_{q1}^{\, \, 2} + x\bar{x}Q^{2}) - x\bar{x}Q^{2}\vec{p}_{3}^{\, \, 2}}\right.  \right. \nonumber\\
& \times & \ln\left(  \frac{x\bar{x}}{e^{2\eta}}\right)  \ln\left(  \frac
{(\vec{p}_{\bar{q2}}^{\, \, 2}+x\bar{x}Q^{2})\left(  \vec{p}_{q1}^{\, \, 2}%
+x\bar{x}Q^{2}\right)  }{x\bar{x}Q^{2}\vec{p_{3}}{}^{2}}\right) \nonumber\\
& - & \left.  \left(  \frac{2x\bar{x}}{Q^{2}+x\bar{x}\vec{p}_{q1}^{\,\,2}}
\ln\left(  \frac{x}{e^{\eta}}\right) \ln\left(\frac{\vec{p_{3}}{}^{2}}{\mu^{2}}\right)  +\left(  q\leftrightarrow\bar{q}\right)  \right)  \right\}
\nonumber\\
& + &\left.  \frac{Q^{2}(p_{0}^{-})^{2}}{p_{\gamma}^{+}s^{2}}tr((C_{2\Vert}%
^{5}+C_{2\Vert}^{6})\hat{p}_{\bar{q}}\gamma^{+}\hat{p}_{q})\right)  +h.c.
\end{eqnarray}%
We will write
\begin{equation}
\frac{(p_{0}^{-})^{2}}{s^{2}p_{\gamma}^{+}}tr(C_{2||}^{n}\hat{p}_{\bar{q}%
}\gamma^{+}\hat{p}_{q})=\int_{0}^{x}dz\left[ (\phi_n)_{LL} \right]_+ +\left(
q\leftrightarrow\bar{q}\right)  \, ,
\end{equation}%
with $n=5, \, 6$. The values for $\phi_n$ are given in appendix A.

\subsubsection{LT photon transition}%

\begin{eqnarray} \label{dsigma2LT}
d\sigma_{2TL}  & = &  \frac{\alpha_{s}Q}{4\pi}\frac{\alpha_{\mathrm{em}} Q_{q}^{2}}{\left(  2\pi\right)  ^{4}N_{c}}dxd\bar{x}d^{d}p_{q\bot}d^{d}p_{\bar{q}\bot}\delta(1-x-\bar{x}) \int d^dp_{1\bot}d^dp_{2\bot}d^dp_{1\bot}^\prime d^dp_{2\bot}^\prime \\
& \times & \int \frac{d^dp_{3\bot}d^dp_{3\bot}^\prime}{(2\pi)^d} \delta(p_{q1\bot}+p_{\bar{q}2\bot}+p_{g3\bot})\delta(p_{11^\prime\bot}+p_{22^\prime\bot}+p_{33^\prime\bot}) \nonumber \\
& \times & \varepsilon_{Ti}^{\ast} \left[ \frac{\delta(p_{3\bot}^\prime)}{\vec{p}_{q1^{\prime}}^{\, \,2}+x\bar{x}Q^{2}} %
\mathbf{\tilde{F}}\left( \frac{p_{12\bot}}{2},p_{3\bot}\right) \mathbf{F}^{\ast}\left( \frac{p_{1^\prime 2^\prime\bot}}{2}\right)  \right. \nonumber\\
& \times & \left(  2(1-2x)p_{q1^{\prime}\bot}^{i}\left\{  \frac{x\bar{x}(\vec
{p_{3}}{}^{2}-\vec{p}_{\bar{q}2}^{\, \, 2}-\vec{p}_{q1}^{\, \, 2}-2x\bar{x}Q^{2})}{(\vec{p}_{\bar{q}2}^{\, \, 2} + x\bar{x}Q^{2})(\vec{p}_{q1}^{\, \, 2}+ x\bar{x} Q^{2}) - x\bar{x}Q^{2}\vec{p_{3}}{}^{2}}\right.
\right. \ln\left(  \frac{x\bar{x}}{e^{2\eta}}\right)   \nonumber\\
& \times & \ln\left(  \frac{(\vec{p}_{\bar{q}2}^{\, \, 2}+x\bar{x}Q^{2})\left(  \vec{p}_{q1}^{\, \, 2} + x\bar{x}Q^{2}\right)  }{x\bar{x}Q^{2}\vec{p_{3}}{}^{2}}\right) \nonumber\\
& - & \left.  \left(  \frac{2x\bar{x}}{Q^{2}+x\bar{x}\vec{p}_{q1}^{\,\,2}}
\ln\left(  \frac{x}{e^{\eta}}\right)  \ln\left(
\frac{\vec{p_{3}}{}^{2}}{\mu^{2}}\right) +\left(  q\leftrightarrow\bar{q}\right)  \right)  \right\}
\nonumber\\
& + & \left.  \frac{(p_{0}^{-})^{2}}{2s^{2}x\bar{x}p_{\gamma}^{+}}tr[(C_{2\Vert
}^{5}+C_{2\Vert}^{6})\hat{p}_{\bar{q}}(p_{q1^{\prime}\bot}^{i}(1-2x)-\frac
{1}{2}[\hat{p}_{q1^{\prime}\bot} \, , \, \gamma^{i}])\gamma^{+}\hat{p}_{q}]\right)
\nonumber\\
& + & \frac{\delta(p_{3\bot})}{\vec{p}_{q1}^{\, \, 2}+x\bar{x}Q^{2}} \mathbf{F}\left( \frac{p_{12\bot}}{2}\right)\mathbf{\tilde{F}}^\ast\left( \frac{p_{1^\prime 2^\prime\bot}}{2},p_{3^{\prime}\bot}\right)\nonumber\\
& \times & \left(  \left\{  2x\bar{x}(1-2x)p_{q1^{\prime}\bot}^{\, \, i}\left(
\frac{-2}{Q^{2}+x\bar{x}\vec{p}_{q1^{\prime}}^{\, \, 2}} \ln\left(  \frac
{x}{e^{\eta}}\right)\ln\left(  \frac{\vec{p_{3}%
}{}^{\prime2}}{\mu^{2}}\right)  \right.
\right.  \right. \nonumber\\
& + & \ln\left(  \frac{x\bar{x}}{e^{2\eta}}\right)  \left[  -\frac{%
\vec{p}_{\bar{q}2^\prime}^{\, \, 2}+x\bar{x}Q^{2}}{( \vec{p}_{q1^{\prime}}^{\, \, 2}
+x\bar{x}Q^{2})  ( \vec{p}_{\bar{q}2^\prime}^{\, \, 2} +x\bar{x} Q^{2})-x\bar{x}Q^{2}\vec{p_{3}}^{\prime2}}\right. \nonumber\\
& \times &\ln\left(  \frac{( \vec{p}_{q1^{\prime}}^{\, \, 2} +x\bar{x} Q^{2})  ( \vec{p}_{\bar{q}2^\prime}^{\, \, 2} + x\bar{x}Q^{2} )  }{x\bar{x}Q^{2}\vec{p_{3}}{}^{\prime2}}\right)
\nonumber\\
& + & \left.  \left.  \left.  \frac{1}{\vec{p}_{q1^{\prime}}^{\, \, 2}}\ln\left(
\frac{\vec{p}_{q1^{\prime}}^{\, \, 2}+x\bar{x}Q^{2}}{x\bar{x}Q^{2}}\right)  \right]  \right)
+(q\leftrightarrow\bar{q})\right\}  +\left.  \left.  \frac{(p_{0}^{-})^{2}%
}{s^{2}}tr((C_{2\bot}^{\prime5i}+C_{2\bot}^{\prime6i})\hat{p}_{\bar{q}}%
\gamma^{+}\hat{p}_{q})^{\ast}\right)  \right]  .\nonumber
\end{eqnarray}%

Again we will write

\begin{equation}
\frac{(p_{0}^{-})^{2}}{s^{2}}tr(C_{2\bot}^{ni}\hat{p}_{\bar{q}}\gamma^{+}%
\hat{p}_{q})=\int_{0}^{x}dz\left[ (\phi_n^i)_{LT} \right]_+dz+\left(
q\leftrightarrow\bar{q}\right)  \, ,
\end{equation}%
and\begin{equation}
\frac{(p_{0}^{-})^{2}}{s^{2}p_{\gamma}^{+}}tr(C_{2||}^{n}\hat{p}_{\bar{q}%
}(p_{q1^{\prime}\bot}^{i}(1-2x)-\frac{1}{2}[\hat{p}_{q1^{\prime}\bot}%
\gamma_{\bot}^{i}])\gamma^{+}\hat{p}_{q})=\int_{0}^{x}dz\left[(\phi_n^i)_{TL}\right]_++\left(  q\leftrightarrow\bar{q}\right)  .
\end{equation}
The values for $(\phi_{5,6})_{LT}$ and $(\phi_{5,6})_{TL}$ are given in appendix A.

\subsubsection{TT photon transition}%

\begin{eqnarray} \label{dsigma2TT}
&& d\sigma_{2TT}  = \frac{\alpha_{s}}{4\pi}\frac{\alpha_{\mathrm{em}} Q_{q}^{2}}{\left(  2\pi\right)  ^{4}N_{c}}dx d\bar{x} d^{d}p_{q\bot}d^{d}p_{\bar{q}\bot} \delta(1-x-\bar{x}) \int d^dp_{1\bot}d^dp_{2\bot}d^dp_{1\bot}^\prime d^dp_{2\bot}^\prime \int\frac{d^{d}p_{3\bot}}{\left(  2\pi\right)  ^{d}} \nonumber\\
& \times & \frac{(\varepsilon_{Ti}\varepsilon_{Tj}^{\ast})}{\vec{p}_{q1^{\prime}}^{\, \,2}+x\bar{x}Q^{2}} \left[  \mathbf{\tilde{F}}\left( \frac{p_{12\bot}}{2},p_{3\bot}\right) \mathbf{F}^{\ast}\left( \frac{p_{1^\prime 2^\prime\bot}}{2}\right) \delta(p_{q1\bot}+p_{\bar{q}2\bot}-p_{3\bot})\delta(p_{11^\prime\bot}+p_{22^\prime\bot}+p_{3\bot}) \right. \nonumber\\
& \times & \left(  \left\{  p_{q1^{\prime}\bot l}p_{q1\bot k}[(1-2x)^{2}g_{\bot
}^{ki}g_{\bot}^{lj}-g_{\bot}^{kj}g_{\bot}^{li}+g_{\bot}^{kl}g_{\bot}%
^{ij}]\left(  \frac{-1}{Q^{2}+x\bar{x}\vec{p}_{q1}^{\, \, 2}}\right.  \right.
\right. \nonumber\\
& \times &  2\ln\left(  \frac{x}{e^{\eta}}\right) \ln\left(  \frac{\vec{p_{3}}{}^{2}}{\mu^{2}}\right)  +\ln\left(  \frac{x\bar{x}}{e^{2\eta}}\right)
\left[  \frac{1}{\vec{p}_{q1}^{\, \, 2}}\ln\left(  \frac{\vec{p}_{q1}^{\, \, 2} + x\bar{x}Q^2%
}{x\bar{x}Q^{2}}\right)  \right. \nonumber\\
& - & \frac{\vec{p}_{\bar{q}2}^{\,\,2}+x\bar{x}Q^{2}}{( \vec{p}_{q1}^{\, \, 2} + x\bar{x} Q^{2})  (\vec{p}_{\bar{q}2}^{\, \, 2}+x\bar{x}Q^{2})-x\bar{x}Q^{2}\vec{p_{3}}{}^{2}}\nonumber\\
& \times & \left.  \left.  \left.  \ln\left(  \frac{( \vec
{p}_{q1}^{\, \, 2} +x\bar{x} Q^{2})  ( \vec{p}_{\bar{q}2}^{\, \, 2} +x\bar{x} Q^{2} )  }{x\bar{x}Q^{2}\vec{p}_{3}^{\, \, 2}}\right)  \right]  \right)
+(q\leftrightarrow\bar{q})\right\} \nonumber\\
& + & \left.  \frac{(p_{0}^{-})^{2}}{2s^{2}x\bar{x}}tr((C_{2\bot}^{5i}+C_{2\bot
}^{6i})\hat{p}_{\bar{q}}[p_{q1^{\prime}\bot}^{j}(1-2x)-\frac{1}{2}[\hat
{p}_{q1^{\prime}\bot} \, , \, \gamma^{j}])\gamma^{+}\hat{p}_{q}]\right)  +\left.
h.c.|_{\substack{p_{1},p_{3}\leftrightarrow p_{1}^{\prime},p_{3}^{\prime
}\\i\leftrightarrow j}}\right]  . \, 
%
\end{eqnarray}%
As for the other contributions, we wrote
\begin{equation}
\frac{(p_{0}^{-})^{2}}{2s^{2}}tr(C_{2\bot}^{ni}\hat{p}_{\bar{q}}%
(p_{q1^{\prime}\bot}^{j}(1-2x)-\frac{1}{2}[\hat{p}_{q1^{\prime}\bot}%
\gamma_{\bot}^{j}])\gamma^{+}\hat{p}_{q})=\int_{0}^{x}dz\left[(\phi_n^{ij})_{TT}\right]_{+}dz+\left(  q\leftrightarrow\bar{q}\right)  .
\end{equation}
$(\phi_5^{ij})_{TT}$ and $(\phi_6^{ij})_{TT}$ can be found in appendix A.

\section{Cross section for $\gamma P\rightarrow q\bar{q}gP^{\prime}$
transition}
As for section 5 we define a reduced matrix element $A_4$ such that the $\gamma P\rightarrow q\bar{q}gP^{\prime}$ cross section reads%
\begin{equation}
d\sigma_{(q\bar{q}g)}=\frac{1}{4s}(2\pi)^{D}\delta^{(D)}(p_{\gamma}+p_{0}%
-p_{q}-p_{\bar{q}}-p_{g}-p_{0}^{\prime})|A_{4}|^{2}d\rho_{4},
\end{equation}
where
\begin{align}
\delta^{(D)}(p_{\gamma}+p_{0}-p_{q}-p_{\bar{q}}-p_{g}-p_{0}^{\prime})  &
=\delta(p_{00^{\prime}}^{-})\delta(p_{q}^{+}+p_{\bar{q}}^{+}+p_{g}%
^{+}-p_{\gamma}^{+})\nonumber\\
&  \times\delta^{(d)}(p_{q\bot}+p_{\bar{q}\bot}+p_{g\bot}-p_{\gamma\bot
}+p_{0^{\prime}0\bot}),
\end{align}
with the 4-body phase space measure
\begin{align}
d\rho_{4}  &  =\frac{dp_{q}^{+}d^{d}p_{q\bot}}{2p_{q}^{+}(2\pi)^{d+1}}%
\frac{dp_{\bar{q}}^{+}d^{d}p_{\bar{q}\bot}}{2p_{\bar{q}}^{+}(2\pi)^{d+1}}%
\frac{dp_{g}^{+}d^{d}p_{g\bot}}{2p_{g}^{+}(2\pi)^{d+1}}\frac{dp_{0}^{\prime
-}d^{d}p_{0\bot}^{\prime}}{2p_{0}^{\prime-}(2\pi)^{d+1}} \, .
\end{align}%
The reduced matrix element can be derived from section 4 and reads
\begin{align}
A_{4}  &  =\frac{-e_{q}2p_{0}^{-}\varepsilon_{\alpha}}{\left(  2\pi\right)
^{D-4}}\sqrt{\frac{2}{N_{c}^{2}-1}}g\mu^{-\epsilon}\int d^{d}p_{1\bot}%
d^{d}p_{2\bot}\left\{  \delta(p_{q1\bot}+p_{\bar{q}2\bot}+p_{g\gamma\bot}%
)\Phi_{3}^{\alpha}\frac{N_{c}^{2}-1}{N_{c}}\mathbf{F(}\frac{p_{12\bot}}%
{2}\mathbf{)}\right. \nonumber\\
&  +\left.  \int\frac{d^{d}p_{3\bot}}{(2\pi)^{d}}\delta(p_{q1\bot}+p_{\bar
{q}2\bot}+p_{g\gamma\bot}-p_{3\bot})\Phi_{4}^{\alpha}\mathbf{\tilde{F}}%
(\frac{p_{12}{}_{\bot}}{2},p_{3\bot})\right\}  .
\end{align}
This cross section has a contribution $d\sigma_3$ with 2 dipole operators, a contribution $d\sigma_4$ with a dipole operator and a double dipole operator, and a contribution $d\sigma_5$ with 2 double dipole operators.
\begin{equation} \label{sigmaR}
d\sigma_{(q\bar{q}g)}=d\sigma_{3}+d\sigma_{4}+d\sigma_{5}.
\end{equation}
The dipole $\times$ dipole contribution reads
\begin{eqnarray}
d\sigma_{3JI} & = &  \frac{\alpha_{s}}{\mu^{2\epsilon}} \left( \frac{N_{c}^{2}-1}{N_{c}} \right) \frac{\alpha_{\mathrm{em}} Q_{q}^{2}}%
{(2\pi)^{4(d-1)}N_c}\frac{(p_{0}^{-})^{2}}{s^{2}x_q x_{\bar{q}}} (\varepsilon_{I\alpha} \varepsilon_{J\beta}^\ast)  \nonumber \\
& \times & \, dx_q \, dx_{\bar{q}} \, d^{d}p_{q\bot} \, d^{d}p_{\bar{q}\bot} \frac{dzd^{d}p_{g\bot}}%
{z(2\pi)^{d}} \delta(1-x_q-x_{\bar{q}}-z)   \nonumber\\
& \times & \int d^dp_{1\bot} d^dp_{2\bot} d^dp_{1\bot}^\prime d^dp_{2\bot}^\prime \delta(p_{q1\bot}+p_{\bar{q}2\bot} + p_{g\bot}) \delta(p_{11^\prime\bot}+p_{22^\prime\bot}) \nonumber \\
& \times & \Phi_3^\alpha (p_{1\bot}, p_{2\bot}) \Phi_3^{\beta\ast} (p_{1\bot}^\prime, p_{2\bot}^\prime)  \textbf{F}\left(\frac{p_{12\bot}}{2}\right) \textbf{F}^\ast \left( \frac{p_{1^\prime 2^\prime \bot}}{2} \right) \, . \label{dsigma3}%
\end{eqnarray}%
The dipole $\times$ double dipole contribution reads
\begin{eqnarray}
d\sigma_{4JI} & =  & \frac{ \alpha_{s} } {\mu^{2\epsilon}}\frac{\alpha_{\mathrm{em}} Q_{q}^{2}}{(2\pi)^{4(d-1)}N_{c}}\frac{(p_{0}%
^{-})^{2}}{s^{2}x_q x_{\bar{q}}} (\varepsilon_{I\alpha} \varepsilon_{J\beta}^\ast) dx_q dx_{\bar{q}} d^{d}p_{q\bot}d^{d}p_{\bar{q}\bot} \frac{dzd^{d}p_{g\bot}}%
{z(2\pi)^{d}} \delta(1-x_q -x_{\bar{q}}-z) \nonumber \\
& \times & \int d^dp_{1\bot}d^dp_{2\bot} d^dp_{1\bot}^\prime d^dp_{2\bot}^\prime \frac{d^dp_{3\bot}d^dp_{3\bot}^\prime}{\left( 2\pi \right)^d} \delta(p_{q1\bot}+p_{\bar{q}2\bot}+p_{g3\bot}) \delta(p_{11^\prime \bot}+p_{22^\prime \bot}+p_{33^\prime \bot}) \nonumber \\
&\times & \left[ \Phi_3^\alpha(p_{1\bot},p_{2\bot}) \Phi_4^{\beta\ast}(p_{1\bot}^\prime, p_{2\bot}^\prime, p_{3\bot}^\prime) \mathbf{F}\left(\frac{p_{12\bot}}{2}\right) \mathbf{\tilde{F}}^\ast \left( \frac{p_{1^\prime 2^\prime \bot}}{2}, p_{3\bot}^\prime \right) \delta(p_{3\bot}) \right. \\
& + & \left. \Phi_4^\alpha (p_{1\bot}, p_{2\bot}, p_{3\bot}) \Phi_3^{\beta\ast} ( \frac{p_{1^\prime 2^\prime \bot}}{2} ) \mathbf{\tilde{F}}\left(\frac{p_{12\bot}}{2}, p_{3\bot}\right) \mathbf{F}^\ast\left(\frac{p_{1^\prime 2^\prime\bot}}{2} \right) \delta(p_{3\bot}^\prime) \right] \nonumber ,
\end{eqnarray}%
and the double dipole $\times$ double dipole contribution is given by
\begin{eqnarray}
d\sigma_{5JI} & = &  \frac{\alpha_{s}}{\mu^{2\epsilon}}\frac{\alpha_{\mathrm{em}} Q_{q}^{2}}{(2\pi)^{4(d-1)}}\frac{(p_{0}^{-})^{2}}{s^{2}x_q x_{\bar{q}}} \frac{(\varepsilon_{I\alpha} \varepsilon_{J\beta}^\prime)}{N_{c}^{2}-1} dx_q dx_{\bar{q}}d^{d}p_{q\bot}d^{d}p_{\bar{q}\bot
} \frac{dzd^{d}p_{g\bot}}%
{z(2\pi)^{d}} \delta(1-x_q -x_{\bar{q}}-z) \\
& \times & \int d^dp_{1\bot}d^dp_{2\bot} d^dp_{1\bot}^\prime d^dp_{2\bot}^\prime \frac{d^dp_{3\bot}d^dp_{3\bot}^\prime}{\left( 2\pi \right)^{2d}} \delta(p_{q1\bot}+p_{\bar{q}2\bot}+p_{g3\bot}) \delta(p_{11^\prime \bot}+p_{22^\prime \bot}+p_{33^\prime \bot}) \nonumber \\
& \times & \Phi_4^\alpha(p_{1\bot},p_{2\bot},p_{3\bot}) \Phi_4^{\beta\ast}(p_{1\bot}^\prime,p_{2\bot}^\prime, p_{3\bot}^\prime) \mathbf{\tilde{F}}\left( \frac{p_{12\bot}}{2}, p_{3\bot} \right) \mathbf{\tilde{F}}^\ast \left(\frac{p_{1^\prime 2^\prime \bot}}{2},p_{3\bot}^\prime\right).
\end{eqnarray}
We present the results for the products $\Phi_{a}\Phi_{b}^{\ast}$ in Appendix B
in $D$-dimensional space. They can be used directly in dimension 4 to describe the exclusive production 	of 3 jets. However, since the main motivation of present paper is to study the production of dijet with NLO accuracy, in the next section we will only extract the soft and collinear divergences in these real terms to construct a well defined cross section for our process.

\section{Cross section for $\gamma P\rightarrow2jets\,P^{\prime}$ exclusive
transition}

The expressions for $\gamma\rightarrow q\bar{q}$ and $\gamma \rightarrow q\bar{q}g$ impact factors can be used to construct IR stable cross sections for dijet production. Whatever the experimental conditions are, one has to combine the $q\bar{q}$ and $q\bar{q}g$ production cross sections obtained above to cancel the soft and collinear singularities in the virtual part. They cancel with the singular
contribution of $q\bar{q}g$ production arising from the emitted gluon phase space area where
the gluon is soft or collinear to the quark or the antiquark.

We will explicitly show this cancellation on the example of the $\gamma
P\rightarrow2jetsP^{\prime}$ exclusive production cross section experimentally
studied in \cite{Abramowicz:2015vnu}. By exclusive production we understand that only two jets and the scattered proton are seen in the detector and there is nothing else. Since we want our result for the cross section to be differential only in the jets' momenta, we integrate over the transverse momentum of the outgoing proton as before.
We define jets using the small cone algorithm, as in \cite{Ivanov:2012ms}.\\
Let us define a jet cone radius $R^2$. For convenience, we will assume that $R^2 \ll 1$. Two given particles will form a jet with a momentum equal to the sum of their momenta if they both satisfy the following condition :
\begin{equation}
\Delta\phi^{2}+\Delta Y^{2}<R^{2}, \label{jetalg}%
\end{equation}
where $\Delta\phi$ is the azimuthal angle difference between the particle and the jet, and $\Delta Y$ is the rapidity difference between the particle and the jet.
Let us consider for example a jet built from the quark and the gluon. Its momentum will be given by
\begin{equation}
p_{j}=x_{j}p_{\gamma}^{+}n_{1}^{\mu}+\frac{\vec{p}_{j}^{\,\,2}}{2p_{\gamma
}^{+}x_{j}}n_{2}^{\mu}+p_{j\bot}^{\mu},\quad\quad x_{j}=x_{q}+z,\quad\vec
{p}_{j}=\vec{p}_{q}+\vec{p}_{g}.
\end{equation}
In the small cone limit, $p_q^- +p_g ^- \sim \frac{\vec{p}_{j}^{\,\,2}}{2p_{\gamma
}^{+}x_{j}}$ up to a $O(R)$ correction so the jet is on-shell in this approximation. The azimuthal angle and rapidity differences read :

\begin{equation}
\Delta\phi=\arccos\frac{\vec{p}_{j}\cdot\vec{p}_{g}}{|\vec{p}_{j}|\left\vert
\vec{p}_{g}\right\vert },\quad\Delta Y=\frac{1}{2}\ln\frac
{x_{j}^{2}\vec{p}_{g}^{\,\,2}}{z^{2}\vec{p}_{j}^{\,\,2}}.
\end{equation}
Introducing the variable%
\begin{equation}
\vec{\Delta}_{q}=\frac{x_{q}}{x_{j}}\vec{p}_{g}-\frac{z}{x_{j}}\vec{p}_{q}
\label{Delta}%
\end{equation}
which approaches 0 when the quark and the gluon are collinear, we get the
condition for the gluon to be inside the cone :
\begin{equation}
\vec{\Delta}_{q}^{2}<R^{2}\frac{\vec{p}_{j}^{\,\,2}z^{2}}{x_{j}^{2}}.
\label{g-jet}%
\end{equation}
The corresponding condition for the quark reads
\begin{equation}
\vec{\Delta}_{q}^{2}<R^{2}\frac{\vec{p}_{j}^{\,\,2}x_{q}^{2}}{x_{j}^{2}}.
\label{q-jet}%
\end{equation}
To obtain the 2-jet exclusive cross section in the small cone limit, we only need the contributions for the (LO + NLO) $q \bar{q}$ production, and the part of the contribution of the $q \bar{q} g$ production where the gluon is collinear to either the quark or the antiquark, so that they both form a single jet. Any non-collinearly divergent contribution will scale as a positive power of $R$ and therefore it will be neglected. \\
We denote the jet variables as $x_j$, $x_{\bar{j}}$, $p_{j\bot}$ and $p_{\bar{j}\bot}$. In the case of the virtual contribution one gets immediately the jet cross section by performing the following change of variables in (\ref{dsigma0}--\ref{dsigma2}) :
\begin{equation} \label{jetrenaming}
(x, p_{q\bot}) \rightarrow (x_j , p_{j\bot}),\quad (\bar{x}, p_{\bar{q}\bot}) \rightarrow (x_{\bar{j}}, p_{\bar{j}\bot}) \, ,
\end{equation}
and by symmetrizing $j \leftrightarrow \bar{j}$. \\
For a given contribution $d\sigma_n$ for partons in (\ref{sigmaR}) or (\ref{sigmaV}), we will denote the corresponding contribution to the cross section for jets as $d\sigma_n^\prime$. \\
One can find the contribution of the collinear real gluons from the quasi-real electron
approximation \cite{ioffe2010quantum}. Indeed, the real contribution with a jet formed by the quark and the gluon and the other jet formed by the antiquark reads%
\begin{equation}
d\sigma_{3JI}^{\prime}(x_{q},\vec{p}_{q})|_{\operatorname{col}}=d\sigma
_{0JI}^{\prime}(x_{j},\vec{p}_{j})\,\alpha_{s}\frac{\Gamma(1-\epsilon)}%
{(4\pi)^{1+\epsilon}}\frac{N_{c}^{2}-1}{2N_{c}}n_{j},
\end{equation}
where $n_{j}$ is proportional to the \textquotedblleft number of jets in the
quark"
\begin{align}
n_{j}  &  =\frac{(4\pi)^{1+\epsilon}}{\Gamma(1-\epsilon)}\int_{\alpha
}^{x_j}\frac{dz}{2z}\int_{\vec{\Delta}%
_{q}^{\,2}<\frac{R^{2}\vec{p}_{j}^{\,\,2}}{x_{j}^{2}}\min(z^{2},(x_{j}%
-z)^{2})}\frac{d^{d}\vec{\Delta}_{q}}{(2\pi)^{d}}\frac{\mu^{-2\epsilon}}{2p_{j}%
^{+}2p_{q}^{+}}\frac{tr(\hat{p}_{q}\gamma^{\mu}\hat{p}_{j}\gamma^{\nu}%
)d_{\mu\nu}(p_{g})}{(p_{q}^{-}+p_{g}^{-}-p_{j}^{-})^{2}}\nonumber\\
&  =4\int_{\alpha}^{x_{j}}\frac{x_{j}dz}{z(x_{j}-z)}\frac{\mu^{-2\epsilon}%
}{\Gamma(1-\epsilon)\pi^{\frac{d}{2}}}\int_{\vec{\Delta}_{q}^{\,2}<\frac
{R^{2}\vec{p}_{j}^{\,\,2}}{x_{j}^{2}}\min(z^{2},(x_{j}-z)^{2})}d^{d}\vec{\Delta}
_{q}\nonumber\\
&  \times\frac{1}{4}\frac{(x_{j}-z){}\left(  dz^{2}+4x_{j}\left(
x_{j}-z\right)  \right)  }{x_{j}^{3}\vec{\Delta}_{q}^{2}}. \label{CollinearFactor}%
\end{align}
This result is obtained by taking the collinear limit in the squares of the real impact factors, which one can find in appendix B. \\
Here we intoduced the jet $j$ by performing the change of variables
\begin{equation}
(\vec{p}_q, \vec{p}_g) \rightarrow (\vec{p}_j, \vec{\Delta}_q), \quad (x,z)\rightarrow (x_j,z) \label{jetcol1}%
\end{equation}
and the jet $\bar{j}$ by 
\begin{equation}
\vec{p}_{\bar{q}}\rightarrow\vec{p}_{\bar{j}},\quad x_{\bar{q}}\rightarrow
x_{\bar{j}}=1-x_{j}, \label{jetcol2}%
\end{equation}
and integrated inside the jet cone (\ref{g-jet}--\ref{q-jet}). The
contribution of the jet built from the antiquark and the gluon is recovered
via the $j\leftrightarrow\bar{j}$ symmetry. In Appendix B we explicitly
show that in the collinear limit the convolutions of the impact factors for
real gluon production reproduce the last line in (\ref{CollinearFactor}), while
all the other factors come from $d\sigma_{3JI}$ (\ref{dsigma3}). The total
collinear contribution reads
\begin{align}
&  n_{j}+n_{\bar{j}}=4\left[  \frac{1}{2}\left(  \ln\left(  \frac{x_{\bar{j}%
}x_{j}}{\alpha^{2}}\right)  -\frac{3}{2}\right)  \ln\left(  \frac{R^{4}%
\vec{p_{j}}{}^{2}\vec{p}_{\bar{j}}{}^{2}}{\mu^{4}}\right)  +\frac{1}{\epsilon
}\left(  \ln\left(  \frac{x_{\bar{j}}x_{j}}{\alpha^{2}}\right)  -\frac{3}%
{2}\right)  \right. \nonumber\\
&  -\left.  \frac{1}{2}\ln^{2}\left(  \frac{x_{\bar{j}}x_{j}}{\alpha^{2}%
}\right)  +\frac{1}{2}\ln\left(  \frac{x_{j}}{x_{\bar{j}}}\right)  \ln\left(
\frac{\vec{p}_{j}{}^{2}}{\vec{p}{}_{\bar{j}}^{\,\,2}}\right)  -\frac{1}{2}%
\ln^{2}\left(  \frac{x_{j}}{x_{\bar{j}}}\right)  -\frac{\pi^{2}}{3}+\frac
{7}{2}+\ln(8)\right]  . \label{collinearContribution}%
\end{align}
In the soft gluon limit, the real cross section has the form 
\begin{equation}
d\sigma_{3JI}^{\prime}|_{soft}=d\sigma_{0JI}^{\prime}\alpha_{s}\frac{N_{c}%
^{2}-1}{2N_{c}}\frac{\Gamma(1-\epsilon)}{(4\pi)^{1+\epsilon}}S,\quad
S \equiv \frac{(4\pi)^{1+\epsilon}}{\Gamma(1-\epsilon)}\int\left\vert \frac
{p_{q}^{\mu}}{(p_{q} \cdot p_{g})}-\frac{p_{\bar{q}}^{\mu}}{(p_{\bar{q}} \cdot p_{g}%
)}\right\vert ^{2}\frac{dz}{z}\frac{d^{d}p_{g}}{(2\pi)^{d}} \, ,
\end{equation}
as shown in appendix B.
We have to integrate this formula over
\begin{equation}
\omega_{g}=\frac{1}{2}\left(zp_{\gamma}^{+}+\frac{\vec{p}_{g}^{\,\,2}}{zp_{\gamma
}^{+}}\right)<E\ll p_{\gamma}^{+},
\end{equation}
where $\omega_{g}$ is the emitted gluon energy and $E$ is the energy
resolution. The small energy limit for the gluon occurs when all the components
of the gluon momentum approach 0 simultaneously. We achieve this by rescaling the
gluon transverse momentum as
\begin{equation}
\vec{p}_{g}=z\vec{u}%
\end{equation}
and going to the limit $z\rightarrow0.$ In this limit the integration area
reads%
\begin{equation}
z\left(1+\frac{\vec{u}^{\,\,2}}{(p_{\gamma}^{+})^{2}}\right)<\frac{2E}{p_{\gamma}^{+}}%
\ll1,
\end{equation}
and we have%
\begin{equation}
S=\,\int_{\alpha}^{\frac{2E}{p_{\gamma}^{+}}}z^{d-3}dz\int\nolimits_{\vec
{u}^{\,\,2}<(p_{\gamma}^{+})^{2}(\frac{2E}{zp_{\gamma}^{+}}-1)}\frac
{\mu^{-2\epsilon}d^{d}u_{\bot}}{\Gamma(1-\epsilon)\pi^{\frac{d}{2}}}%
\frac{4(\frac{\vec{p}_{j}}{x_{j}}-\frac{\vec{p}_{\bar{j}}}{x_{\bar{j}}})^{2}%
}{(\vec{u}-\frac{\vec{p}_{\bar{j}}}{x_{\bar{j}}})^{2}(\vec{u}-\frac{\vec
{p}_{j}}{x_{j}})^{2}},
\end{equation}
as shown explicitely in appendix B.
We have restored the rapidity cutoff $\alpha$ which of course will play a role to regularize the soft divergence. \\
However, in the sum $n_{j}+n_{\bar{j}}+S$ the region with a gluon both soft and collinear to the quark or to the antiquark is calculated twice. To avoid double counting we restrict the integration in $S$
so that the gluons sit outside the cones (\ref{g-jet}). The new integration region reads %
\begin{eqnarray}
\Omega & = & \left\{\vec{u}^{\,\,2}<(p_{\gamma}^{+})^{2}\left(\frac{2E}{zp_{\gamma}^{+}%
}-1\right)\right\}\cap \, \Omega_{nc}  \, ,\\ 
\Omega_{nc} & \equiv & \left\{\left(\vec{u}-\frac{\vec{p}_{j}}{x_{j}%
}\right)^{2}>\frac{R^{2}\vec{p}_{j}^{\,\,2}}{x_{j}^{2}}\right\}\cup\left\{\left(\vec{u}-\frac
{\vec{p}_{\bar{j}}}{x_{\bar{j}}}\right)^{2}>\frac{R^{2}\vec{p}_{\bar{j}}^{\,\,2}%
}{x_{\bar{j}}^{2}}\right\}.
\end{eqnarray}
Let us denote $S^\prime$ the new definition of $S$ with this integration area :
\begin{eqnarray}
S^{\prime} & \equiv & 4\int_{\alpha}^{\frac{2E}{p_{\gamma}^{+}}}\frac{dz}{z}%
\int_{\Omega}\frac{d\vec{u}}{\pi}\frac{(\frac{\vec{p}_{j}}{x_{j}}-\frac
{\vec{p}_{\bar{j}}}{x_{\bar{j}}})^{2}}{(\vec{u}-\frac{\vec{p}_{\bar{j}}%
}{x_{\bar{j}}})^{2}(\vec{u}-\frac{\vec{p}_{j}}{x_{j}})^{2}}, \\ \nonumber
& = & 4 \int_{\alpha}^{\frac
{2E}{p_{\gamma}^{+}}}\frac{dz}{z}\int_{\Omega_{nc}}\frac{d\vec{u}}{\pi}%
\frac{(\frac{\vec{p}_{j}}{x_{j}}-\frac{\vec{p}_{\bar{j}}}{x_{\bar{j}}})^{2}%
}{(\vec{u}-\frac{\vec{p}_{\bar{j}}}{x_{\bar{j}}})^{2}(\vec{u}-\frac{\vec
{p}_{j}}{x_{j}})^{2}}+4 I(R,E) \\ \nonumber
& = & 4 \ln\left(  \frac{2E}{\alpha p_{\gamma}^{+}}\right)  \ln\left(  \frac
{(\vec{p}_{j}x_{\bar{j}}-x_{j}\vec{p}_{\bar{j}})^{4}}{(R^{2}\vec{p}%
_{j}^{\,\,2}x_{j}^{2})(R^{2}\vec{p}_{\bar{j}}^{\,\,2}x_{\bar{j}}^{2})}\right)
+4 I(R,E) \, ,
\end{eqnarray}
where we defined
\begin{equation}
I(R,E) \equiv -\int_{0}^{\frac{2E}{p_{\gamma}^{+}}}\frac{dz}{z}\int_{\{\vec
{u}^{\,\,2}>(p_{\gamma}^{+})^{2}(\frac{2E}{zp_{\gamma}^{+}}-1)\}\cap
\Omega_{nc}}\frac{d\vec{u}}{\pi}\frac{(\frac{\vec{p}_{j}}{x_{j}}-\frac{\vec
{p}_{\bar{j}}}{x_{\bar{j}}})^{2}}{(\vec{u}-\frac{\vec{p}_{\bar{j}}}{x_{\bar
{j}}})^{2}(\vec{u}-\frac{\vec{p}_{j}}{x_{j}})^{2}}.
\end{equation}
The integral $I(R,E)$ is convergent and depends neither on $\alpha$ nor on
$\epsilon$. Finally,%
\begin{equation}
S^{\prime}=\,4\left[  \ln\left(  \frac{2E}{\alpha p_{\gamma}^{+}}\right)
\ln\left(  \frac{(\vec{p}_{j}x_{\bar{j}}-x_{j}\vec{p}_{\bar{j}})^{4}}%
{(R^{2}\vec{p}_{j}^{\,\,2}x_{j}^{2})(R^{2}\vec{p}_{\bar{j}}^{\,\,2}x_{\bar{j}%
}^{2})}\right)  +I(R,E)\right]  .\label{Soft_Contribution}%
\end{equation}
Combining (\ref{collinearContribution}) and (\ref{Soft_Contribution}) we have%
\begin{align}
S^{\prime}+n_{j}+n_{\bar{j}} & = \,2\left[  \ln\left(  \frac{{}%
(x_{\bar{j}}\vec{p}_{j}-x_{j}\vec{p}_{\bar{j}})^{4}}{x_{\bar{j}}^{2}x_{j}%
^{2}R^{4}\vec{p}_{\bar{j}}^{\,\,2}\vec{p}_{j}^{\,\,2}}\right)  \ln\left(
\frac{4E^{2}}{x_{\bar{j}}x_{j}(p_{\gamma}^{+}{})^{2}}\right)  \right.
\nonumber\\
&  +2I(R,E)+2\ln\left(  \frac{x_{\bar{j}}x_{j}}{\alpha^{2}}\right)  \left(
\frac{1}{\epsilon}-\ln\left(  \frac{x_{\bar{j}}x_{j}\mu^{2}}{(x_{\bar{j}}%
\vec{p}_{j}-x_{j}\vec{p}_{\bar{j}}){}^{2}}\right)  \right)  -\ln^{2}\left(
\frac{x_{\bar{j}}x_{j}}{\alpha^{2}}\right)  \nonumber\\
&  +\left.  \frac{3}{2}\ln\left(  \frac{16\mu^{4}}{R^{4}{}\vec{p}_{j}%
^{\,\,2}{}\vec{p}_{\bar{j}}^{\,\,2}}\right)  -\ln\left(  \frac{x_{j}}%
{x_{\bar{j}}}\right)  \ln\left(  \frac{x_{j}{}\vec{p}_{\bar{j}}^{\,\,2}%
}{x_{\bar{j}}\vec{p}_{j}^{\,\,2}}\right)  -\frac{3}{\epsilon}-\frac{2\pi^{2}%
}{3}+7\right]  .
\end{align}
Adding the singular part of the virtual correction (\ref{SV}), one finally cancels the $\log(\alpha)$ and $\frac{1}{\epsilon}$ divergences and gets :
\begin{align}
S_{R} &  =S^{\prime}+n_{j}+n_{\bar{j}}+S_{V}+S_{V}^{\ast}=4\left[  \frac{1}%
{2}\ln\left(  \frac{{}(x_{\bar{j}}\vec{p}_{j}-x_{j}\vec{p}_{\bar{j}}){}^{4}%
}{x_{\bar{j}}^{2}x_{j}^{2}R^{4}\vec{p}_{\bar{j}}^{\,\,2}\vec{p}_{j}^{\,\,2}%
}\right)  \left(  \ln\left(  \frac{4E^{2}}{x_{\bar{j}}x_{j}(p_{\gamma}^{+}%
{})^{2}}\right)  +\frac{3}{2}\right)  \right.  \nonumber\\
&  +\left.  I(R,E)+\ln\left(  8\right)  -\frac{1}{2}\ln\left(  \frac{x_{j}%
}{x_{\bar{j}}}\right)  \ln\left(  \frac{x_{j}{}\vec{p}_{\bar{j}}^{\,\,2}%
}{x_{\bar{j}}\vec{p}_{j}^{\,\,2}}\right)  +\frac{13-\pi^{2}}{2}\right]  .
\end{align}
This demonstration of the IR finiteness of this cross section is the main result of present paper. \\
To get the IR-safe exclusive diffractive dijet production cross section in the small
cone approximation one has to take the $q\bar{q}$ production cross section
from section 6, rename the quark momenta via (\ref{jetrenaming}), and
substitute $S_{V}+S_{V}^{\ast}\rightarrow S_{R}$ in the sum of (\ref{dsigma1LL}) and (\ref{dsigma2LL}) for the LL transition, in the sum of (\ref{dsigma1LT}) and (\ref{dsigma2LT}) for the LT transition, and in the sum of (\ref{dsigma1TT}) and (\ref{dsigma2TT}) for the TT transition.

\section{Summary and prospects for further studies}

Using the QCD shock-wave approach~\cite{Balitsky:1995ub,Balitsky:2010ze,Balitsky:2012bs}, we have obtained for the first time 
the $\gamma^* \rightarrow q\bar{q}$ impact factor with one loop accuracy. 
Combined with our previous study of the  $\gamma^* \rightarrow q\bar{q} g$ impact factor~\cite{Boussarie:2014lxa},
 we calculated the cross section for exclusive
diffractive dijet electroproduction off the proton.
For this specific example, we have shown in a detailed way the cancellation of UV and IR soft, collinear and rapidity divergencies. All presented results were obtained without any collinear of soft approximations, in an arbitrary kinematics:  {\it i.e.} for nonzero
incoming photon virtuality, arbitrary $t-$channel momentum transfer and invariant
mass of the produced state .

There are several theoretical developments to be persued based on our study.

\no
First, after applying a suitable Fierz projection, one can obtain the
NLO impact factor for the $\gamma^{(*)} \to \rho-$meson transition in arbitrary kinematics, therefore extending the forward result of Ref.~\cite{Ivanov:2004pp}. At leading twist, this process is dominated by the $\gamma^*_L \to \rho_L$ transition, while transitions with other polarizations start at twist 3.
The impact factor for the transition $\gamma^*_T \to \rho_T$ in the
forward limit was obtained at LO in Ref.~\cite{Anikin:2009hk,Anikin:2009bf}, including both the kinematical twist 3 (the so-called Wandzura Wilczek (WW)~\cite{Wandzura:1977qf} contribution, where the produced meson Fock state is only made of a $q \bar{q}$) and the genuine twist 3 contributions ({\it i.e.} including a $q\bar{q}g$ Fock state).
The present result opens the way to a computation of LO 
$\gamma^* \to \rho$ transitions for arbitrary polarizations and kinematics
(using our $\gamma^{(*)} \rightarrow q\bar{q} g$ Born order result), as well as of the 
NLO $\gamma^{(*)} \to \rho$ impact factor in the WW approximation, using our one-loop $\gamma^{(*)} \rightarrow q\bar{q}$ result.

\no
Second, one could extend the results of our studies to massive quarks. This would allow
for a study of diffractive open charm production, measured at HERA~\cite{Aktas:2006up}, and studied in the large $M$ limit based on the direct coupling between a Pomeron and a $q \bar{q}$ or a $q\bar{q}g$ state, with massive quarks~\cite{Bartels:2002ri}. After applying an appropriate Fierz projection, the NLO $\gamma^{(*)} \to J/\Psi$ impact factor could then be obtained.

\no
Third, there are two ways to apply our result for phenomenological applications. A linearization procedure of the $U$ operators allows one to make connection with the linear BFKL regime. On the other hand, one can also construct a phenomenological model for the matrix elements of the Wilson operators acting on the proton states to approach the saturated regime of the proton or nucleus target. 

\no
Fourth, although we here restricted ourselves to a color-singlet exchange in the $t-$channel, and thus to diffractive processes, an extension to the octet case can be performed, 
{\it i.e.} to the inclusive case. By integrating our results for the $q\bar{q}$ and $q\bar{q}g$ cross sections w.r.t. the external momenta, one can directly obtain the results for NLO $\gamma^* \to \gamma^*$ which was presented in Refs.~\cite{Balitsky:2010ze,Balitsky:2012bs}. A detailed comparison is left for further studies.

On the phenomenological side,
the applications of our results are multiple, and are expected to improve essentially the precision of models based on the $k_T-$factorization picture, since several observables could now be made accessible theoretically with a NLO precision. 
Indeed, it is known that passing from LO to NLO impact factors can have major effects in BFKL type of predictions. The only available process for which such a complete NLO description was obtained~\cite{Colferai:2010wu,Ducloue:2013hia,Caporale:2012ih,Caporale:2013uva,Caporale:2015uva,Celiberto:2015yba} is Mueller-Navelet dijet production~\cite{Mueller:1986ey}. In particular,
the azimuthal decorrelation was recently extracted by CMS~\cite{Khachatryan:2016udy} and confronted with its very good theoretical
description of Refs.~\cite{Ducloue:2013bva,Caporale:2014gpa}. Furthermore, the fact that the $t-$channel exchanged state in our present computation is very general allows one to study not only the linear BFKL regime, but also saturation effects in a proton or a nucleus, here with a NLO precision.

\no
First, the NLO impact factor of the present study could be used to describe 
exclusive dijet diffractive electroproduction~\cite{Abramowicz:2015vnu}, as well as non-exclusive dijet diffractive electroproduction, available at HERA~\cite{Aaron:2011mp}. In the limit $Q^2 \to 0,$
our general result could be also applied to photo-production of diffractive jets~\cite{Chekanov:2007rh,Aaron:2010su}, with a hard scale given by the invariant mass of the produced state, and a precise comparison, now at NLO in the BFKL framework, could be performed with the NLO collinear factorization approach~\cite{Klasen:2004qr,Klasen:2008ah}. 
More generally, at future $ep$ and $eA$ colliders, like EIC~\cite{Boer:2011fh} and LHeC~\cite{AbelleiraFernandez:2012cc}, a large set of
observables will give a possibility to enter the saturation regime in a controllable way, since the saturation scale becomes perturbative for large center of mass energy and/or large values of $A$. 
This includes photoproduction of heavy quarkonia, exclusive diffractive production of light mesons, e.g. $\rho-$meson, either in electroproduction or in large $t$ 
photoproduction. 
In particular, our result allows one to use diffractive dijet production, now considered as a very promising observable for probe the color glass condensate and more generally to perform proton and nucleus tomography at low $x$, now beyond some recent LO analysis~\cite{Altinoluk:2015dpi,Hatta:2016dxp}.

\no
Second, before the advent of future high energy and high luminosity $ep$ and $eA$ colliders, ultraperipheral collisions (UPCs) at high energy, which provide a source of photon from a projectile proton or nucleus, are perfect playgrounds in order to probe the high-energy partonic
content of the target proton or nucleus.
These  are already accessible at the LHC. 
In particular, during the Run I of the LHC, the LHCb collaboration have measured exclusive photoproduction of $J/\psi$ and $\psi(2S)$ mesons~\cite{Aaij:2013jxj,Aaij:2014iea} in $pp$ collision (later extended to $\Upsilon$ in Ref.~\cite{Aaij:2015kea}), while the ALICE collaboration
measured this process in $Pb$~\cite{TheALICE:2014dwa} and 
$PbPb$~\cite{Abbas:2013oua,Abelev:2012ba,Adam:2015sia}
collisions. CMS very recently released a similar analysis for $PbPb$~\cite{Khachatryan:2016qhq}. 
The physics potential of UPCs will improve very significantly thanks to several very forward detectors which are installed, under test or planned in each of the four LHC experiments, in particular the CMS-TOTEM Proton Spectrometer, AD-ALICE, HERSCHEL at LHCb and AFP at ATLAS~\cite{N.Cartiglia:2015gve}.
For example, the protoproduction of large invariant mass diffractive dijet could be studied in UPC during Run II at LHC\footnote{In the usual collinear picture, a recent study of this process has been performed in Ref.~\cite{Guzey:2016tek}.} 
\\
Note : we are aware of a simultaneous computation based on old-fashioned perturbation theory~\cite{Beuf:2016new} for the NLO $\gamma^\ast \rightarrow q\bar{q}$ wavefunction. The results of this study should match our results for $\Phi_{23}^\alpha$ when setting the shockwave momenta $p_{1\bot}$ and $p_{2\bot}$ to 0. Previous results by the same author~\cite{Beuf:2011xd} were confirmed by our previous study~\cite{Boussarie:2014lxa}.

\section*{Acknowledgements}

We thank Ian Balitsky, Guillaume Beuf, Michel Fontannaz, Cyrille Marquet and St\'ephane Munier for discussions.
Andrey V. Grabovsky thanks Andrey L. Feldman, Roman E. Gerasimov, Mikhail G. Kozlov, Alexander I. Milstein, Aleksey V. Reznichenko, and especially Victor S. Fadin for helpful discussions.
This work is partly supported by the grant No 2015/17/B/ST2/01838 of the National Science Center in Poland, by the French grant ANR PARTONS (Grant No. ANR-12-MONU-0008-01), the COPIN-IN2P3 Agreement and the Th\'eorie-LHC France Initiative.
Renaud Boussarie thanks RFBR for financial support
via grant 15-32-50219, and Novosibirsk State University and Budker Institute of Nuclear Physics for their hospitality. Andrey~V.~Grabovsky acknowledges support of president scholarship 171.2015.2,
RFBR grant 16-02-00888, Dynasty foundation, Metchnikov grant and University Paris Sud. He 
is also
grateful to LPT Orsay for hospitality  while part of the
presented work was being done. Lech Szymanowski was supported by a grant from the French Government.

\appendix

\section{Finite part of the virtual correction}

\subsection{Building-block integrals}

Throughout this section, we will need the following integrals

\begin{eqnarray}
I_{1}^{k}(\vec{q}_1,\, \vec{q}_2,\, \Delta_1,\, \Delta_2) & \equiv & \frac{1}{\pi}\int\frac{d^{d}\vec{l}\left(l_{\perp}^{k}\right)}{\left[(\vec{l}-\vec{q}_{1})^{2}+\Delta_{1}\right]\left[(\vec{l}-\vec{q}_{2})^{2}+\Delta_{2}\right]\vec{l}^{^{\, \, 2}}} \label{I1k}, \\
I_2(\vec{q}_1,\, \vec{q}_2,\, \Delta_1,\, \Delta_2) & \equiv & \frac{1}{\pi}\int \frac{d^d \vec{l}}{\left[ (\vec{l}-\vec{q}_1)^2+\Delta_1 \right] \left[ (\vec{l}-\vec{q}_2)^2 +\Delta_2 \right]} \label{I2}, \\
I_3^k(\vec{q}_1,\, \vec{q}_2,\, \Delta_1,\, \Delta_2) & \equiv & \frac{1}{\pi}\int \frac{d^d \vec{l}\left( l_\bot^k \right)}{\left[ (\vec{l}-\vec{q}_1)^2+\Delta_1 \right] \left[ (\vec{l}-\vec{q}_2)^2 +\Delta_2 \right]} \label{I3k}, \\
I^{jk}(\vec{q}_1,\, \vec{q}_2,\, \Delta_1,\, \Delta_2) & \equiv & \frac{1}{\pi}\int\frac{d^{d}\vec{l}\left( l_{\perp}^j l_{\perp}^{k}\right)}{\left[(\vec{l}-\vec{q}_{1})^{2}+\Delta_{1}\right]\left[(\vec{l}-\vec{q}_{2})^{2}+\Delta_{2}\right]\vec{l}^{^{\, \, 2}}} \label{Ijk} \, .
\end{eqnarray}
The arguments of these integrals will be different for each diagram so we will write them explicitly before giving the expression of each diagram, but we will ommit them in the equations for reader's convenience. \\
Explicit results for the first 3 integrals in (\ref{I1k}-\ref{Ijk}) are obtained by a straightforward Feynman parameter integration. We will express them using the following variables :

\begin{eqnarray}
\rho_{1} & \equiv & \frac{\left(\vec{q}_{12}^{\, \, 2}+\Delta_{12}\right)-\sqrt{\left(\vec{q}_{12}^{\, \, 2}+\Delta_{12}\right)^{2}+4\vec{q}_{12}^{\, \, 2}\Delta_{2}}}{2\vec{q}_{12}^{\, \, 2}} ,\\
\rho_{2} & \equiv & \frac{\left(\vec{q}_{12}^{\, \, 2}+\Delta_{12}\right)+\sqrt{\left(\vec{q}_{12}^{\, \, 2}+\Delta_{12}\right)^{2}+4\vec{q}_{12}^{\, \, 2}\Delta_{2}}}{2\vec{q}_{12}^{\, \, 2}} \, , \label{rhovar}
\end{eqnarray}
where $\Delta_{ij} = \Delta_i - \Delta_j$ . \\
One gets :

\begin{eqnarray}
I_{1}^{k} & = & \frac{q_{1\perp}^{k}}{2\left[\vec{q}_{12}^{\, \, 2}\left(\vec{q}_{1}^{\, \, 2}+\Delta_{1}\right)\left(\vec{q}_{2}^{\, \, 2}+\Delta_{2}\right)-\left(\vec{q}_{1}^{\, \, 2}-\vec{q}_{2}^{\, \, 2}+\Delta_{12}\right)\left(\vec{q}_{1}^{\, \, 2}\Delta_{2}-\vec{q}_{2}^{\, \, 2}\Delta_{1}\right)\right]}\\ \nonumber
 & \times & \left\{ \frac{\left(\vec{q}_{2}^{\, \, 2}+\Delta_{2}\right)\vec{q}_{12}^{\, \, 2}+\vec{q}_{2}^{\, \, 2}\left(\Delta_{1}+\Delta_{2}\right)+\Delta_{2}\left(\Delta_{21}-2\vec{q}_{1}^{\, \, 2}\right)}{\left(\rho_{1}-\rho_{2}\right)\vec{q}_{12}^{\, \, 2}}\ln\left[\left(\frac{-\rho_{1}}{1-\rho_{1}}\right)\left(\frac{1-\rho_{2}}{-\rho_{2}}\right)\right]\right.\\ \nonumber
 & \times & \left.\left(\vec{q}_{2}^{\, \, 2}+\Delta_{2}\right)\ln\left[\frac{\Delta_{2}\left(\vec{q}_{1}^{\, \, 2}+\Delta_{1}\right)^{2}}{\Delta_{1}\left(\vec{q}_{2}^{\, \, 2}+\Delta_{2}\right)^{2}}\right]+\left(1\leftrightarrow2\right)\right\} \, ,
\end{eqnarray}

\begin{eqnarray}
I_{2} & = & \frac{1}{\vec{q}_{12}^{\, \, 2}\left(\rho_{1}-\rho_{2}\right)}\ln\left[\left(\frac{-\rho_{1}}{1-\rho_{1}}\right)\left(\frac{1-\rho_{2}}{-\rho_{2}}\right)\right] \, ,
\end{eqnarray}
and

\begin{eqnarray}
I_{3}^{k} & = & \frac{\left(\vec{q}_{12}^{\, \, 2}+\Delta_{12}\right)q_{1}^{k}+\left(\vec{q}_{21}^{\, \, 2}+\Delta_{21}\right)q_{2}^{k}}{2\left(\rho_{1}-\rho_{2}\right)(\vec{q}_{12}^{\, \, 2})^2}\ln\left[\left(\frac{-\rho_{1}}{1-\rho_{1}}\right)\left(\frac{1-\rho_{2}}{-\rho_{2}}\right)\right] \nonumber \\ &-& \frac{q_{12}^{k}}{2\vec{q}_{12}^{\, \, 2}}\ln\left(\frac{\Delta_{1}}{\Delta_{2}}\right) \, .
\end{eqnarray}
Please note that in some cases the real part of $\Delta_1$ or $\Delta_2$ will be negative so the previous results can acquire an imaginary part from the imaginary part $\pm \, i0$ of the arguments. \\ 
The last integral in (\ref{Ijk}) can be expressed in terms of the other ones by writing 
\begin{equation}
I^{jk} = I_{11}\left(q_{1\perp}^{j}q_{1\perp}^{k}\right)+I_{12}\left(q_{1\perp}^{j}q_{2\perp}^{k}+q_{2\perp}^{j}q_{1\perp}^{k}\right)+I_{22}\left(q_{2\perp}^{j}q_{2\perp}^{k}\right) \, ,
\end{equation}
with

\begin{eqnarray}
&& I_{11} = -\frac{1}{2}\frac{\left[\vec{q}_{2}^{\, \, 2}q_{1\perp k}-\left(\vec{q}_{1}\cdot\vec{q}_{2}\right)q_{2\perp k}\right]}{\left[\vec{q}_{1}^{\, \, 2}\vec{q}_{2}^{\, \, 2}-\left(\vec{q}_{1} \cdot \vec{q}_{2}\right)^{2}\right]^{2}} \\ \nonumber
& \times & \left[\left(\frac{\vec{q}_{1}\cdot\vec{q}_{2}}{\vec{q}_{1}^{\, \, 2}}\right)\ln\left(\frac{\vec{q}_{1}^{\, \, 2}+\Delta_{1}}{\Delta_{1}}\right)q_{1\perp}^{k}-\left(\vec{q}_{2}\cdot\vec{q}_{12}\right)I_{3}^{k}+\left\{ \vec{q}_{2}^{\, \, 2}\left(\vec{q}_{1}\cdot\vec{q}_{12}\right)+\Delta_{1}\vec{q}_{2}^{\, \, 2}-\Delta_{2}\left(\vec{q}_{1}\cdot\vec{q}_{2}\right)\right\} I_{1}^{k}\right]\\
&& I_{12} = -\frac{1}{2}\frac{\left[\vec{q}_{1}^{\, \, 2}q_{2\perp k}-\left(\vec{q}_{1}\cdot\vec{q}_{2}\right)q_{1\perp k}\right]}{\left[\vec{q}_{1}^{\, \, 2}\vec{q}_{2}^{\, \, 2}-\left(\vec{q}_{1}\cdot\vec{q}_{2}\right)^{2}\right]^{2}} \\ \nonumber
& \times & \left[-\ln\left(\frac{\vec{q}_{2}^{\, \, 2}+\Delta_{2}}{\Delta_{2}}\right)q_{2\perp}^{k}-\left(\vec{q}_{2}\cdot\vec{q}_{12}\right)\tilde{I}_{3}^{k}+\left\{ \vec{q}_{2}^{\, \, 2}\left(\vec{q}_{1}\cdot\vec{q}_{12}\right)+\Delta_{1}\vec{q}_{2}^{\, \, 2}-\Delta_{2}\left(\vec{q}_{1}\cdot\vec{q}_{2}\right)\right\} \tilde{I}_{1}^{k}\right]\\
&& I_{22} = -\frac{1}{2}\frac{\left[\vec{q}_{1}^{\, \, 2}q_{2\perp k}-\left(\vec{q}_{1}\cdot\vec{q}_{2}\right)q_{1\perp k}\right]}{\left[\vec{q}_{1}^{\, \, 2}\vec{q}_{2}^{\, \, 2}-\left(\vec{q}_{1}\cdot\vec{q}_{2}\right)^{2}\right]^{2}} \\ \nonumber
& \times & \left[\left(\frac{\vec{q}_{1}\cdot\vec{q}_{2}}{\vec{q}_{2}^{\, \, 2}}\right)\ln\left(\frac{\vec{q}_{2}^{\, \, 2}+\Delta_{2}}{\Delta_{2}}\right)q_{2\perp}^{k}+\left(\vec{q}_{1}\cdot\vec{q}_{12}\right)I_{3}^{k}-\left\{ \vec{q}_{1}^{\, \, 2}\left(\vec{q}_{2}\cdot\vec{q}_{12}\right)-\Delta_{2}\vec{q}_{1}^{\, \, 2}+\Delta_{1}\left(\vec{q}_{1}\cdot\vec{q}_{2}\right)\right\} I_{1}^{k}\right] \, .
\end{eqnarray}
This last expression makes it seem that there is a singularity when $\vec{q}_1$ and $\vec{q}_2$ are collinear or anticollinear. However this singularity is non physical and only appears when projecting on the particular basis of 2-dimensional symmetric tensors $(q_1^j q_1^k, \, q_1^j q_2^k + q_2^j q_1^k, \, q_2^j q_2^k)$. One can show that it disappears when projecting on the non-minimal basis $(q_1^j q_1^k, \, q_1^j q_2^k + q_2^j q_1^k, \, q_2^j q_2^k, \, g^{jk}_\bot)$. For a further study, the reader is referred to~\cite{Bogdan:2007qj}

\subsection{Diagram 4}

Here the integrals from section (A.1) will have the following arguments
:
\begin{eqnarray*}
\vec{q}_{1} & = & \vec{p}_{1}-\left(\frac{x-z}{x}\right)\vec{p}_{q}, \quad \, \, \, \, \,
\vec{q}_{2}  =  \left(\frac{x-z}{x}\right)\left(x\vec{p}_{\bar{q}}-\bar{x}\vec{p}_{q}\right) \, ,\\
\Delta_{1} & = & \left(x-z\right)\left(\bar{x}+z\right)Q^{2}, \quad
\Delta_{2}  =  -\frac{x\left(\bar{x}+z\right)}{\bar{x}\left(x-z\right)}\vec{q}^{2}-i0\,.
\end{eqnarray*}

Let us write the impact factors in terms of these variables. They read : \\
(longitudinal NLO) $\times$ (longitudinal LO) contribution :
\begin{equation}
\left(\phi_{4}\right)_{LL}=-\frac{4(x-z)(\bar{x}+z)}{z}[-\bar{x}(x-z)(z+1)I_{2}+q_{2\bot k}(2x^{2}-(2x-z)(z+1))I_{1}^{k}] \, ,
\end{equation}
(longitudinal NLO) $\times$ (transverse LO) contribution :
\begin{equation}
\left(\phi_{4}\right)_{LT}^{j}=(1-2x)p_{q1^{\prime}}{}_{\bot}^{j}\left(\phi_{4}\right)_{LL}-4(x-z)(\bar{x}+z)(1-2x+z)[(\vec{q}\cdot\vec{p}_{q1^{\prime}})g_{\bot k}^{j}+q_{2\bot}^{j}p_{q1^{\prime}\bot k}]I_{1}^{k} \, ,
\end{equation}
(transverse NLO) $\times$ (longitudinal LO) contribution :
\begin{align}
\left(\phi_{4}\right)_{TL}^{i} & =2\{[(x-\bar{x}-z)q_{2\bot}^{i}q_{1\bot k}+(-8x\bar{x}-6xz+2z^{2}+3z+1)q_{1\perp}^{j}q_{2\bot k}]I_{1}^{k}\nonumber \\
 & -2[4x^{2}-x(3z+5)+(z+1)^{2}]q_{2\bot k}I^{ik}+(x-\bar{x}-z)\left(\vec{q}_{2}\cdot\vec{q}_{1}\right)I^{i}\nonumber \\
 & +I_{2}[(x-\bar{x}-z)q_{2\bot}^{i}+\bar{x}(2(x-z)^{2}-5x+3z+1)q_{1\perp}^{i}]\nonumber \\
 & -\bar{x}[2(x-z)^{2}-5x+3z+1]I_{3}^{i}\nonumber \\
 & +\frac{x\bar{x}(1-2x)}{z}[2q_{2\bot k}I^{ik}+I_{3}^{i}-q_{1\perp}^{i}(2q_{2\bot k}I_{1}^{k}+I_{2})]\} \, ,
\end{align}
(transverse NLO) $\times$ (transverse LO) contribution :
\begin{eqnarray} \nonumber
\left(\phi_{4}\right)_{TT}^{ij} & = & \left[(x-\bar{x}-2z)(x-\bar{x}-z)(\vec{q}_{2}\cdot\vec{p}_{q1^{\prime}})q_{1\perp}^{i}+(z+1)(\left(\vec{q}_{1}\cdot\vec{q}_{2}\right)p_{q1^{\prime}\perp}^{i}-(\vec{q}_{1}\cdot\vec{p}_{q1^{\prime}})q_{2\bot}^{i})\right]I_{1}^{j}\\ \nonumber
 &+& 2\bar{x}[q_{2\bot k}-(x-z)q_{1\perp k}](p_{q1^{\prime}\bot}^{i}I^{jk}-g_{\bot}^{ij}p_{q1^{\prime}\bot l}I^{kl}) \\ \nonumber
 &+& 2(x-z)[(2\bar{x}+z)(\vec{q}_{2}\cdot\vec{p}_{q1^{\prime}})-\bar{x}(\vec{q}_{1}\cdot\vec{p}_{q1^{\prime}})]I^{ij}\\ \nonumber
 &+& [(1-z)((\vec{q}_{1}\cdot\vec{p}_{q1^{\prime}})q_{2\bot}^{j}-(\vec{q}_{2}\cdot\vec{p}_{q1^{\prime}})q_{1\perp}^{j})-(1-2x)(\bar{x}-x+z)\left(\vec{q}_{1}\cdot\vec{q}_{2}\right)p_{q1^{\prime}\perp}^{j}]I_{1}^{i}\\ \nonumber
 &-& 2\left[(x-z)(\bar{x}q_{1\perp}^{j}-(2\bar{x}+z)q_{2\bot}^{j})p_{q1^{\prime}\perp k} \right. \\ \nonumber 
 &+& \left. (1-2x)\left(4x^{2}-(3z+5)x+(z+1)^{2}\right)q_{2\bot k}p_{q1^{\prime}}{}_{\bot}^{j}\right]I^{ik}\\ \nonumber
 &-& \bar{x}\left(\bar{x}-x\right)\left(2(x-z)^{2}-5x+3z+1\right)p_{q1^{\prime}\perp}^{j}I_{3}^{i} \\ \nonumber
 &+& \bar{x}\left(\bar{x}+z\right)(p_{q1^{\prime}\perp}^{i}I_{3}^{j}-g_{\bot}^{ij}p_{q1^{\prime}\perp k}I_{3}^{k})\\ \nonumber
 &+& I_{2}\left[g_{\bot}^{ij}\left((1-z)(\vec{q}_{2}\cdot\vec{p}_{q1^{\prime}})-\bar{x}(1+x-z)(\vec{q}_{1}\cdot\vec{p}_{q1^{\prime}})\right) \right. \\ \nonumber 
 &+& \left.((1-z)q_{2\bot}^{j}-\bar{x}(1+x-z)q_{1\perp}^{j})p_{q1^{\prime}}{}_{\bot}^{i}\right.\\ \nonumber
 &-& \left.(\bar{x}-x)\left((\bar{x}-x+z)q_{2\bot}^{i}-\bar{x}\left(2(x-z)^{2}-5x+3z+1\right)q_{1\perp}^{i}\right)p_{q1^{\prime}}{}_{\bot}^{j}\right]\\ \nonumber
 &+& I_{1}^{k}\left[g_{\bot}^{ij}\left((x-\bar{x}+z)(\vec{q}_{1}\cdot\vec{p}_{q1^{\prime}})q_{2\bot k}+(1-z)(\vec{q}_{2}\cdot\vec{p}_{q1^{\prime}})q_{1\bot k}-(z+1)\left(\vec{q}_{1}\cdot\vec{q}_{2}\right)p_{q1^{\prime}}{}_{\bot k}\right)\right.\\ \nonumber
 &+& q_{1\perp}^{j}((x-\bar{x}+z)q_{2\bot k}p_{q1^{\prime}\perp}^{i}-(z+1)q_{2\bot}^{i}p_{q1^{\prime}}{}_{\bot k})\\ \nonumber
 &+& q_{2\bot}^{j}((x-\bar{x}-2z)(x-\bar{x}-z)q_{1\perp}^{i}p_{q1^{\prime}\perp k}+(1-z)q_{1\perp k}p_{q1^{\prime}}{}_{\bot}^{i})\\ \nonumber
&-&\left.(1-2x)((1-2x+z)q_{2\bot}^{i}q_{1\perp k}-(2z^{2}+3z-x(8\bar{x}+6z)+1)q_{1\perp}^{i}q_{2\bot k})p_{q1^{\prime}}{}_{\bot}^{j}\right]\\ \nonumber
 &+& \frac{x\bar{x}}{z}\left[(x-\bar{x})^{2}p_{q1^{\prime}\perp}^{j}(2q_{2\bot k}I^{ik}+I_{3}^{i}-q_{1\perp}^{i}(I_{2}+2q_{2\bot k}I_{1}^{k}))\right.\\ \nonumber
 &+& p_{q1^{\prime}\perp}^{i}(q_{1\perp}^{j}(I_{2}+2q_{2\bot k}I_{1}^{k})-2q_{2\bot k}I^{jk}-I_{3}^{j})\\
 &+& \left.g_{\bot}^{ij}((\vec{q}_{1}\cdot\vec{p}_{q1^{\prime}})(I_{2}+2q_{2\bot k}I_{1}^{k})+p_{q1^{\prime}\perp k}(2q_{2\bot l}I^{kl}+I_{3}^{k}))\right]\, .
\end{eqnarray}

\subsection{Diagram 5}
Here the integrals from section (A.2) will have the following arguments :

\begin{equation}
\vec{q}_1 = \left( \frac{x-z}{x} \right) \vec{p}_3 -\frac{z}{x}\vec{p}_1, \quad \vec{q}_2 = \vec{p}_{q1} - \frac{z}{x}\vec{p}_q \, ,\label{var1D5}
\end{equation}
\begin{equation}
 \Delta_1 = \frac{z(x-z)}{x^2\bar{x}} (\vec{p}_{\bar{q}2}^{\, \, 2}+ x\bar{x}Q^2), \quad  \Delta_2 = (x-z)(\bar{x}+z)Q^2 \label{var2D5}\, ,
\end{equation}
With such variables, it is easy to see that the argument in the square roots in (\ref{rhovar}) are full squares.
In terms of the variables in (\ref{var1D5}), the impact factors read : \\
(longitudinal NLO) $\times$ (longitudinal LO) : 
\begin{eqnarray}
\left(\phi_{5}\right)_{LL}=\frac{4(x-z)(-2x(\bar{x}+z)+z^{2}+z)}{xz}\left[\bar{x}(x-z)I_{2}-\left(zq_{1\perp k}-x\left(\bar{x}+z\right)q_{2\bot k}\right)I_{1}^{k}\right] \, ,
\end{eqnarray}
(longitudinal NLO) $\times$ (transverse LO) : 
\begin{align}
\left(\phi_{5}\right)_{LT}^{j} & =(\bar{x}-x)p_{q1^{\prime}\bot}^{j}\left(\phi_{5}\right)_{LL} \\ \nonumber &+\frac{4(x-z)(x-\bar{x}-z)}{x}\left(zq_{1\perp}^{k}-x(\bar{x}+z)q_{2\perp}^{k}\right)p_{q1^{\prime}\perp l}\left(g_{\perp k}^{j}I_{1}^{l}+I_{1}^{j}\right)\, ,
\end{align}
(transverse NLO) $\times$ (longitudinal LO) : 
\begin{align}
\left(\phi_{5}\right)_{TL}^{i} & =2\left[(x-\bar{x}-z)\left(\vec{q}_{1}\cdot\vec{q}_{2}\right)-\bar{x}(x-z)^{2}Q^{2}+(\frac{z}{x}-x)\vec{q}_{1}^{\,\,2}\right]I_{1}^{i}\nonumber \\
 & +\frac{2}{x}\left[xq_{2\bot k}(-8x\bar{x}-6xz+2z^{2}+3z+1)+2q_{1\bot k}(2xz-2x^{2}+x-z^{2})\right]q_{1\perp}^{i}I_{1}^{k}\nonumber \\
 & +2q_{2\bot}^{i}q_{1\perp k}(x-\bar{x}-z)I_{1}^{k}+2\frac{\bar{x}}{x}(x(8x-3)-6xz+2z^{2}+z)I_{1}^{i}\nonumber \\
 & +\frac{2}{x}\left[xq_{2\bot}^{i}(x-\bar{x}-z)+q_{1\perp}^{i}(8x^{3}-6x^{2}(z+2)+x(z+3)(2z+1)-2z^{2})\right]I_{2}\nonumber \\
 & -\frac{4}{x}\left[(x-z)(\bar{x}+z)q_{1\perp k}+x(4x^{2}-x(3z+5)+(z+1)^{2})q_{2\perp k}\right]I^{ik}\nonumber \\
 & -\frac{4}{z}x\bar{x}(x-\bar{x})\left[q_{2\perp k}I^{ik}+I_{3}^{i}-q_{1\perp}^{i}\left(q_{2\perp k}I_{1}^{k}+I_{2}\right)\right] \, ,
\end{align}
(transverse NLO) $\times$ (transverse LO) : 

\begin{eqnarray*}
&& \left(\phi_{5}\right)_{TT}^{ij}  =  -2(x-z)\left[\frac{z}{x}(\vec{q}_{1}\cdot\vec{p}_{q1^{\prime}})-(2\bar{x}+z)(\vec{q}_{2}\cdot\vec{p}_{q1^{\prime}})\right]I^{ij}\\
 & + & \left[-\bar{x}(x-z)^{2}Q^{2}p_{q1^{\prime}\perp}^{i}+(\bar{x}-x+2z)(\bar{x}-x+z)(\vec{q}_{2}\cdot\vec{p}_{q1^{\prime}})q_{1\perp}^{i}\right.\\
 & - & \left.(\vec{q}_{1}\cdot\vec{p}_{q1^{\prime}})((z+1)q_{2\bot}^{i}-2\frac{z}{x}(2x-z)q_{1\perp}^{i}) \right. \\ \nonumber
 &+& \left. ((z+1)\left(\vec{q}_{1}\cdot\vec{q}_{2}\right)-\left(x+\frac{z}{x}\right)\vec{r}^{\,\,2})p_{q1^{\prime}\bot}^{i}\right]I_{1}^{j}\\
 & - & 2\frac{\bar{x}}{x}(xq_{2\perp k}+(x-z)q_{1\perp k})\left(g_{\bot}^{ij}p_{q1^{\prime}\perp l}I^{kl}-p_{q1^{\prime}}{}_{\bot}^{i}I^{jk}\right)\\
 & + & \left[\bar{x}\left(x-\bar{x}\right)(x-z)^{2}Q^{2}p_{q1^{\prime}\perp}^{j}-(z-1)(\vec{q}_{1}\cdot\vec{p}_{q1^{\prime}})q_{2\bot}^{j}\right.\\
 & + & \left.(z-1)(\vec{q}_{2}\cdot\vec{p}_{q1^{\prime}})q_{1\perp}^{j}+\frac{x-\bar{x}}{x}\left((x^{2}-z)\vec{q}_{1}^{\,\,2}+x(\bar{x}-x+z)(\vec{q}_{1}\cdot\vec{q}_{2})\right)p_{q1^{\prime}\perp}^{k}\right]I_{1}^{i}\\
 & + & 2\left[\frac{x-\bar{x}}{x}\left(x(4x^{2}-(3z+5)x+(z+1)^{2})q_{2\bot k}+(x-z)(\bar{x}+z)q_{1\perp k}\right)p_{q1^{\prime}\perp}^{j}\right.\\
 & - & \left.\frac{x-z}{x}\left(x(2x-z-2)q_{2\bot}^{j}+zq_{1\perp}^{j}\right)p_{q1^{\prime}\perp k}\right]I^{ik} \\ \nonumber 
 & + & \frac{\bar{x}\left(\bar{x}-x\right)}{x}\left(2z^{2}-6xz+z+x(8x-3)\right)p_{q1^{\prime}\perp}^{j}I_{3}^{i}\\
 & + & \left[(x-\bar{x})\left((\bar{x}-x+z)q_{2\bot}^{i}+\left(6(z+2)x-8x^{2}-(z+3)(2z+1)+2\frac{z^{2}}{x}\right)q_{1\perp}^{i}r_{\bot}^{i}\right)p_{q1^{\prime}\perp}^{j}\right.\\
 & + & \left.(1-z)(g_{\bot}^{ij}(\vec{q}_{2}\cdot\vec{p}_{q1^{\prime}})+q_{2\bot}^{k}p_{q1^{\prime}\perp}^{i})+(2x+z-3)(g_{\bot}^{ik}(\vec{q}_{1}\cdot\vec{p}_{q1^{\prime}})+q_{1\perp}^{k}p_{q1^{\prime}\perp}^{i})\right]I_{2}\\
 & + & \left(3\bar{x}+z-\frac{z}{x}\right)p_{q1^{\prime}\perp}^{i}I_{3}^{k}-\frac{\bar{x}}{x}(3x-z)g_{\bot}^{ij}p_{q1^{\prime}\perp k}I_{3}^{k}\\
 & + & \left[(x-\bar{x})p_{q1^{\prime}\perp}^{j}\left\{ (\bar{x}-x+z)q_{2\bot}^{i}q_{1\perp k}-(2z^{2}-6xz+3z-8x\bar{x}+1)q_{2\perp k}q_{1\perp}^{i}\right.\right.\\
 & - & \left. 2(\bar{x}-x+2z-\frac{z^{2}}{x})q_{1\perp k}q_{1\perp}^{i}\right\} +\bar{x}(x-z)^{2}Q^{2}g_{\bot}^{ij}p_{q1^{\prime}\perp k} \\ \nonumber 
 &+& (1-z)q_{1\perp k}(g_{\bot}^{ij}(\vec{q}_{2}\cdot\vec{p}_{q1^{\prime}})+q_{2\bot}^{j}p_{q1^{\prime}\perp}^{i})\\
 & + & \left((x-\bar{x}+z)q_{2\bot k}-2q_{1\perp k}\right)(g_{\bot}^{ij}(\vec{q}_{1}\cdot\vec{p}_{q1^{\prime}})+q_{1\perp}^{j}p_{q1^{\prime}\perp}^{i}) \\ \nonumber 
 &+& g_{\bot}^{ij}\left(\left(x+\frac{z}{x}\right)\vec{q}_{1}^{\,\,2}-(z+1)(\vec{q}_{1}\cdot\vec{q}_{2})\right)p_{q1^{\prime}\perp k}\\
 & + & \left.\left((x-\bar{x}-2z)(x-\bar{x}-z)q_{1\perp}^{i}q_{2\perp}^{j}-(z+1)q_{2\perp}^{i}q_{1\perp}^{j}+2(2x-z)\frac{z}{x}q_{1\perp}^{i}q_{1\perp}^{j}\right)p_{q1^{\prime}\perp k}\right]I_{1}^{k}\\
 & + & \frac{2x\bar{x}}{z}\left[(x-\bar{x})^{2}p_{q1^{\prime}\perp}^{j}(q_{2\bot k}I^{ik}+I_{3}^{i})-p_{q1^{\prime}\perp}^{i}(q_{2\bot k}I^{jk}+I_{3}^{k})+g_{\bot}^{ij}p_{q1^{\prime}\bot k}(q_{2\bot l}I^{kl}+I_{3}^{k})\right.\\
 & + & \left.(I_{2}+q_{2\perp k}I_{1}^{k})\left(g_{\bot}^{ij}(\vec{q}_{1}\cdot\vec{p}_{q1^{\prime}})+q_{1\perp}^{j}p_{q1^{\prime}\perp}^{i}-(1-2x)^{2}q_{1\perp}^{i}p_{q1^{\prime}\perp}^{j}\right)\right].
\end{eqnarray*}

\subsection{Diagram 6}

For this diagram we will use the variable
\begin{equation}
\vec{q}=\left( \frac{x-z}{x} \right) \vec{p}_{3}-\frac{z}{x}\vec{p}_1 \, .
\end{equation}%
Then the impact factors read : \\
(longitudinal NLO) $\times $ (longitudinal NLO) :
\begin{equation}
(\phi_6)_{LL}=-4x\bar{x}^2 J_0 \, ,
\end{equation}
(longitudinal NLO) $\times $ (transverse NLO) :
\begin{equation}
(\phi_6)_{LT}^j = (1-2x)p_{q1^\prime\bot}^j(\phi_6)_{LL}\, ,
\end{equation}
(transverse NLO) $\times $ (longitudinal NLO) :
\begin{equation}
(\phi_6)_{TL}^i = 2\bar{x}\left[ (1-2x) p_{\bar{q}2\bot}^{i} J_0 - J_{1\bot}^i \right] \, ,
\end{equation}
(transverse NLO) $\times $ (transverse NLO) :
\begin{align}
(\phi_6)_{TT}^{ij} &  =\bar{x}\left[(x-\bar{x})^{2}p_{\bar{q}2\bot}^{i}%
p_{q1^{\prime}\bot}^{j}-g_{\bot}^{ij}(\vec{p}_{\bar{q}2}
\cdot\vec{p}_{q1^{\prime}})-p_{q1^{\prime}\bot}^{i}p_{\bar{q}2\bot}^{j}\right]J_0 \nonumber\\
&  +\bar{x} \left[(x-\bar{x})p_{q1^{\prime}\bot}^{j}g_{\bot k}^i - p_{q1^\prime\bot k}g_{\bot}^{ij}+p_{q1^{\prime
}\bot}^{i}g_{\bot k}^j\right]J_{1\bot}^k \, .
\end{align}%
We introduced
\begin{align}
J_{1\bot}^{k}  &  =\frac{(x-z)^{2}}{x^{2}}\frac{q_{\bot}^{k}}{\vec
{q}^{\,\,2}}\ln\left(  \frac{\vec{p}_{\bar{q}2}^{\,\,2}+x\bar{x}Q^{2}}{\vec{p}_{\bar{q}2}^{\,\,2}+x\bar{x}Q^{2}+	\frac{x^2 \bar{x}}{z(x-z)}\vec{q}^{\,\,2}%
}\right)  ,\\ \nonumber
\mathrm{and} \\
J_0 &  =\frac{z}{x(\vec{p}_{\bar{q}2}^{\,\,2}+x\bar{x}Q^{2})}  -\frac{2x(x-z)+z^{2}}{xz(\vec{p}_{\bar{q}2}^{\,\,2}+x\bar{x}Q^{2})}
\ln\left(  \frac{x^2\bar{x}\mu^{2}}{z(x-z)(\vec{p}_{\bar{q}2}^{\,\,2}+x\bar{x}Q^{2})+x^{2}\bar{x}\vec{q}^{\,\,2}}\right)  .
\end{align}

\section{Real correction}

Here we present the convoluted impact factors from section 5.

\subsection{LL photon transition}%

\begin{align}
\Phi &  _{4}^{+}(p_{1\bot},p_{2\bot},p_{3\bot})\Phi_{4}^{+}(p_{1\bot}^{\prime
},p_{2\bot}^{\prime},p_{3\bot}^{\prime})^{\ast}=\frac{8p_{\gamma}^{+}{}^{4}%
}{z^{2}\left(  \frac{\vec{p}_{\bar{q}2^\prime}^{\, \, 2}%
}{x_{\bar{q}}\left( 1 - x_{\bar{q}} \right) }+Q^{2}\right)  \left( Q^2+\frac{\vec{p_{q1^{\prime}}}{}^{2}}{x_{q}%
} + \frac{\vec{p}_{\bar{q}2^\prime}^{\, \, 2}}{x_{\bar{q}}}+\frac
{\vec{p_{g3^{\prime}}}{}^{2}}{z}\right)  }\nonumber\\
\times &  \left[  \frac{x_{\bar{q}}\left(  dz^{2}+4x_{q}\left(  x_{q}%
+z\right)  \right)  \left(  x_{q}\vec{p}_{g3}-z\vec{p}_{q1})(x_{q}\vec
{p}_{g3^{\prime}}-z\vec{p}_{q1^{\prime}}\right)  }{x_{q}\left(  x_{q}%
+z\right)  ^{2}{}\left(  \frac{(\vec{p}_{g3}+\vec{p}_{q1}){}^{2}}{x_{\bar{q}%
}\left(  x_{q}+z\right)  }+Q^{2}\right)  \left(  \frac{(\vec{p}_{g3}+\vec
{p}_{q1}){}^{2}}{x_{\bar{q}}}+\frac{\vec{p_{g3}}{}^{2}}{z}+\frac{\vec{p_{q1}%
}{}^{2}}{x_{q}}+Q^{2}\right)  }\right. \nonumber\\
-  &  \left.  \frac{(4x_{q}x_{\bar{q}}+2z-dz^{2})(x_{\bar{q}}\vec{p}%
_{g3}-z\vec{p}_{\bar{q}2})(x_{q}\vec{p}_{g3^{\prime}}-z\vec{p}_{q1^{\prime}}%
)}{\left(  x_{\bar{q}}+z\right)  \left(  x_{q}+z\right)  \left(  \frac
{(\vec{p}_{\bar{q}2}+\vec{p}_{g3}){}^{2}}{x_{q}\left(  x_{\bar{q}}+z\right)
}+Q^{2}\right)  \left(  \frac{(\vec{p}_{\bar{q}2}+\vec{p}_{g3}){}^{2}}{x_{q}%
}+\frac{\vec{p_{g3}}{}^{2}}{z}+\frac{\vec{p}_{\bar{q}2}^{\,\,2}}{x_{\bar{q}}%
}+Q^{2}\right)  }\right]  +(q\leftrightarrow\bar{q}).
\end{align}
Now $(q\leftrightarrow\bar{q})$ stands for $p_{q}\leftrightarrow p_{\bar{q}%
},\,p_{1}^{(\prime)}\leftrightarrow p_{2}^{(\prime)},\,x_{q}\leftrightarrow
x_{\bar{q}}.$%
\begin{equation}
\Phi_{3}^{+}(p_{1\bot},p_{2\bot})\Phi_{3}^{+}(p_{1\bot}^{\prime},p_{2\bot
}^{\prime})^{\ast}=\Phi_{4}^{+}(p_{1\bot},p_{2\bot},0)\Phi_{4}^{+}(p_{1\bot
}^{\prime},p_{2\bot}^{\prime},0)^{\ast}+A^{++}+B^{++}.
\end{equation}
Here the first term in the r.h.s. is responsible for the emission of the gluon before crossing the shockwave, $A$ describes the emission after the shockwave and $B$ is
the interference term. $A$ and $B$ are given by :
\begin{align}
A^{++}=  &  \frac{8x_{\bar{q}}{}p_{\gamma}^{+}{}^{4}\left(  dz^{2}%
+4x_{q}\left(  x_{q}+z\right)  \right)  }{x_{q}{}(\vec{p}_{g}-\frac{z\vec
{p}_{q}}{x_{q}})^{2}\left(  \frac{\vec{p}_{\bar{q}2}{}^{2}}{x_{\bar{q}}\left(
x_{q}+z\right)  }+Q^{2}\right)  \left(  \frac{\vec{p}_{\bar{q}2^{\prime}}%
{}^{2}}{x_{\bar{q}}\left(  x_{q}+z\right)  }+Q^{2}\right)  }\nonumber\\
-  &  \frac{8p_{\gamma}^{+}{}^{4}\left(  2z-dz^{2}+4x_{q}x_{\bar{q}}\right)
(\vec{p}_{g}-\frac{z\vec{p}_{q}}{x_{q}})(\vec{p}_{g}-\frac{z\vec{p}_{\bar{q}}%
}{x_{\bar{q}}})}{{}(\vec{p}_{g}-\frac{z\vec{p}_{q}}{x_{q}})^{2}({}\vec{p}%
_{g}-\frac{z\vec{p}_{\bar{q}}}{x_{\bar{q}}})^{2}\left(  \frac{\vec{p}_{\bar
{q}2^{\prime}}{}^{2}}{x_{\bar{q}}\left(  x_{q}+z\right)  }+Q^{2}\right)
\left(  \frac{\vec{p}_{q1}{}^{2}}{x_{q}\left(  x_{\bar{q}}+z\right)  }%
+Q^{2}\right)  }+(q\leftrightarrow\bar{q}) \, , \label{App}
\end{align}
and
\begin{align}
B^{++}  &  =\left[  \frac{8p_{\gamma}^{+}{}^{4}}{z\left(  x_{q}+z\right)
\left(  \frac{\vec{p}{}_{\bar{q}2^{\prime}}^{\,\,2}}{x_{\bar{q}}\left(
x_{q}+z\right)  }+Q^{2}\right)  \left(  \frac{\vec{p}{}_{q1^{\prime}}^{\,\,2}%
}{x_{q}}+\frac{\vec{p}{}_{\bar{q}2^{\prime}}^{\,\,2}}{x_{\bar{q}}}+\frac
{\vec{p_{g}}{}^{2}}{z}+Q^{2}\right)  }\right. \nonumber\\
&  \times\left\{  \frac{\left(  4x_{q}x_{\bar{q}}+z(2-dz)\right)  (\vec{p_{g}%
}{}-\frac{z}{x_{\bar{q}}}\vec{p}_{\bar{q}})(x_{q}\vec{p}_{g}-z\vec
{p}_{q1^{\prime}})}{(\vec{p}_{g}-\frac{z\vec{p}_{\bar{q}}}{x_{\bar{q}}}){}%
^{2}\left(  \frac{\vec{p}{}_{q1}^{\,\,2}}{x_{q}\left(  x_{\bar{q}}+z\right)
}+Q^{2}\right)  }\right. \nonumber\\
&  -\left.  \left.  \frac{x_{\bar{q}}\left(  dz^{2}+4x_{q}\left(
x_{q}+z\right)  \right)  ({}p_{g}-\frac{z}{x_{q}}\vec{p}_{q})(\vec{p_{g}%
}-\frac{z}{x_{q}}\vec{p}_{q1^{\prime}})}{(\vec{p}_{g}-\frac{z\vec{p}_{q}%
}{x_{q}}){}^{2}\left(  \frac{\vec{p}{}_{\bar{q}2}^{\,\,2}}{x_{\bar{q}}\left(
x_{q}+z\right)  }+Q^{2}\right)  }\right\}  +(q\leftrightarrow\bar{q})\right]
\nonumber\\
&  +(1\leftrightarrow1^{\prime},2\leftrightarrow2^{\prime}).
\end{align}
On one hand in the collinear region (\ref{jetalg}) only the first term of $A^{++}$ in (\ref{App}) gives
a nonvanishing contribution in the small cone approximation. Using the variables defined in
(\ref{Delta}), (\ref{jetcol1}), (\ref{jetcol2}), the first line of
$A^{++}$ becomes
\begin{align}
\frac{8x_{\bar{q}}{}p_{\gamma}^{+}{}^{4}\left(  dz^{2}+4x_{q}\left(
x_{q}+z\right)  \right)  }{x_{q}{}(\vec{p}_{g}-\frac{z\vec{p}_{q}}{x_{q}}%
)^{2}\left(  \frac{\vec{p}_{\bar{q}2}{}^{2}}{x_{\bar{q}}\left(  x_{q}%
+z\right)  }+Q^{2}\right)  \left(  \frac{\vec{p}_{\bar{q}2^{\prime}}{}^{2}%
}{x_{\bar{q}}\left(  x_{q}+z\right)  }+Q^{2}\right)  }  &  =\frac
{8(x_{j}-z)(1-x_{j}){}p_{\gamma}^{+}{}^{4}\left(  dz^{2}+4x_{j}\left(
x_{j}-z\right)  \right)  }{x_{j}^{2}\Delta_{q}^{2}\left(  \frac{\vec{p}%
_{\bar{j}2}{}^{2}}{x_{j}\left(  1-x_{j}\right)  }+Q^{2}\right)  \left(
\frac{\vec{p}_{\bar{j}2^{\prime}}{}^{2}}{x_{j}\left(  1-x_{j}\right)  }%
+Q^{2}\right)  }\nonumber\\
&  =\Phi_{0}^{+}\Phi_{0}^{+\ast}\frac{1}{4}\frac{(x_{j}-z){}\left(
dz^{2}+4x_{j}\left(  x_{j}-z\right)  \right)  }{x_{j}^{3}\Delta_{q}^{2}},
\end{align}
which coincides with the integrand for $n_{j}$ defined in (\ref{CollinearFactor}). \\
On the other hand if we use the soft gluon approximation by renaming
\begin{equation}
\vec{p}_{g}=z\vec{u} \label{p=zu}%
\end{equation}
and taking $z\rightarrow0,$ we obtain
\begin{equation}
\Phi_{3}^{+}(p_{1\bot},p_{2\bot})\Phi_{3}^{+}(p_{1\bot}^{\prime},p_{2\bot
}^{\prime})^{\ast}|_{z\rightarrow0}=\frac{1}{z^{2}}\Phi_{0}^{+}(p_{1\bot
},p_{2\bot})\Phi_{0}^{+}(p_{1\bot}^{\prime},p_{2\bot}^{\prime})^{\ast}%
\frac{(\frac{\vec{p}_{q}}{x_{q}}-\frac{\vec{p}_{\bar{q}}}{x_{\bar{q}}})^{2}%
}{(\vec{u}-\frac{\vec{p}_{q}}{x_{q}})^{2}(\vec{u}-\frac{\vec{p}_{\bar{q}}%
}{x_{\bar{q}}})^{2}}+O(z^{-1}),
\end{equation}
which is the soft gluon emission factor.

\subsection{TL photon transition}%

\begin{align}
&  \Phi_{4}^{i}(p_{1\bot},p_{2\bot},p_{3\bot})\Phi_{4}^{+}(p_{1\bot}^{\prime
},p_{2\bot}^{\prime},p_{3\bot}^{\prime})^{\ast}=\frac{-4p_{\gamma}^{+}{}^{3}%
}{\left(  Q^{2}+\frac{\vec{p}{}_{g3}^{\,\,2}}{z}+\frac{\vec{p}{}_{q1}^{\,\,2}%
}{x_{q}}+\frac{\vec{p}{}_{{\bar{q}}2}^{\,\,2}}{x_{\bar{q}}}\right)  \left(
Q^{2}+\frac{\vec{p}{}_{g3^{\prime}}^{\,\,2}}{z}+\frac{\vec{p}{}_{q1^{\prime}%
}^{\,\,2}}{x_{q}}+\frac{\vec{p}{}_{{\bar{q}}2^{\prime}}^{\,\,2}}{x_{\bar{q}}%
}\right)  }\nonumber\\
\times &  \left(  \frac{z\left(  (\vec{P}\vec{p}_{q1})G_{\bot}^{i}-(\vec
{G}\vec{p}_{q1})P_{\bot}^{i}\right)  \left(  dz+4x_{q}-4\right)  -(\vec{G}%
\vec{P})p_{q1}^{i}{}_{\bot}\left(  2x_{q}-1\right)  \left(  4\left(
x_{q}-1\right)  x_{\bar{q}}-dz^{2}\right)  }{z^{2}x_{\bar{q}}\left(
z+x_{\bar{q}}\right)  {}^{3}\left(  Q^{2}+\frac{\vec{p}{}_{q1}^{\,\,2}}%
{x_{q}\left(  z+x_{\bar{q}}\right)  }\right)  \left(  Q^{2}+\frac{\vec{p}%
{}_{q1^{\prime}}^{\,\,2}}{x_{q}\left(  z+x_{\bar{q}}\right)  }\right)
}\right. \nonumber\\
+  &  \frac{z\left(  (\vec{P}\vec{p}_{q1})H_{\bot}^{i}-(\vec{H}\vec{p}%
_{q1})P_{\bot}^{i}\right)  \left(  dz+4x_{q}-2\right)  -(\vec{H}\vec{P}%
)p_{q1}^{i}{}_{\bot}\left(  2x_{q}-1\right)  \left(  z(2-dz)+4x_{q}x_{\bar{q}%
}\right)  }{z^{2}x_{q}\left(  z+x_{q}\right)  \left(  z+x_{\bar{q}}\right)
{}^{2}\left(  Q^{2}+\frac{\vec{p}{}_{\bar{q}2^{\prime}}^{\,\,2}}{\left(
z+x_{q}\right)  x_{\bar{q}}}\right)  \left(  Q^{2}+\frac{\vec{p}{}%
_{q1}^{\,\,2}}{x_{q}\left(  z+x_{\bar{q}}\right)  }\right)  }\nonumber\\
+  &  \left.  \frac{H_{\bot}^{i}\left(  z(zd+d-2)+x_{q}\left(  2-4x_{\bar{q}%
}\right)  \right)  x_{\bar{q}}}{z\left(  z+x_{q}\right)  {}^{2}\left(
z+x_{\bar{q}}\right)  \left(  Q^{2}+\frac{\vec{p}{}_{\bar{q}2^{\prime}%
}^{\,\,2}}{\left(  z+x_{q}\right)  x_{\bar{q}}}\right)  }\right)
+(q\leftrightarrow\bar{q}).
\end{align}
Here%
\begin{equation}
G_{\bot}^{i}=x_{\bar{q}}p_{g3^{\prime}\bot}^{i}-zp_{\bar{q}2^{\prime}\bot}%
^{i},\quad H_{\bot}^{i}=x_{q}p_{g3^{\prime}\bot}^{i}-zp_{q1^{\prime}\bot}%
^{i},\quad P_{\bot}^{i}=x_{\bar{q}}p_{g3\bot}^{i}-zp_{\bar{q}2\bot}^{i}.
\end{equation}
Similarly to the longitudinal to longitudinal photon transition, we write
\begin{equation}
\Phi_{3}^{i}(p_{1\bot},p_{2\bot})\Phi_{3}^{+}(p_{1\bot}^{\prime},p_{2\bot
}^{\prime})^{\ast}=\Phi_{4}^{i}(p_{1\bot},p_{2\bot},0)\Phi_{4}^{+}(p_{1\bot
}^{\prime},p_{2\bot}^{\prime},0)^{\ast}+A^{i+}+B^{i+} \, ,
\end{equation}%
where $A$ and $B$ are now given by
\begin{align}
&  A^{i+}=\frac{4p_{\gamma}^{+}{}^{3}x_{q}}{\vec{\Delta}{}_{q}^{2}\vec{\Delta
}{}_{\bar{q}}^{2}\left(  x_{q}+z\right)  {}^{2}\left(  x_{\bar{q}}+z\right)
\left(  \frac{\vec{p}{}_{\bar{q}2}^{\,\,2}}{x_{\bar{q}}\left(  x_{q}+z\right)
}+Q^{2}\right)  \left(  \frac{\vec{p}{}_{q1^{\prime}}^{\,\,2}}{x_{q}\left(
x_{\bar{q}}+z\right)  }+Q^{2}\right)  }\nonumber\\
&  \times\left(  z\left(  4x_{\bar{q}}+dz-2\right)  \left(  \Delta_{q}^{i}%
{}_{\bot}(\vec{p}_{\bar{q}2}\vec{\Delta}_{\bar{q}})-\Delta_{\bar{q}}^{i}%
{}_{\bot}(\vec{p}_{\bar{q}2}\vec{\Delta}_{q})\right)  +\left(  2x_{\bar{q}%
}-1\right)  (\vec{\Delta}_{q}\vec{\Delta}_{\bar{q}})p_{\bar{q}2}^{i}{}_{\bot
}\left(  4x_{q}x_{\bar{q}}+z(2-dz)\right)  \right) \nonumber\\
&  -\frac{4p_{\gamma}^{+}{}^{3}x_{q}\left(  2x_{\bar{q}}-1\right)  \left(
dz^{2}+4x_{q}\left(  x_{q}+z\right)  \right)  p_{\bar{q}2}^{i}{}_{\bot}}%
{\vec{\Delta}{}_{q}^{2}\left(  x_{q}+z\right)  {}^{3}\left(  \frac{\vec{p}%
{}_{\bar{q}2}^{\,\,2}}{x_{\bar{q}}\left(  x_{q}+z\right)  }+Q^{2}\right)
\left(  \frac{\vec{p}{}_{\bar{q}2^{\prime}}^{\,\,2}}{x_{\bar{q}}\left(
x_{q}+z\right)  }+Q^{2}\right)  }+(q\leftrightarrow\bar{q}).
\end{align}

\begin{align}
&  B^{i+}=4p_{\gamma}^{+}{}^{3}\left(  \frac{\Delta_{q}{}_{\bot}^{i}%
x_{q}x_{\bar{q}}\left(  dz^{2}+dz-2z+2x_{q}-4x_{q}x_{\bar{q}}\right)  }%
{\vec{\Delta}{}_{q}^{2}\left(  z+x_{q}\right)  {}^{2}\left(  z+x_{\bar{q}%
}\right)  \left(  Q^{2}+\frac{\vec{p}_{g}^{\,\,2}}{z}+\frac{\vec{p}{}%
_{q1}^{\,\,2}}{x_{q}}+\frac{\vec{p}{}_{\bar{q}2}^{\,\,2}}{x_{\bar{q}}}\right)
\left(  Q^{2}+\frac{\vec{p}{}_{\bar{q}2^{\prime}}^{\,\,2}}{\left(
z+x_{q}\right)  x_{\bar{q}}}\right)  }\right. \nonumber\\
&  -\frac{(\vec{J}\vec{\Delta}_{q})p_{\bar{q}2}^{i}{}_{\bot}\left(
dz^{2}+4x_{q}\left(  z+x_{q}\right)  \right)  \left(  1-2x_{\bar{q}}\right)
+z\left(  (\vec{J}\vec{p}_{\bar{q}2})\Delta_{q}^{i}{}_{\bot}-(\vec{p}_{\bar
{q}2}\vec{\Delta}_{q})J_{\bot}^{i}\right)  \left(  dz+4x_{\bar{q}}-4\right)
}{z\left(  z+x_{q}\right)  {}^{3}\vec{\Delta}{}_{q}^{2}\left(  Q^{2}%
+\frac{\vec{p}_{g}^{\,\,2}}{z}+\frac{\vec{p}{}_{q1^{\prime}}^{\,\,2}}{x_{q}%
}+\frac{\vec{p}{}_{\bar{q}2^{\prime}}^{\,\,2}}{x_{\bar{q}}}\right)  \left(
Q^{2}+\frac{\vec{p}{}_{\bar{q}2}^{\,\,2}}{\left(  z+x_{q}\right)  x_{\bar{q}}%
}\right)  \left(  Q^{2}+\frac{\vec{p}{}_{\bar{q}2^{\prime}}^{\,\,2}}{\left(
z+x_{q}\right)  x_{\bar{q}}}\right)  }\nonumber\\
&  -\frac{x_{q}\left(  z\left(  (\vec{K}\vec{p}_{\bar{q}2})\Delta_{q}^{i}%
{}_{\bot}-(\vec{p}_{\bar{q}2}\vec{\Delta}_{q})K_{\bot}^{i}\right)  \left(
dz+4x_{\bar{q}}-2\right)  +(\vec{K}\vec{\Delta}_{q})p_{\bar{q}2}^{i}{}_{\bot
}\left(  1-2x_{\bar{q}}\right)  \left(  z(dz-2)-4x_{q}x_{\bar{q}}\right)
\right)  }{z\left(  z+x_{q}\right)  {}^{2}x_{\bar{q}}\left(  z+x_{\bar{q}%
}\right)  \vec{\Delta}{}_{q}^{2}\left(  Q^{2}+\frac{\vec{p}_{g}^{\,\,2}}%
{z}+\frac{\vec{p}{}_{q1^{\prime}}^{\,\,2}}{x_{q}}+\frac{\vec{p}{}_{\bar
{q}2^{\prime}}^{\,\,2}}{x_{\bar{q}}}\right)  \left(  Q^{2}+\frac{\vec{p}%
{}_{\bar{q}2}^{\,\,2}}{\left(  z+x_{q}\right)  x_{\bar{q}}}\right)  \left(
Q^{2}+\frac{\vec{p}{}_{q1^{\prime}}^{\,\,2}}{x_{q}\left(  z+x_{\bar{q}%
}\right)  }\right)  }\nonumber\\
&  -\frac{z\left(  (\vec{p}_{q1}\vec{\Delta}_{q})X_{\bot}^{i}-(\vec{X}\vec
{p}_{q1})\Delta_{q}^{i}{}_{\bot}\right)  \left(  dz+4x_{q}-2\right)  +(\vec
{X}\vec{\Delta}_{q})p_{q1}^{i}{}_{\bot}\left(  1-2x_{q}\right)  \left(
z(dz-2)-4x_{q}x_{\bar{q}}\right)  }{z\vec{\Delta}{}_{q}^{2}\left(
z+x_{q}\right)  \left(  z+x_{\bar{q}}\right)  {}^{2}\left(  Q^{2}+\frac
{\vec{p}_{g}^{\,\,2}}{z}+\frac{\vec{p}{}_{q1}^{\,\,2}}{x_{q}}+\frac{\vec{p}%
{}_{\bar{q}2}^{\,\,2}}{x_{\bar{q}}}\right)  \left(  Q^{2}+\frac{\vec{p}%
{}_{\bar{q}2^{\prime}}^{\,\,2}}{\left(  z+x_{q}\right)  x_{\bar{q}}}\right)
\left(  Q^{2}+\frac{\vec{p}{}_{q1}^{\,\,2}}{x_{q}\left(  z+x_{\bar{q}}\right)
}\right)  }\nonumber\\
&  +\left.  \frac{z\left(  (\vec{X}\vec{p}_{q1})\Delta_{\bar{q}}^{i}{}_{\bot
}-(\vec{p}_{q1}\vec{\Delta}_{\bar{q}})X_{\bot}^{i}\right)  \left(
dz+4x_{q}-4\right)  -(\vec{X}\vec{\Delta}_{\bar{q}})p_{q1}^{i}{}_{\bot}\left(
2x_{q}-1\right)  \left(  4\left(  x_{q}-1\right)  x_{\bar{q}}-dz^{2}\right)
}{z\left(  z+x_{\bar{q}}\right)  {}^{3}\vec{\Delta}{}_{\bar{q}}^{2}\left(
Q^{2}+\frac{\vec{p}_{g}^{\,\,2}}{z}+\frac{\vec{p}{}_{q1}^{\,\,2}}{x_{q}}%
+\frac{\vec{p}{}_{\bar{q}2}^{\,\,2}}{x_{\bar{q}}}\right)  \left(  Q^{2}%
+\frac{\vec{p}{}_{q1}^{\,\,2}}{x_{q}\left(  z+x_{\bar{q}}\right)  }\right)
\left(  Q^{2}+\frac{\vec{p}{}_{q1^{\prime}}^{\,\,2}}{x_{q}\left(  z+x_{\bar
{q}}\right)  }\right)  }\right) \nonumber\\
&  +(q\leftrightarrow\bar{q}).
\end{align}
Here we used the following variables :
\begin{align}
X_{\bot}^{i}  &  =x_{\bar{q}}p_{g\bot}^{i}-zp_{\bar{q}2\bot}^{i}=P_{\bot}%
^{i}|_{p_{3}=0},\quad J_{\bot}^{i}=x_{q}p_{g\bot}^{i}-zp_{q1^{\prime}\bot}%
^{i}=H_{\bot}^{i}|_{p_{3}^{\prime}=0},\nonumber\\
K_{\bot}^{i}  &  =x_{\bar{q}}p_{g\bot}^{i}-zp_{\bar{q}2^{\prime}\bot}%
^{i}=G_{\bot}^{i}|_{p_{3}^{\prime}=0}.
\end{align}
Similarly to the case of a longitudinal photon, in the collinear region {\it i.e.} when $\Delta
_{q}\rightarrow0$ only the last line of $A^{i+}$ gives a divergent contribution, which will be
proportional to the square of the Born impact factor. Using the jet
variables (\ref{jetcol1}), (\ref{jetcol2})\ we write
\begin{equation}
\Phi_{3}^{i}(p_{1\bot},p_{2\bot})\Phi_{3}^{k}(p_{1\bot}^{\prime},p_{2\bot
}^{\prime})^{\ast}=\Phi_{0}^{i}\Phi_{0}^{k\ast}\frac{1}{4}\frac{(x_{j}%
-z){}\left(  dz^{2}+4x_{j}\left(  x_{j}-z\right)  \right)  }{x_{j}^{3}%
\Delta_{q}^{2}}+O\left(  \Delta_{q}^{-1}\right)  .
\end{equation}
Again the integrand of $n_{j}$ appears as in (\ref{CollinearFactor}).
In the soft gluon region we can also write, using (\ref{p=zu}) :
\begin{equation}
\Phi_{3}^{i}(p_{1\bot},p_{2\bot})\Phi_{3}^{+}(p_{1\bot}^{\prime},p_{2\bot
}^{\prime})^{\ast}|_{z\rightarrow0}=\frac{1}{z^{2}}\Phi_{0}^{i}\Phi_{0}%
^{+\ast}\frac{(\frac{\vec{p}_{q}}{x_{q}}-\frac{\vec{p}_{\bar{q}}}{x_{\bar{q}}%
})^{2}}{(\vec{u}-\frac{\vec{p}_{q}}{x_{q}})^{2}(\vec{u}-\frac{\vec{p}_{\bar
{q}}}{x_{\bar{q}}})^{2}}+O(z^{-1}),
\end{equation}
which is the soft gluon emission factor, coming from $A^{i+}$.

\subsection{TT photon transition}%

\begin{align}
&  \Phi_{4}^{i}(p_{1\bot},p_{2\bot},p_{3\bot})\Phi_{4}^{k}(p_{1\bot}^{\prime
},p_{2\bot}^{\prime},p_{3\bot}^{\prime})^{\ast}=\left(  \frac{p_{\gamma}^{+}%
{}^{2}}{\left(  Q^{2}+\frac{\vec{p_{g3}}{}^{2}}{z}+\frac{\vec{p_{q1}}{}^{2}%
}{x_{q}}+\frac{\vec{p}{}_{\bar{q}2}^{\,\,2}}{x_{\bar{q}}}\right)  \left(
Q^{2}+\frac{\vec{p_{g3^{\prime}}}{}^{2}}{z}+\frac{\vec{p_{q1^{\prime}}}{}^{2}%
}{x_{q}}+\frac{\vec{p}{}_{\bar{q}2^{\prime}}^{\,\,2}}{x_{\bar{q}}}\right)
}\right. \nonumber\\
&  \times\left[  -\frac{g_{\bot}^{ik}x_{q}x_{\bar{q}}\left(  zd+d-2+2x_{\bar
{q}}\right)  }{\left(  z+x_{q}\right)  {}^{2}\left(  z+x_{\bar{q}}\right)
}-\frac{2P_{\bot}^{k}p_{q1\bot}{}^{i}\left(  1-2x_{q}\right)  }{z\left(
z+x_{\bar{q}}\right)  {}^{2}\left(  Q^{2}+\frac{\vec{p}_{q1}{}^{2}}%
{x_{q}\left(  z+x_{\bar{q}}\right)  }\right)  }\left(  \frac{(d-2)z-2x_{\bar
{q}}}{z+x_{\bar{q}}}+\frac{dz+2x_{\bar{q}}}{z+x_{q}}\right)  \right.
\nonumber\\
&  -\frac{2\left(  g_{\bot}^{ik}(\vec{P}\vec{p}_{q1})+P_{\bot}^{i}p_{q1\bot}%
{}^{k}\right)  }{z\left(  z+x_{\bar{q}}\right)  {}^{2}\left(  Q^{2}+\frac
{\vec{p}_{q1}{}^{2}}{x_{q}\left(  z+x_{\bar{q}}\right)  }\right)  }\left(
\frac{(d-4)z-2x_{\bar{q}}}{z+x_{q}}+\frac{(d-2)z-2x_{\bar{q}}}{z+x_{\bar{q}}%
}\right) \nonumber
\end{align}%
\begin{align}
&  -\frac{1}{z^{2}x_{q}\left(  z+x_{q}\right)  {}^{2}x_{\bar{q}}\left(
z+x_{\bar{q}}\right)  {}^{2}\left(  Q^{2}+\frac{\vec{p}{}_{\bar{q}2^{\prime}%
}^{\,\,2}}{\left(  z+x_{q}\right)  x_{\bar{q}}}\right)  \left(  Q^{2}%
+\frac{\vec{p}_{q1}{}^{2}}{x_{q}\left(  z+x_{\bar{q}}\right)  }\right)
}\left\{  (\vec{H}\vec{P})\left[  p_{q1\bot}^{i}{}p_{\bar{q}2^{\prime}\bot
}^{k}{}\left(  1-2x_{q}\right)  \right.  \right. \nonumber\\
&  \times\left.  \left(  1-2x_{\bar{q}}\right)  \left(  z(2-dz)+4x_{q}%
x_{\bar{q}}\right)  +(g_{\bot}^{ik}(\vec{p}_{q1}\vec{p}_{\bar{q}2^{\prime}%
})+p_{q1\bot}{}^{k}p_{\bar{q}2^{\prime}\bot}{}^{i})\left(  z(2-(d-4)z)+4x_{q}%
x_{\bar{q}}\right)  \right] \nonumber\\
&  +((d-4)z-2)\left[  z(\vec{H}\vec{p}_{\bar{q}2^{\prime}})(g_{\bot}^{ik}%
(\vec{P}\vec{p}_{q1})+P_{\bot}^{i}p_{q1\bot}{}^{k})+zH_{\bot}^{k}\left(
(\vec{P}\vec{p}_{q1})p_{\bar{q}2^{\prime}\bot}{}^{i}-(\vec{p}_{q1}\vec
{p}_{\bar{q}2^{\prime}})P_{\bot}^{i}\right)  \right] \nonumber\\
&  +((d-4)z+2)\left[  zH^{i}\left(  (\vec{P}\vec{p}_{\bar{q}2^{\prime}%
})p_{q1\bot}{}^{k}-(\vec{p}_{q1}\vec{p}_{\bar{q}2^{\prime}})P_{\bot}%
^{k}\right)  +z(\vec{H}\vec{p}_{q1})(g_{\bot}^{ik}(\vec{P}\vec{p}_{\bar
{q}2^{\prime}})+P_{\bot}^{k}p_{\bar{q}2^{\prime}\bot}{}^{i})\right]
\nonumber\\
&  +\left.  2z\left(  (\vec{H}\vec{p}_{\bar{q}2^{\prime}})P_{\bot}^{k}%
-(\vec{P}\vec{p}_{\bar{q}2^{\prime}})H_{\bot}^{k}\right)  p_{q1\bot}{}%
^{i}\left(  1-2x_{q}\right)  \left(  dz+4x_{\bar{q}}-2\right)  \right\}
\nonumber
\end{align}%
\begin{align}
&  -\frac{1}{z^{2}x_{q}x_{\bar{q}}\left(  z+x_{\bar{q}}\right)  {}^{4}\left(
Q^{2}+\frac{\vec{p}_{q1}{}^{2}}{x_{q}\left(  z+x_{\bar{q}}\right)  }\right)
\left(  Q^{2}+\frac{\vec{p}_{q1^{\prime}}{}^{2}}{x_{q}\left(  z+x_{\bar{q}%
}\right)  }\right)  }\left\{  z\left(  (d-4)z-4x_{\bar{q}}\right)  \frac{{}%
}{{}}\right. \nonumber\\
&  \times\left[  g_{\bot}^{ik}\left(  (\vec{G}\vec{p}_{q1^{\prime}})(\vec
{P}\vec{p}_{q1})-(\vec{G}\vec{p}_{q1})(\vec{P}\vec{p}_{q1^{\prime}})\right)
+(\vec{p}_{q1}\vec{p}_{q1^{\prime}})\left(  G_{\bot}^{i}P_{\bot}^{k}-G_{\bot
}^{k}P_{\bot}^{i}\right)  \right. \nonumber\\
&  +\left.  2(\vec{G}\vec{p}_{q1^{\prime}})\left(  P_{\bot}^{i}p_{q1\bot}%
{}^{k}+P_{\bot}^{k}p_{q1\bot}{}^{i}\left(  1-2x_{q}\right)  \right)
-2(\vec{G}\vec{p}_{q1})\left(  P_{\bot}^{k}p_{q1^{\prime}\bot}{}^{i}+P_{\bot
}^{i}p_{q1^{\prime}\bot}{}^{k}\left(  1-2x_{q}\right)  \right)  \right]
\nonumber\\
&  +\left.  \left.  (\vec{G}\vec{P})\left[  p_{q1\bot}{}^{k}p_{q1^{\prime}%
\bot}{}^{i}-p_{q1\bot}{}^{i}p_{q1^{\prime}\bot}{}^{k}\left(  1-2x_{q}\right)
{}^{2}+g_{\bot}^{ik}\left(  \vec{p}_{q1}\vec{p}_{q1^{\prime}}\right)  \right]
\left(  dz^{2}+4x_{\bar{q}}\left(  z+x_{\bar{q}}\right)  \right)  \right\}
\right] \nonumber\\
&  +\left.  \frac{{}}{{}}(1\leftrightarrow1^{\prime},2\leftrightarrow
2^{\prime},3\leftrightarrow3^{\prime},i\leftrightarrow k)\right)
+(q\leftrightarrow\bar{q}).
\end{align}
Once more we write
\begin{equation}
\Phi_{3}^{i}(p_{1\bot},p_{2\bot})\Phi_{3}^{k}(p_{1\bot}^{\prime},p_{2\bot
}^{\prime})^{\ast}=\Phi_{4}^{i}(p_{1\bot},p_{2\bot},0)\Phi_{4}^{k}(p_{1\bot
}^{\prime},p_{2\bot}^{\prime},0)^{\ast}+A^{ik}+B^{ik}.
\end{equation}%
Then
\begin{align}
A^{ik}=  &  \frac{-2p_{\gamma}^{+}{}^{2}}{\vec{\Delta}{}_{q}^{2}\vec{\Delta}%
{}_{\bar{q}}^{2}\left(  x_{q}+z\right)  {}^{2}\left(  x_{\bar{q}}+z\right)
{}^{2}\left(  \frac{\vec{p}{}_{\bar{q}2}^{\,\,2}}{x_{\bar{q}}\left(
x_{q}+z\right)  }+Q^{2}\right)  \left(  \frac{\vec{p}{}_{q1^{\prime}}^{\,\,2}%
}{x_{q}\left(  x_{\bar{q}}+z\right)  }+Q^{2}\right)  }\left\{
z((d-4)z-2)\frac{{}}{{}}\right. \nonumber\\
\times &  \left[  (\vec{p}_{q1^{\prime}}\vec{\Delta}_{\bar{q}})\left(
(\vec{p}_{\bar{q}2}\vec{\Delta}_{q})g_{\bot}^{ik}+\Delta_{q\bot}^{i}{}%
p_{\bar{q}2\bot}^{k}{}\right)  -(\vec{\Delta}_{q}\vec{\Delta}_{\bar{q}%
})\left(  (\vec{p}_{q1^{\prime}}\vec{p}_{\bar{q}2})g_{\bot}^{ik}%
+p_{q1^{\prime}\bot}^{i}{}p_{\bar{q}2\bot}^{k}{}\right)  \right. \nonumber\\
+  &  \left.  \Delta_{\bar{q}\bot}^{k}{}p_{q1^{\prime}\bot}^{i}{}(\vec
{p}_{\bar{q}2}\vec{\Delta}_{q})-\Delta_{q\bot}^{i}{}\Delta_{\bar{q}\bot}^{k}%
{}(\vec{p}_{q1^{\prime}}\vec{p}_{\bar{q}2})\right]  +(\vec{\Delta}_{q}%
\vec{\Delta}_{\bar{q}})\nonumber\\
\times &  \left[  \left(  2x_{q}-1\right)  \left(  2x_{\bar{q}}-1\right)
p_{q1^{\prime}\bot}^{k}{}p_{\bar{q}2\bot}^{i}{}\left(  4x_{q}x_{\bar{q}%
}+z(2-dz)\right)  +4x_{q}x_{\bar{q}}\left(  (\vec{p}_{q1^{\prime}}\vec
{p}_{\bar{q}2})g_{\bot}^{ik}+p_{q1^{\prime}\bot}^{i}{}p_{\bar{q}2\bot}^{k}%
{}\right)  \right] \nonumber\\
+  &  \left(  (\vec{p}_{q1^{\prime}}\vec{\Delta})\left(  (\vec{p}_{\bar{q}%
2}\vec{\Delta}_{\bar{q}})g_{\bot}^{ik}+\Delta_{\bar{q}\bot}^{i}{}p_{\bar
{q}2\bot}^{k}{}\right)  +\Delta_{q\bot}^{k}{}p_{q1^{\prime}\bot}^{i}{}(\vec
{p}_{\bar{q}2}\vec{\Delta}_{\bar{q}})-\Delta_{q\bot}^{k}{}\Delta_{\bar{q}\bot
}^{i}(\vec{p}_{q1^{\prime}}\vec{p}_{\bar{q}2}){}\right) \nonumber\\
\times &  z((d-4)z+2)+z\left(  2x_{\bar{q}}-1\right)  \left(  dz+4x_{q}%
-2\right)  p_{\bar{q}2\bot}^{i}{}\left(  \Delta_{\bar{q}\bot}^{k}{}(\vec
{p}_{q1^{\prime}}\vec{\Delta}_{q})-\Delta_{q\bot}^{k}{}(\vec{p}_{q1^{\prime}%
}\vec{\Delta}_{\bar{q}})\right) \nonumber\\
+  &  \left.  z\left(  2x_{q}-1\right)  p_{q1^{\prime}\bot}^{k}{}\left(
4x_{\bar{q}}+dz-2\right)  \left(  \Delta_{q\bot}^{i}{}(\vec{p}_{\bar{q}2}%
\vec{\Delta}_{\bar{q}})-\Delta_{\bar{q}\bot}^{i}{}(\vec{p}_{\bar{q}2}%
\vec{\Delta}_{q})\right)  \right\} \nonumber\\
-  &  \frac{2x_{q}p_{\gamma}^{+}{}^{2}\left(  dz^{2}+4x_{q}\left(
x_{q}+z\right)  \right)  \left(  (\vec{p}_{\bar{q}2}\vec{p}_{\bar{q}2^{\prime
}})g_{\bot}^{ik}-\left(  1-2x_{\bar{q}}\right)  {}^{2}p_{\bar{q}2\bot}^{i}%
{}p_{\bar{q}2^{\prime}\bot}^{k}{}+p_{\bar{q}2^{\prime}\bot}^{i}{}p_{\bar
{q}2\bot}^{k}{}\right)  }{x_{\bar{q}}\vec{\Delta}{}_{q}^{2}\left(
x_{q}+z\right)  {}^{4}\left(  \frac{\vec{p}{}_{\bar{q}2}^{\,\,2}}{x_{\bar{q}%
}\left(  x_{q}+z\right)  }+Q^{2}\right)  \left(  \frac{\vec{p}{}_{\bar
{q}2^{\prime}}^{\,\,2}}{x_{\bar{q}}\left(  x_{q}+z\right)  }+Q^{2}\right)
}+(q\leftrightarrow\bar{q}).
\end{align}
\begin{align*}
B^{ik}  &  =\left(  \frac{2p_{\gamma}^{+}{}^{2}}{\vec{\Delta}{}_{q}^{2}\left(
Q^{2}+\frac{\vec{p}_{g}^{\,\,2}}{z}+\frac{\vec{p}{}_{q1}^{\,\,2}}{x_{q}}%
+\frac{\vec{p}{}_{\bar{q}2}^{\,\,2}}{x_{\bar{q}}}\right)  \left(  Q^{2}%
+\frac{\vec{p}{}_{\bar{q}2^{\prime}}^{\,\,2}}{\left(  z+x_{q}\right)
x_{\bar{q}}}\right)  }\right. \\
&  \times\left[  \frac{\left(  (d-2)z-2x_{q}\right)  x_{q}}{\left(
z+x_{q}\right)  {}^{3}}\left(  g_{\bot}^{ik}(\vec{p}_{\bar{q}2^{\prime}}%
\vec{\Delta}_{q})+p_{\bar{q}2^{\prime}}{}_{\bot}^{i}\Delta_{q\bot}{}%
^{k}+p_{\bar{q}2^{\prime}\bot}{}^{k}\Delta_{q\bot}{}^{i}\left(  1-2x_{\bar{q}%
}\right)  \right)  \right. \\
&  +\frac{x_{q}\left(  \left(  (d-4)z-2x_{q}\right)  \left(  g_{\bot}%
^{ik}(\vec{p}_{\bar{q}2^{\prime}}\vec{\Delta}_{q})+p_{\bar{q}2^{\prime}\bot
}^{i}{}\Delta_{q\bot}^{k}{}\right)  +p_{\bar{q}2^{\prime}\bot}^{k}{}%
\Delta_{q\bot}^{i}{}\left(  dz+2x_{q}\right)  \left(  1-2x_{\bar{q}}\right)
\right)  }{\left(  z+x_{q}\right)  {}^{2}\left(  z+x_{\bar{q}}\right)  } \\
&   -\frac{1}{z\left(  z+x_{q}\right)  {}^{2}x_{\bar{q}}\left(  z+x_{\bar{q}%
}\right)  {}^{2}\left(  Q^{2}+\frac{\vec{p_{q1}}{}^{2}}{x_{q}\left(
z+x_{\bar{q}}\right)  }\right)  }\left\{  z((d-4)z+2)\frac{{}}{{}}\right. \\
&  \times\left[  p_{q1}{}_{\bot}^{i}\left(  (\vec{p}_{\bar{q}2^{\prime}}%
\vec{\Delta}_{q})X_{\bot}^{k}-(\vec{X}\vec{p}_{\bar{q}2^{\prime}})\Delta_{q}%
{}_{\bot}^{k}\right)  \left(  2x_{q}-1\right)  -(\vec{X}\vec{p}_{\bar
{q}2^{\prime}})\left(  g_{\bot}^{ik}(\vec{p}_{q1}\vec{\Delta}_{q})+p_{q1}%
{}_{\bot}^{k}\Delta_{q}{}_{\bot}^{i}\right)  \right. \\
&  -\left.  X_{\bot}^{k}\left(  (\vec{p}_{q1}\vec{\Delta}_{q})p_{\bar
{q}2^{\prime}}{}_{\bot}^{i}-(\vec{p}_{q1}\vec{p}_{\bar{q}2^{\prime}}%
)\Delta_{q}{}_{\bot}^{i}\right)  \right]  +4x_{q}z\left(  1-2x_{q}\right)
p_{q1}{}_{\bot}^{i}\left(  (\vec{p}_{\bar{q}2^{\prime}}\vec{\Delta}%
_{q})X_{\bot}^{k}-(\vec{X}\vec{p}_{\bar{q}2^{\prime}})\Delta_{q}{}_{\bot}%
^{k}\right) \\
&  +z\left(  1-2x_{\bar{q}}\right)  \left(  dz+4x_{q}-2\right)  p_{\bar
{q}2^{\prime}}{}_{\bot}^{k}\left(  (\vec{p}_{q1}\vec{\Delta}_{q})X_{\bot}%
^{i}-(\vec{X}\vec{p}_{q1})\Delta_{q}{}_{\bot}^{i}\right)  -z((d-4)z-2)\\
&  \times\left[  \left(  g_{\bot}^{ik}(\vec{X}\vec{p}_{q1})+X_{\bot}^{i}%
p_{q1}{}_{\bot}^{k}\right)  (\vec{p}_{\bar{q}2^{\prime}}\vec{\Delta}%
_{q})+\left(  (\vec{X}\vec{p}_{q1})p_{\bar{q}2^{\prime}}{}_{\bot}^{i}-(\vec
{p}_{q1}\vec{p}_{\bar{q}2^{\prime}})X_{\bot}^{i}\right)  \Delta_{q}{}_{\bot
}^{k}\right] \\
&  +(\vec{X}\vec{\Delta}_{q})p_{q1}{}_{\bot}^{i}p_{\bar{q}2^{\prime}}{}_{\bot
}^{k}\left(  1-2x_{q}\right)  \left(  1-2x_{\bar{q}}\right)  \left(
z(dz-2)-4x_{q}x_{\bar{q}}\right) \\
&  -\left.  (\vec{X}\vec{\Delta}_{q})\left(  g_{\bot}^{ik}(\vec{p}_{q1}\vec
{p}_{\bar{q}2^{\prime}})+p_{q1}{}_{\bot}^{k}p_{\bar{q}2^{\prime}}{}_{\bot}%
^{i}\right)  \left(  z(2-(d-4)z)+4x_{q}x_{\bar{q}}\right)  \right\}\\
-  &  \frac{1}{z\left(  z+x_{q}\right)  {}^{4}\left(  Q^{2}+\frac{\vec{p}%
{}_{\bar{q}2}^{\,\,2}}{\left(  z+x_{q}\right)  x_{\bar{q}}}\right)  x_{\bar
{q}}}\left\{  z\left(  dz+4x_{\bar{q}}-4\right)  \left[  \left(  1-2x_{\bar
{q}}\right)  \frac{{}}{{}}\right.  \right. \nonumber\\
\times &  \left(  p_{\bar{q}2^{\prime}}{}_{\bot}^{k}\left(  (\vec{p}_{\bar
{q}2}\vec{\Delta}_{q})V_{\bot}^{i}-(\vec{V}\vec{p}_{\bar{q}2})\Delta_{q}%
{}_{\bot}^{i}\right)  +p_{\bar{q}2}{}_{\bot}^{i}\left(  (\vec{V}\vec{p}%
_{\bar{q}2^{\prime}})\Delta_{q}{}_{\bot}^{k}-(\vec{p}_{\bar{q}2^{\prime}}%
\vec{\Delta}_{q})V_{\bot}^{k}\right)  \right) \nonumber\\
+  &  V_{\bot}^{k}\left(  (\vec{p}_{\bar{q}2}\vec{\Delta}_{q})p_{\bar
{q}2^{\prime}}{}_{\bot}^{i}-(\vec{p}_{\bar{q}2}\vec{p}_{\bar{q}2^{\prime}%
})\Delta_{q}{}_{\bot}^{i}\right)  +\left(  (\vec{p}_{\bar{q}2}\vec{p}_{\bar
{q}2^{\prime}})V_{\bot}^{i}-(\vec{V}\vec{p}_{\bar{q}2})p_{\bar{q}2^{\prime}}%
{}_{\bot}^{i}\right)  \Delta_{q}{}_{\bot}^{k}\nonumber\\
+  &  \left.  g_{\bot}^{ik}\left(  (\vec{V}\vec{p}_{\bar{q}2^{\prime}}%
)(\vec{p}_{\bar{q}2}\vec{\Delta}_{q})-(\vec{V}\vec{p}_{\bar{q}2})(\vec
{p}_{\bar{q}2^{\prime}}\vec{\Delta}_{q})\right)  +p_{\bar{q}2}{}_{\bot}%
^{k}\left(  (\vec{V}\vec{p}_{\bar{q}2^{\prime}})\Delta_{q}{}_{\bot}^{i}%
-(\vec{p}_{\bar{q}2^{\prime}}\vec{\Delta}_{q})V_{\bot}^{i}\right)  \right]
\nonumber\\
+  &  \left.  \left.  (\vec{V}\vec{\Delta}_{q})\left(  p_{\bar{q}2}{}_{\bot
}^{i}p_{\bar{q}2^{\prime}}{}_{\bot}^{k}\left(  1-2x_{\bar{q}}\right)  {}%
^{2}-g_{\bot}^{ik}(\vec{p}_{\bar{q}2}\vec{p}_{\bar{q}2^{\prime}})-p_{\bar{q}%
2}{}_{\bot}^{k}p_{\bar{q}2^{\prime}}{}_{\bot}^{i}\right)  \left(
dz^{2}-4x_{q}\left(  x_{\bar{q}}-1\right)  \right)  \right\}  \frac{{}}{{}%
}\right] \nonumber\\
+  &  \left.  \frac{{}}{{}}(1\leftrightarrow1^{\prime},2\leftrightarrow
2^{\prime},i\leftrightarrow k)\right)  +(q\leftrightarrow\bar{q}).
\end{align*}
Here we introduced
\begin{equation}
V_{\bot}^{i}=x_{q}p_{g\bot}^{i}-zp_{q1\bot}^{i}.
\end{equation}
Once again in the collinear region, i.e. when $\Delta_{q}\rightarrow0$, only the
last line of $A^{ik}$ gives a divergent contribution, which is proportional to the
square of the Born impact factors. Using the jet variables (\ref{jetcol1}%
), (\ref{jetcol2})\ we obtain
\begin{equation}
\Phi_{3}^{i}(p_{1\bot},p_{2\bot})\Phi_{3}^{k}(p_{1\bot}^{\prime},p_{2\bot
}^{\prime})^{\ast}=\Phi_{0}^{i}\Phi_{0}^{k\ast}\frac{1}{4}\frac{(x_{j}%
-z){}\left(  dz^{2}+4x_{j}\left(  x_{j}-z\right)  \right)  }{x_{j}^{3}%
\Delta_{q}^{2}}+O\left(  \Delta_{q}^{-1}\right)  .
\end{equation}
Again the integrand of $n_{j}$ appears as in (\ref{CollinearFactor}). \\
In the soft gluon region we can also write, using (\ref{p=zu}) :
\begin{equation}
\Phi_{3}^{i}(p_{1\bot},p_{2\bot})\Phi_{3}^{k}(p_{1\bot}^{\prime},p_{2\bot
}^{\prime})^{\ast}|_{z\rightarrow0}=\frac{1}{z^{2}}\Phi_{0}^{i}\Phi_{0}%
^{k\ast}\frac{(\frac{\vec{p}_{q}}{x_{q}}-\frac{\vec{p}_{\bar{q}}}{x_{\bar{q}}%
})^{2}}{(\vec{u}-\frac{\vec{p}_{q}}{x_{q}})^{2}(\vec{u}-\frac{\vec{p}_{\bar
{q}}}{x_{\bar{q}}})^{2}}+O(z^{-1}),
\end{equation}
which is the soft gluon emission factor.

\providecommand{\href}[2]{#2}\begingroup\raggedright\endgroup


\begin{thebibliography}{10}

\bibitem{Wusthoff:1999cr}
M.~Wusthoff and A.~D. Martin, {\it {The QCD description of diffractive
  processes}},  {\em J. Phys.} {\bf G25} (1999) R309--R344,
  [\href{http://xxx.lanl.gov/abs/hep-ph/9909362}{{\tt hep-ph/9909362}}].

\bibitem{Wolf:2009jm}
G.~Wolf, {\it {Review of High Energy Diffraction in Real and Virtual Photon
  Proton scattering at HERA}},  {\em Rept. Prog. Phys.} {\bf 73} (2010) 116202,
  [\href{http://xxx.lanl.gov/abs/0907.1217}{{\tt arXiv:0907.1217}}].

\bibitem{Aktas:2006hx}
{\bf H1} Collaboration, A.~Aktas {\em et.~al.}, {\it {Diffractive
  deep-inelastic scattering with a leading proton at HERA}},  {\em Eur. Phys.
  J.} {\bf C48} (2006) 749--766,
  [\href{http://xxx.lanl.gov/abs/hep-ex/0606003}{{\tt hep-ex/0606003}}].

\bibitem{Aktas:2006hy}
{\bf H1} Collaboration, A.~Aktas {\em et.~al.}, {\it {Measurement and {QCD}
  analysis of the diffractive deep- inelastic scattering cross-section at
  HERA}},  {\em Eur. Phys. J.} {\bf C48} (2006) 715--748,
  [\href{http://xxx.lanl.gov/abs/hep-ex/0606004}{{\tt hep-ex/0606004}}].

\bibitem{Chekanov:2004hy}
{\bf ZEUS Collaboration} Collaboration, S.~Chekanov {\em et.~al.}, {\it
  {Dissociation of virtual photons in events with a leading proton at HERA}},
  {\em Eur. Phys. J.} {\bf C38} (2004) 43--67,
  [\href{http://xxx.lanl.gov/abs/hep-ex/0408009}{{\tt hep-ex/0408009}}].

\bibitem{Chekanov:2005vv}
{\bf ZEUS} Collaboration, S.~Chekanov {\em et.~al.}, {\it {Study of deep
  inelastic inclusive and diffractive scattering with the ZEUS forward plug
  calorimeter}},  {\em Nucl. Phys.} {\bf B713} (2005) 3--80,
  [\href{http://xxx.lanl.gov/abs/hep-ex/0501060}{{\tt hep-ex/0501060}}].

\bibitem{Aaron:2010aa}
F.~Aaron, C.~Alexa, V.~Andreev, S.~Backovic, A.~Baghdasaryan, {\em et.~al.},
  {\it {Measurement of the cross section for diffractive deep-inelastic
  scattering with a leading proton at HERA}},  {\em Eur. Phys. J.} {\bf C71}
  (2011) 1578, [\href{http://xxx.lanl.gov/abs/1010.1476}{{\tt
  arXiv:1010.1476}}].

\bibitem{Aaron:2012ad}
{\bf H1 Collaboration} Collaboration, F.~Aaron {\em et.~al.}, {\it {Inclusive
  Measurement of Diffractive Deep-Inelastic Scattering at HERA}},  {\em Eur.
  Phys. J.} {\bf C72} (2012) 2074,
  [\href{http://xxx.lanl.gov/abs/1203.4495}{{\tt arXiv:1203.4495}}].

\bibitem{Chekanov:2008fh}
{\bf ZEUS Collaboration} Collaboration, S.~Chekanov {\em et.~al.}, {\it {Deep
  inelastic scattering with leading protons or large rapidity gaps at HERA}},
  {\em Nucl. Phys.} {\bf B816} (2009) 1--61,
  [\href{http://xxx.lanl.gov/abs/0812.2003}{{\tt arXiv:0812.2003}}].

\bibitem{Aaron:2012hua}
{\bf H1 Collaboration, ZEUS Collaboration} Collaboration, F.~Aaron {\em
  et.~al.}, {\it {Combined inclusive diffractive cross sections measured with
  forward proton spectrometers in deep inelastic $ep$ scattering at HERA}},
  {\em Eur. Phys. J.} {\bf C72} (2012) 2175,
  [\href{http://xxx.lanl.gov/abs/1207.4864}{{\tt arXiv:1207.4864}}].

\bibitem{Collins:1997sr}
J.~C. Collins, {\it {Proof of factorization for diffractive hard scattering}},
  {\em Phys. Rev.} {\bf D57} (1998) 3051--3056,
  [\href{http://xxx.lanl.gov/abs/hep-ph/9709499}{{\tt hep-ph/9709499}}].

\bibitem{Bartels:1998ea}
J.~Bartels, J.~R. Ellis, H.~Kowalski, and M.~Wusthoff, {\it An analysis of
  diffraction in deep-inelastic scattering},  {\em Eur. Phys. J.} {\bf C7}
  (1999) 443--458, [\href{http://xxx.lanl.gov/abs/hep-ph/9803497}{{\tt
  hep-ph/9803497}}].

\bibitem{Wusthoff:1995hd}
M.~Wusthoff, {\it {Photon diffractive dissociation in deep inelastic
  scattering}}, .

\bibitem{Gotsman:1996ix}
E.~Gotsman, E.~Levin, and U.~Maor, {\it {Diffractive leptoproduction of small
  masses in QCD}},  {\em Nucl. Phys.} {\bf B493} (1997) 354--396,
  [\href{http://xxx.lanl.gov/abs/hep-ph/9606280}{{\tt hep-ph/9606280}}].

\bibitem{Wusthoff:1997fz}
M.~Wusthoff, {\it {Large rapidity gap events in deep inelastic scattering}},
  {\em Phys. Rev.} {\bf D56} (1997) 4311--4321,
  [\href{http://xxx.lanl.gov/abs/hep-ph/9702201}{{\tt hep-ph/9702201}}].

\bibitem{Bartels:1999tn}
J.~Bartels, H.~Jung, and M.~Wusthoff, {\it {Quark - anti-quark gluon jets in
  DIS diffractive dissociation}},  {\em Eur. Phys. J.} {\bf C11} (1999)
  111--125, [\href{http://xxx.lanl.gov/abs/hep-ph/9903265}{{\tt
  hep-ph/9903265}}].

\bibitem{Bartels:2002ri}
J.~Bartels, H.~Jung, and A.~Kyrieleis, {\it {Massive c anti-c g: Calculation in
  diffractive DIS and diffractive D* production at HERA}},  {\em Eur. Phys. J.}
  {\bf C24} (2002) 555--560,
  [\href{http://xxx.lanl.gov/abs/hep-ph/0204269}{{\tt hep-ph/0204269}}].

\bibitem{Marquet:2007nf}
C.~Marquet, {\it {A Unified description of diffractive deep inelastic
  scattering with saturation}},  {\em Phys. Rev.} {\bf D76} (2007) 094017,
  [\href{http://xxx.lanl.gov/abs/0706.2682}{{\tt arXiv:0706.2682}}].

\bibitem{Balitsky:1995ub}
I.~Balitsky, {\it Operator expansion for high-energy scattering},  {\em Nucl.
  Phys.} {\bf B463} (1996) 99--160,
  [\href{http://xxx.lanl.gov/abs/hep-ph/9509348}{{\tt hep-ph/9509348}}].

\bibitem{Balitsky:1998kc}
I.~Balitsky, {\it Factorization for high-energy scattering},  {\em Phys. Rev.
  Lett.} {\bf 81} (1998) 2024--2027,
  [\href{http://xxx.lanl.gov/abs/hep-ph/9807434}{{\tt hep-ph/9807434}}].

\bibitem{Balitsky:1998ya}
I.~Balitsky, {\it Factorization and high-energy effective action},  {\em Phys.
  Rev.} {\bf D60} (1999) 014020,
  [\href{http://xxx.lanl.gov/abs/hep-ph/9812311}{{\tt hep-ph/9812311}}].

\bibitem{Balitsky:2001re}
I.~Balitsky, {\it Effective field theory for the small-x evolution},  {\em
  Phys. Lett.} {\bf B518} (2001) 235--242,
  [\href{http://xxx.lanl.gov/abs/hep-ph/0105334}{{\tt hep-ph/0105334}}].

\bibitem{Boussarie:2014lxa}
R.~Boussarie, A.~Grabovsky, L.~Szymanowski, and S.~Wallon, {\it {Impact factor
  for high-energy two and three jets diffractive production}},  {\em JHEP} {\bf
  1409} (2014) 026, [\href{http://xxx.lanl.gov/abs/1405.7676}{{\tt
  arXiv:1405.7676}}].

\bibitem{Boussarie:2015qet}
R.~Boussarie, A.~V. Grabovsky, L.~Szymanowski, and S.~Wallon, {\it {Diffractive
  production of jets at high-energy in the QCD shock-wave approach}},  in {\em
  {Photon 2015: International Conference on the Structure and Interactions of
  the Photon and the 21th International Workshop on Photon-Photon Collisions
  and International Workshop on High Energy Photon Linear Colliders
  Novosibirsk, Russia, June 15-19, 2015}}, 2015.
\newblock \href{http://xxx.lanl.gov/abs/1511.0278}{{\tt arXiv:1511.0278}}.

\bibitem{Boussarie:2015acw}
R.~Boussarie, A.~Grabovsky, L.~Szymanowski, and S.~Wallon, {\it {Photon
  dissociation into two and three jets: initial and final state corrections}},
  {\em Acta Phys. Polon. Supp.} {\bf 8} (2015) 897,
  [\href{http://xxx.lanl.gov/abs/1512.0077}{{\tt arXiv:1512.0077}}].

\bibitem{Fadin:1975cb}
V.~S. Fadin, E.~A. Kuraev, and L.~N. Lipatov, {\it {On the {P}omeranchuk
  Singularity in Asymptotically Free Theories}},  {\em Phys. Lett.} {\bf B60}
  (1975) 50--52.

\bibitem{Kuraev:1976ge}
E.~A. Kuraev, L.~N. Lipatov, and V.~S. Fadin, {\it {Multi - Reggeon Processes
  in the {Y}ang-{M}ills Theory}},  {\em Sov. Phys. JETP} {\bf 44} (1976)
  443--450.

\bibitem{Kuraev:1977fs}
E.~A. Kuraev, L.~N. Lipatov, and V.~S. Fadin, {\it {The Pomeranchuk Singularity
  in Nonabelian Gauge Theories}},  {\em Sov. Phys. JETP} {\bf 45} (1977)
  199--204.

\bibitem{Balitsky:1978ic}
I.~I. Balitsky and L.~N. Lipatov, {\it {The Pomeranchuk Singularity in Quantum
  Chromodynamics}},  {\em Sov. J. Nucl. Phys.} {\bf 28} (1978) 822--829.

\bibitem{Fadin:1998py}
V.~S. Fadin and L.~N. Lipatov, {\it {BFKL pomeron in the next-to-leading
  approximation}},  {\em Phys. Lett.} {\bf B429} (1998) 127--134,
  [\href{http://xxx.lanl.gov/abs/hep-ph/9802290}{{\tt hep-ph/9802290}}].

\bibitem{Ciafaloni:1998gs}
M.~Ciafaloni and G.~Camici, {\it {Energy scale(s) and next-to-leading BFKL
  equation}},  {\em Phys. Lett.} {\bf B430} (1998) 349--354,
  [\href{http://xxx.lanl.gov/abs/hep-ph/9803389}{{\tt hep-ph/9803389}}].

\bibitem{Kovchegov:1999yj}
Y.~V. Kovchegov, {\it {Small-$x$ $F_2$ structure function of a nucleus
  including multiple pomeron exchanges}},  {\em Phys. Rev.} {\bf D60} (1999)
  034008, [\href{http://xxx.lanl.gov/abs/hep-ph/9901281}{{\tt
  hep-ph/9901281}}].

\bibitem{Kovchegov:1999ua}
Y.~V. Kovchegov, {\it {Unitarization of the BFKL pomeron on a nucleus}},  {\em
  Phys. Rev.} {\bf D61} (2000) 074018,
  [\href{http://xxx.lanl.gov/abs/hep-ph/9905214}{{\tt hep-ph/9905214}}].

\bibitem{GolecBiernat:1998js}
K.~J. Golec-Biernat and M.~Wusthoff, {\it Saturation effects in deep inelastic
  scattering at low ${Q}^2$ and its implications on diffraction},  {\em Phys.
  Rev.} {\bf D59} (1999) 014017,
  [\href{http://xxx.lanl.gov/abs/hep-ph/9807513}{{\tt hep-ph/9807513}}].

\bibitem{GolecBiernat:1999qd}
K.~Golec-Biernat and M.~Wusthoff, {\it {Saturation in diffractive deep
  inelastic scattering}},  {\em Phys. Rev.} {\bf D60} (1999) 114023,
  [\href{http://xxx.lanl.gov/abs/hep-ph/9903358}{{\tt hep-ph/9903358}}].

\bibitem{Chirilli:2013kca}
G.~A. Chirilli and Y.~V. Kovchegov, {\it {Solution of the NLO BFKL Equation and
  a Strategy for Solving the All-Order BFKL Equation}},  {\em JHEP} {\bf 06}
  (2013) 055, [\href{http://xxx.lanl.gov/abs/1305.1924}{{\tt
  arXiv:1305.1924}}].

\bibitem{Grabovsky:2013gta}
A.~V. Grabovsky, {\it {On the solution to the NLO forward BFKL equation}},
  {\em JHEP} {\bf 09} (2013) 098,
  [\href{http://xxx.lanl.gov/abs/1307.3152}{{\tt arXiv:1307.3152}}].

\bibitem{Abramowicz:2015vnu}
{\bf ZEUS} Collaboration, H.~Abramowicz {\em et.~al.}, {\it {Production of
  exclusive dijets in diffractive deep inelastic scattering at HERA}},  {\em
  Eur. Phys. J.} {\bf C76} (2016), no.~1 16,
  [\href{http://xxx.lanl.gov/abs/1505.0578}{{\tt arXiv:1505.0578}}].

\bibitem{Caron-Huot:2013fea}
S.~Caron-Huot, {\it {When does the gluon reggeize?}},  {\em JHEP} {\bf 05}
  (2015) 093, [\href{http://xxx.lanl.gov/abs/1309.6521}{{\tt
  arXiv:1309.6521}}].

\bibitem{Mueller:1993rr}
A.~H. Mueller, {\it {Soft gluons in the infinite momentum wave function and the
  BFKL pomeron}},  {\em Nucl. Phys.} {\bf B415} (1994) 373--385.

\bibitem{Fadin:2011jg}
V.~Fadin, R.~Fiore, A.~Grabovsky, and A.~Papa, {\it {Connection between
  complete and Moebius forms of gauge invariant operators}},  {\em Nucl. Phys.}
  {\bf B856} (2012) 111--124, [\href{http://xxx.lanl.gov/abs/1109.6634}{{\tt
  arXiv:1109.6634}}].

\bibitem{Ivanov:2012ms}
D.~Y. Ivanov and A.~Papa, {\it {The next-to-leading order forward jet vertex in
  the small-cone approximation}},  {\em JHEP} {\bf 1205} (2012) 086,
  [\href{http://xxx.lanl.gov/abs/1202.1082}{{\tt arXiv:1202.1082}}].

\bibitem{ioffe2010quantum}
B.~L. Ioffe, V.~S. Fadin, and L.~N. Lipatov, {\em Quantum chromodynamics:
  Perturbative and nonperturbative aspects}, vol.~30.
\newblock Cambridge University Press, 2010.

\bibitem{Balitsky:2010ze}
I.~Balitsky and G.~A. Chirilli, {\it {Photon impact factor in the
  next-to-leading order}},  {\em Phys. Rev.} {\bf D83} (2011) 031502,
  [\href{http://xxx.lanl.gov/abs/1009.4729}{{\tt arXiv:1009.4729}}].

\bibitem{Balitsky:2012bs}
I.~Balitsky and G.~A. Chirilli, {\it {Photon impact factor and
  $k_T$-factorization for DIS in the next-to-leading order}},  {\em Phys. Rev.}
  {\bf D87} (2013) 014013, [\href{http://xxx.lanl.gov/abs/1207.3844}{{\tt
  arXiv:1207.3844}}].

\bibitem{Ivanov:2004pp}
D.~Y. Ivanov, M.~I. Kotsky, and A.~Papa, {\it The impact factor for the virtual
  photon to light vector meson transition},  {\em Eur. Phys. J.} {\bf C38}
  (2004) 195--213, [\href{http://xxx.lanl.gov/abs/hep-ph/0405297}{{\tt
  hep-ph/0405297}}].

\bibitem{Anikin:2009hk}
I.~V. Anikin, D.~Y. Ivanov, B.~Pire, L.~Szymanowski, and S.~Wallon, {\it {On
  the description of exclusive processes beyond the leading twist
  approximation}},  {\em Phys. Lett.} {\bf B682} (2010) 413--418,
  [\href{http://xxx.lanl.gov/abs/0903.4797}{{\tt arXiv:0903.4797}}].

\bibitem{Anikin:2009bf}
I.~V. Anikin, D.~Y. Ivanov, B.~Pire, L.~Szymanowski, and S.~Wallon, {\it {QCD
  factorization of exclusive processes beyond leading twist: $\gamma^*_T \to
  \rho_T$ impact factor with twist three accuracy}},  {\em Nucl. Phys.} {\bf
  B828} (2010) 1--68, [\href{http://xxx.lanl.gov/abs/0909.4090}{{\tt
  arXiv:0909.4090}}].

\bibitem{Wandzura:1977qf}
S.~Wandzura and F.~Wilczek, {\it {Sum Rules for Spin Dependent
  Electroproduction: Test of Relativistic Constituent Quarks}},  {\em Phys.
  Lett.} {\bf B72} (1977) 195--198.

\bibitem{Aktas:2006up}
{\bf H1 Collaboration} Collaboration, A.~Aktas {\em et.~al.}, {\it {Diffractive
  open charm production in deep-inelastic scattering and photoproduction at
  HERA}},  {\em Eur. Phys. J.} {\bf C50} (2007) 1--20,
  [\href{http://xxx.lanl.gov/abs/hep-ex/0610076}{{\tt hep-ex/0610076}}].

\bibitem{Colferai:2010wu}
D.~Colferai, F.~Schwennsen, L.~Szymanowski, and S.~Wallon, {\it {Mueller
  Navelet jets at LHC - complete NLL BFKL calculation}},  {\em JHEP} {\bf 12}
  (2010) 026, [\href{http://xxx.lanl.gov/abs/1002.1365}{{\tt
  arXiv:1002.1365}}].

\bibitem{Ducloue:2013hia}
B.~Duclou\'e, L.~Szymanowski, and S.~Wallon, {\it {Confronting Mueller-Navelet
  jets in NLL BFKL with LHC experiments at 7 TeV}},  {\em JHEP} {\bf 1305}
  (2013) 096, [\href{http://xxx.lanl.gov/abs/1302.7012}{{\tt
  arXiv:1302.7012}}].

\bibitem{Caporale:2012ih}
F.~Caporale, D.~Y. Ivanov, B.~Murdaca, and A.~Papa, {\it {Mueller-Navelet
  small-cone jets at LHC in next-to-leading BFKL}},  {\em Nucl. Phys.} {\bf
  B877} (2013) 73--94, [\href{http://xxx.lanl.gov/abs/1211.7225}{{\tt
  arXiv:1211.7225}}].

\bibitem{Caporale:2013uva}
F.~Caporale, B.~Murdaca, A.~Sabio~Vera, and C.~Salas, {\it {Scale choice and
  collinear contributions to Mueller-Navelet jets at LHC energies}},  {\em
  Nucl. Phys.} {\bf B875} (2013) 134--151,
  [\href{http://xxx.lanl.gov/abs/1305.4620}{{\tt arXiv:1305.4620}}].

\bibitem{Caporale:2015uva}
F.~Caporale, D.~{\relax Yu}. Ivanov, B.~Murdaca, and A.~Papa, {\it
  {Brodsky-Lepage-Mackenzie optimal renormalization scale setting for semihard
  processes}},  {\em Phys. Rev.} {\bf D91} (2015), no.~11 114009,
  [\href{http://xxx.lanl.gov/abs/1504.0647}{{\tt arXiv:1504.0647}}].

\bibitem{Celiberto:2015yba}
F.~G. Celiberto, D.~{\relax Yu}. Ivanov, B.~Murdaca, and A.~Papa, {\it
  {Mueller-Navelet jets at LHC: BFKL versus high-energy DGLAP}},  {\em Eur.
  Phys. J.} {\bf C75} (2015), no.~6 292,
  [\href{http://xxx.lanl.gov/abs/1504.0823}{{\tt arXiv:1504.0823}}].

\bibitem{Mueller:1986ey}
A.~H. Mueller and H.~Navelet, {\it {An Inclusive Minijet Cross-Section and the
  Bare Pomeron in QCD}},  {\em Nucl. Phys.} {\bf B282} (1987) 727.

\bibitem{Khachatryan:2016udy}
{\bf CMS} Collaboration, V.~Khachatryan {\em et.~al.}, {\it {Azimuthal
  decorrelation of jets widely separated in rapidity in pp collisions at
  $\sqrt{s} =$ 7 TeV}},  {\em Submitted to: JHEP} (2016)
  [\href{http://xxx.lanl.gov/abs/1601.0671}{{\tt arXiv:1601.0671}}].

\bibitem{Ducloue:2013bva}
B.~Duclou\'e, L.~Szymanowski, and S.~Wallon, {\it {Evidence for high-energy
  resummation effects in Mueller-Navelet jets at the LHC}},  {\em
  Phys.Rev.Lett.} {\bf 112} (2014) 082003,
  [\href{http://xxx.lanl.gov/abs/1309.3229}{{\tt arXiv:1309.3229}}].

\bibitem{Caporale:2014gpa}
F.~Caporale, D.~{\relax Yu}. Ivanov, B.~Murdaca, and A.~Papa, {\it
  {Mueller–Navelet jets in next-to-leading order BFKL: theory versus
  experiment}},  {\em Eur. Phys. J.} {\bf C74} (2014), no.~10 3084,
  [\href{http://xxx.lanl.gov/abs/1407.8431}{{\tt arXiv:1407.8431}}]. [Erratum:
  Eur. Phys. J.C75,no.11,535(2015)].

\bibitem{Aaron:2011mp}
{\bf H1 Collaboration} Collaboration, F.~Aaron {\em et.~al.}, {\it {Measurement
  of Dijet Production in Diffractive Deep-Inelastic Scattering with a Leading
  Proton at HERA}},  {\em Eur. Phys. J.} {\bf C72} (2012) 1970,
  [\href{http://xxx.lanl.gov/abs/1111.0584}{{\tt arXiv:1111.0584}}].

\bibitem{Chekanov:2007rh}
{\bf ZEUS Collaboration} Collaboration, S.~Chekanov {\em et.~al.}, {\it
  {Diffractive photoproduction of dijets in ep collisions at HERA}},  {\em Eur.
  Phys. J.} {\bf C55} (2008) 177--191,
  [\href{http://xxx.lanl.gov/abs/0710.1498}{{\tt arXiv:0710.1498}}].

\bibitem{Aaron:2010su}
{\bf H1 Collaboration} Collaboration, F.~Aaron {\em et.~al.}, {\it {Diffractive
  Dijet Photoproduction in ep Collisions at HERA}},  {\em Eur. Phys. J.} {\bf
  C70} (2010) 15--37, [\href{http://xxx.lanl.gov/abs/1006.0946}{{\tt
  arXiv:1006.0946}}].

\bibitem{Klasen:2004qr}
M.~Klasen and G.~Kramer, {\it {Factorization breaking in diffractive dijet
  photoproduction}},  {\em Eur. Phys. J.} {\bf C38} (2004) 93--104,
  [\href{http://xxx.lanl.gov/abs/hep-ph/0408203}{{\tt hep-ph/0408203}}].

\bibitem{Klasen:2008ah}
M.~Klasen and G.~Kramer, {\it {Review of factorization breaking in diffractive
  photoproduction of dijets}},  {\em Mod. Phys. Lett.} {\bf A23} (2008)
  1885--1907, [\href{http://xxx.lanl.gov/abs/0806.2269}{{\tt
  arXiv:0806.2269}}].

\bibitem{Boer:2011fh}
D.~Boer, M.~Diehl, R.~Milner, R.~Venugopalan, W.~Vogelsang, {\em et.~al.}, {\it
  {Gluons and the quark sea at high energies: Distributions, polarization,
  tomography}},  \href{http://xxx.lanl.gov/abs/1108.1713}{{\tt
  arXiv:1108.1713}}.

\bibitem{AbelleiraFernandez:2012cc}
{\bf LHeC Study Group} Collaboration, J.~Abelleira~Fernandez {\em et.~al.},
  {\it {A Large Hadron Electron Collider at CERN: Report on the Physics and
  Design Concepts for Machine and Detector}},  {\em J.Phys.} {\bf G39} (2012)
  075001, [\href{http://xxx.lanl.gov/abs/1206.2913}{{\tt arXiv:1206.2913}}].

\bibitem{Altinoluk:2015dpi}
T.~Altinoluk, N.~Armesto, G.~Beuf, and A.~H. Rezaeian, {\it {Diffractive Dijet
  Production in Deep Inelastic Scattering and Photon-Hadron Collisions in the
  Color Glass Condensate}},  {\em Phys. Lett.} {\bf B758} (2016) 373--383,
  [\href{http://xxx.lanl.gov/abs/1511.0745}{{\tt arXiv:1511.0745}}].

\bibitem{Hatta:2016dxp}
Y.~Hatta, B.-W. Xiao, and F.~Yuan, {\it {Probing the Small-$x$ Gluon Tomography
  in Correlated Hard Diffractive Dijet Production in DIS}},  {\em Phys. Rev.
  Lett.} {\bf 116} (2016), no.~20 202301,
  [\href{http://xxx.lanl.gov/abs/1601.0158}{{\tt arXiv:1601.0158}}].

\bibitem{Aaij:2013jxj}
{\bf LHCb} Collaboration, R.~Aaij {\em et.~al.}, {\it {Exclusive $J/\psi$ and
  $\psi$(2S) production in pp collisions at $ \sqrt{s} = 7$ TeV}},  {\em J.
  Phys.} {\bf G40} (2013) 045001,
  [\href{http://xxx.lanl.gov/abs/1301.7084}{{\tt arXiv:1301.7084}}].

\bibitem{Aaij:2014iea}
{\bf LHCb} Collaboration, R.~Aaij {\em et.~al.}, {\it {Updated measurements of
  exclusive $J/\psi$ and $\psi$(2S) production cross-sections in pp collisions
  at $\sqrt{s}=7$ TeV}},  {\em J. Phys.} {\bf G41} (2014) 055002,
  [\href{http://xxx.lanl.gov/abs/1401.3288}{{\tt arXiv:1401.3288}}].

\bibitem{Aaij:2015kea}
{\bf LHCb} Collaboration, R.~Aaij {\em et.~al.}, {\it {Measurement of the
  exclusive Υ production cross-section in pp collisions at $ \sqrt{s}=7 $ TeV
  and 8 TeV}},  {\em JHEP} {\bf 09} (2015) 084,
  [\href{http://xxx.lanl.gov/abs/1505.0813}{{\tt arXiv:1505.0813}}].

\bibitem{TheALICE:2014dwa}
{\bf ALICE} Collaboration, B.~B. Abelev {\em et.~al.}, {\it {Exclusive
  $\mathrm{J/}\psi$ photoproduction off protons in ultra-peripheral p-Pb
  collisions at $\sqrt{s_{\rm NN}}=5.02$ TeV}},  {\em Phys. Rev. Lett.} {\bf
  113} (2014), no.~23 232504, [\href{http://xxx.lanl.gov/abs/1406.7819}{{\tt
  arXiv:1406.7819}}].

\bibitem{Abbas:2013oua}
{\bf ALICE} Collaboration, E.~Abbas {\em et.~al.}, {\it {Charmonium and
  $e^+e^-$ pair photoproduction at mid-rapidity in ultra-peripheral Pb-Pb
  collisions at $\sqrt{s_{\rm NN}}$=2.76 TeV}},  {\em Eur. Phys. J.} {\bf C73}
  (2013), no.~11 2617, [\href{http://xxx.lanl.gov/abs/1305.1467}{{\tt
  arXiv:1305.1467}}].

\bibitem{Abelev:2012ba}
{\bf ALICE} Collaboration, B.~Abelev {\em et.~al.}, {\it {Coherent $J/\psi$
  photoproduction in ultra-peripheral Pb-Pb collisions at $\sqrt{s_{NN}} =
  2.76$ TeV}},  {\em Phys. Lett.} {\bf B718} (2013) 1273--1283,
  [\href{http://xxx.lanl.gov/abs/1209.3715}{{\tt arXiv:1209.3715}}].

\bibitem{Adam:2015sia}
{\bf ALICE} Collaboration, J.~Adam {\em et.~al.}, {\it {Coherent $\psi$(2S)
  photo-production in ultra-peripheral Pb Pb collisions at $\sqrt{s}_{\rm NN}$
  = 2.76 TeV}},  {\em Phys. Lett.} {\bf B751} (2015) 358--370,
  [\href{http://xxx.lanl.gov/abs/1508.0507}{{\tt arXiv:1508.0507}}].

\bibitem{Khachatryan:2016qhq}
{\bf CMS} Collaboration, V.~Khachatryan {\em et.~al.}, {\it {Coherent J/Psi
  photoproduction in ultra-peripheral PbPb collisions at sqrt(s[NN]) = 2.76 TeV
  with the CMS experiment}},  \href{http://xxx.lanl.gov/abs/1605.0696}{{\tt
  arXiv:1605.0696}}.

\bibitem{N.Cartiglia:2015gve}
{\bf LHC Forward Physics Working Group} Collaboration, e.~Royon, C. {\em
  et.~al.}, {\it {LHC Forward Physics}}, .

\bibitem{Guzey:2016tek}
V.~Guzey and M.~Klasen, {\it {Diffractive dijet photoproduction in
  ultraperipheral collisions at the LHC in next-to-leading order QCD}},  {\em
  JHEP} {\bf 04} (2016) 158, [\href{http://xxx.lanl.gov/abs/1603.0605}{{\tt
  arXiv:1603.0605}}].

\bibitem{Beuf:2016new}
G.~Beuf, {\it {in preparation}},  \href{http://xxx.lanl.gov/abs/1606.xxxx}{{\tt
  arXiv:1606.xxxx}}.

\bibitem{Beuf:2011xd}
G.~Beuf, {\it {NLO corrections for the dipole factorization of DIS structure
  functions at low x}},  {\em Phys. Rev.} {\bf D85} (2012) 034039,
  [\href{http://xxx.lanl.gov/abs/1112.4501}{{\tt arXiv:1112.4501}}].

\bibitem{Bogdan:2007qj}
A.~V. Bogdan and A.~V. Grabovsky, {\it {Radiative corrections to the Reggeized
  quark - Reggeized quark - gluon effective vertex}},  {\em Nucl. Phys.} {\bf
  B773} (2007) 65--83, [\href{http://xxx.lanl.gov/abs/hep-ph/0701144}{{\tt
  hep-ph/0701144}}].

\end{thebibliography}
\end{document}